\documentclass[letterpaper,12pt,leqno]{article}
\usepackage{paper}
\usepackage{bigfoot}
\DeclareRobustCommand{\tightrows}{\def\arraystretch{0.75}}
\DeclareRobustCommand{\fndouble}{\setstretch{1.655}}
\hypersetup{pdftitle={Factor-Biased Efficiency Gains from Exporting: Evidence from Colombia}}

\begin{document}
\setlength{\droptitle}{-4em}
\title{\Large Factor-Biased Efficiency Gains from Exporting: Evidence from Colombia}
\author{\tightrows\begin{tabular}[t]{c}Joonkyo Hong\thanks{\fndouble Joonkyo Hong, Email: \href{mailto:jkhong@ntu.edu.tw}{jkhong@ntu.edu.tw}. Davide Luparello, Email: \href{mailto:dluparello@fsu.edu}{dluparello@fsu.edu}. We are grateful to Paul Grieco, Mark Roberts, Jim Tybout, and Devesh Raval for their insightful suggestions, and to Sofronis Clerides, Mark Roberts, and Jim Tybout for generously sharing their data. We thank the organizers and participants of the 94th SEA Meeting, APIOC 2024, TER 2024, V ISCEMR, Macroeconometric Modelling Workshop 2024, the National Cheng-Kung University seminar, the Penn State Trade and Development and Applied Microeconomics Brownbag seminars, and the NTU Brownbag seminar for their valuable feedback, and Yu Hsuan Chou for outstanding research assistance. Hong acknowledges support from Taiwan's National Science and Technology Council (grant 114-2410-H-002-285-MY2). Additionally, we acknowledge the use of OpenAI's ChatGPT and Anthropic's Claude LLMs, within the guidelines outlined in \citet{korinek2023generative}. This article was previously circulated under the title \textit{In Search of (Factor-Biased) Learning by Exporting}.}\\
{\normalsize\it National Taiwan University}\end{tabular}\hspace{3em}%
\begin{tabular}[t]{c}Davide Luparello\\
{\normalsize\it Florida State University and BAFFI}\end{tabular}}
\date{October 5, 2026}
{\singlespacing\maketitle}
\thispagestyle{empty}
\setcounter{page}{0}

\vspace{-1.5em}
\begin{abstract}

New exporters adopt new technologies, which may reorganize production rather than uniformly increase output, so efficiency gains can vary across inputs. We examine such gains across worker types in Colombian manufacturing, 1981--1991. Developing and applying a model of production and export entry, we find export entry raises unskilled-labor efficiency by 22\% on entry, with no significant change in skilled-labor or neutral efficiency. Total factor productivity rises by about 8\% after entry. We estimate that the two worker types are complements, so exporters produce more with relatively less unskilled labor, raising skill intensity.

\end{abstract}

\vspace{0.5em}


\doublespacing
\section{Introduction}\label{s:introduction}
Exporting plants invest in new machinery, upgrade product quality, and adopt more sophisticated technologies \citetext{e.g., \citealp{lileeva2010improved, aw2011r, bustos2011trade, aghion2018impact, giorcelli2019long}; \citealp{rodrigue2019price}; \citealp{bajgar2020climbing, alfaro2022effects}; see \citealp{verhoogen2023firm} for a review}. Such upgrading should raise productivity, yet the gains have proven hard to detect \citep{clerides1998learning, bernard1999exceptional, aw2000productivity}. Part of the reason may be that this upgrading raises the efficiency of some inputs more than that of others, so that the efficiency gains are \emph{factor-biased}, whereas most studies measure productivity as total factor productivity, a single gain common to all inputs.

In this article we ask whether the efficiency gains from exporting are factor-biased and, in particular, whether they favor skilled or unskilled labor. Because total factor productivity weights each input's efficiency gain by that input's output elasticity, a substantial efficiency gain concentrated in some inputs can translate into only a modest increase in total factor productivity. Measuring these gains solely through total factor productivity can therefore obscure both their magnitude at the input level and their factor bias. Beyond measurement, the distinction matters: at given input prices, an efficiency gain common to all inputs raises output without changing the cost-minimizing input mix, whereas a factor-biased gain changes that mix in a direction set by the elasticity of substitution between the inputs. For skilled and unskilled labor, the bias and this elasticity together determine their relative demand. Export-induced technical change can therefore reorganize production within plants, reallocate inputs across them, and shift the skill composition of employment.

To answer this question, we develop an empirical framework, applicable to standard production datasets, to detect factor-biased technical progress from export entry. The production technology features two factor-augmenting efficiency components and a Hicks-neutral one (a single efficiency gain common to all inputs), all evolving with past exporting and investment decisions. In the spirit of \citet{doraszelski2018measuring}, we recover each factor-augmenting component by inverting the plant's relative demand for a different pair of inputs: relative-unskilled efficiency from the ratio of skilled to unskilled employment given the skill premium, and labor-augmenting efficiency from the ratio of materials to labor given their relative price. The Hicks-neutral component instead enters the production residual together with output measurement error. We recover the residual from deflated revenue after controlling for input choices, then separate the two by casting productivity dynamics as a state-space model and applying the Kalman filter.

Firms decide whether to export by weighing expected future gains against the sunk entry and fixed operating costs, so the conditional choice probability (CCP) of exporting depends on past export status and other payoff-relevant state variables. Under the model's assumptions, conditional on these state variables, export entry depends only on an i.i.d. sunk-cost draw and is therefore as good as random. We exploit this through a microfounded, propensity-score-matched difference-in-differences design that pairs new exporters with never-exporting controls. Rather than estimating the CCPs, we use them to select the state variables on which the two groups are balanced. We then compare how each productivity component evolves after export entry across the matched groups. Under parallel trends, this comparison identifies the average treatment effect on the treated (ATT): the effect of export entry on new exporters that continue to export.

We apply this framework to 19 major manufacturing sectors in Colombia between 1981 and 1991, which together account for the bulk of the country's manufactured exports \citep{roberts1997decision}.\footnote{This period suits our question for two reasons. First, \citet{clerides1998learning} find no evidence of productivity gains from exporting in Colombian manufacturing over these years. Second, commercial policy sheltered import-competing producers through 1989 \citep{roberts1997decision}, so we can study export entry apart from import liberalization. In Section~\ref{s:robustness}, we show that our results are robust to excluding export entrants whose post-entry years overlap the import liberalization of 1990--91, at the end of our sample.} Export entry is associated with factor-biased efficiency gains relative to matched never-exporters. It raises unskilled-labor productivity by about 22\% in the year of entry, while the changes in skilled-labor and Hicks-neutral productivity are statistically insignificant. Total factor productivity rises only modestly, by about 8\% on average after entry. The gain is thus factor-biased: it originates in the rising efficiency of unskilled labor, not in a neutral improvement common to all inputs.

Beyond this modest shift in the level of productivity, exporting reorganizes production within plants, reshaping their input \emph{composition}. Because skilled and unskilled labor are gross complements in our estimates, the disproportionate rise in unskilled-labor productivity is \emph{unskilled-labor-saving}: making unskilled labor more effective lowers its cost-minimizing relative use, raising the skill intensity of production holding relative wages fixed. This delivers the same skill upgrading as the canonical account of skill-biased technical change \citep{katz1992changes, acemoglu2011skills}, but through a different mechanism, operating on the unskilled margin under complementarity rather than the skilled margin under substitutability.

A counterfactual that switches off new exporters' post-entry gains and re-aggregates implies that exporting raises the manufacturing skilled-labor share by $0.18$ percentage points, 0.70\% of its 1981 level, through within-plant upgrading partially offset by cross-plant reallocation. Employment of both worker types still expands as output grows along its demand curve, though by less for unskilled labor, so the skilled-labor share climbs even as unskilled employment rises in absolute terms. This within-plant upgrading is consistent with a separate event-study finding: among new exporters, machinery and equipment investment surges after entry, pointing to mechanization as a channel.

We make three contributions to the literature. First, we contribute to the literature linking international trade to skill upgrading. Existing studies document changes in exporters' occupational composition, skill intensity, and innovation \citep{bernard1997exporters, brambilla2012exports, brambilla2019exporting}, and identify mechanisms linking trade to greater demand for skilled labor. Quality upgrading raises skill demand because producing higher-quality goods requires more skilled labor \citep{verhoogen2008, fieler2018trade}, while access to skill-intensive foreign technology provides another channel \citep{waddle2021trade}. We add a mechanism operating through the factor bias of export-associated technical change, showing how skill upgrading can arise from improvements in unskilled-labor efficiency without requiring either higher product quality or a higher skill premium.

Our work is closest to \citet{harrigan2018techies}, who also study factor-biased technical change associated with trade. Their findings for French firms combine skill-augmenting productivity gains with substitutability between skilled and unskilled labor.\footnote{In related work, \citet{harrigan2026techies} find that technically trained workers raise the Hicks-neutral productivity of French firms. Taken together, their evidence for France points to gains in both skill-augmenting and Hicks-neutral productivity.} We identify a different configuration: unskilled-labor efficiency gains coupled with complementarity. Together, these findings show that unpacking exporters' skill upgrading requires identifying both the direction of factor augmentation and the substitution relationship between workers. We further build on their analysis by examining the timing of these gains. Comparing new exporters with matched never-exporters allows us to trace the adjustment around export entry, extending their evidence on trade-related shifts in productivity processes.

Second, we contribute to research on the direction of technological change. Distinguishing Hicks-neutral from factor-augmenting productivity has proved important for understanding firm performance, labor demand, and measured markups \citep{doraszelski2018measuring, zhang2019non, demirer2026production}. As \citet{raval2019micro} and \citet{jo2025substitution} illustrate, measuring the direction of technological change requires an estimate of the elasticity of substitution between inputs. We focus on the elasticity between skilled and unskilled labor and use it to connect export entry, a potential source of productivity growth, to the efficiency gains of each worker type. Methodologically, our state-space approach departs from \citet{doraszelski2018measuring} in how it recovers Hicks-neutral productivity. Whereas they invert the first-order condition for labor, we use the Kalman filter to separate Hicks-neutral productivity from measurement error in the production residual by their dynamics: productivity is persistent and measurement error is serially independent. Unlike proxy-variable methods \citep{olley1996dynamics, levinsohn2003estimating, ackerberg2015identification}, it inverts no input demand and needs no monotonicity condition.

Third, we speak to the literature on exporting and firm performance. Early influential studies find limited evidence that exporting raises productivity, whereas subsequent work reports positive gains in various settings \citep{bigsten2004african, blalock2004learning, alvarez2005exporting, van2005exporting, de2007exports, park2010exporting, atkin2017exporting, garcia2019exporting}. In particular, \citet{clerides1998learning} find no statistically significant gains in the same Colombian data when productivity is proxied by average variable costs or output per worker. Our results help explain this puzzle: similar scalar measures would capture only the modest aggregate gain and miss the factor-biased one, which raises unskilled-labor productivity alone.

The remainder of this article is structured as follows. Section \ref{s:data} describes the data used in the analysis and highlights pertinent descriptive patterns. Section \ref{s:theory} presents our empirical framework and derives the core implications. Section \ref{s:empirics} details how we take these implications to the data to obtain the structural estimates of interest. Section \ref{s:results} presents and discusses the key findings. Section \ref{s:robustness} reports robustness checks. Section \ref{s:conclusion} concludes.

\section{Data and Descriptive Statistics}\label{s:data}
We employ data from the Colombian Annual Manufacturing Survey (Encuesta Anual Manufacturera) conducted by the Departamento Administrativo Nacional de Estadística (DANE) for 1981--1991. Despite extensive analysis since \citet{roberts1996industrial} and \citet{roberts1997decision}, factor-specific productivity gains from exporting in this context have, to our knowledge, received limited attention, particularly for labor inputs. We address this gap by exploiting a distinctive feature of the dataset: employment and payrolls are reported separately by worker category, which lets us identify productivity differentials between skilled and unskilled labor. Following \citet{fieler2018trade}, we classify skilled labor as encompassing management (\textit{directivos}), technicians (\textit{técnicos}), and employees (\textit{empleados}), whereas we designate workers (\textit{obreros}) as unskilled labor.

DANE defines the four categories as follows. \textit{Directivos} (management) direct the establishment's economic, financial, and administrative functions and are responsible for formulating general firm policy. \textit{Técnicos} (technicians) engage directly in productive activities or production-related tasks.\footnote{DANE's examples include engineers across specializations (mechanical, chemical, industrial, electrical, mining, petroleum) as well as technicians and technologists working in the production area.} \textit{Empleados} (employees) are administrative and sales personnel beyond management, including administrative supervisors, security staff, non-production service workers, sales representatives, and distributors whose workplace and compensation are attributed to the establishment. \textit{Obreros} (unskilled workers) are engaged in fabrication, processing, assembly, installation, maintenance, inspection, storage, packing, and loading operations.\footnote{DANE's examples include shop-floor messengers, boiler operators, machinery-cleaning staff, foremen who perform manual tasks, internal drivers transporting inputs and products, and workers maintaining industrial machinery and equipment.}

We construct an unbalanced panel of 10,023 plants covering 19 key exporting industries \citep{roberts1997decision}.\footnote{These industries correspond to SIC-3 codes: food processing (311/312), textiles (321), clothing (322), leather products (323/324), paper (341), printing (342), chemicals (351/352), plastics (356), glass (362), nonmetal products (369), iron and steel (371), metal products (381), machinery (382/383), transportation equipment (384), and miscellaneous manufacturing (390).} We follow the standard cleaning procedure of \citet{roberts1996industrial} and \citet{raval2023testing}, described in Appendix \ref{a:data_construction}. We measure revenue as deflated aggregate sales across domestic and export markets, skilled and unskilled labor by headcounts, intermediate inputs as expenditures on raw materials, electricity, and fuels (each component deflated by its own price index), and capital as four asset classes (land, structures, equipment, and transportation equipment) accumulated by the perpetual inventory method, with capital service expenditure measured at the user cost of these stocks.

Table \ref{tab:descriptive} compares pooled sample averages of revenue per worker, capital stocks, material expenditures, workforce size, skill ratios, and skill premia between exporters and non-exporters. Exporters comprise 12.6\% of the sample and are 1.8 times as productive as non-exporters. They also operate at a substantially larger scale: capital stocks are 6.5 times as large, material expenditures 5.2 times as high, and workforces 4.2 times the size of those at non-exporting plants. Exporters exhibit 24\% higher skill ratios and pay 21\% higher skill premia.

\FloatBarrier

\subsection{Descriptive Patterns for Exporting Plants}

Colombian exporting plants exhibit systematic factor-biased trajectories in employment, skill intensity, and labor productivity after export entry: patterns that standard TFP measures cannot capture. We document these using an age-cohort-period design on new export entrants, reporting average trajectories that incorporate both within-plant changes and compositional shifts from selective exit.\footnote{For this event-study exercise, we further restrict the sample to plants with complete panels over their observed years, so that event-time comparisons are not confounded by missing-year gaps.}

\begin{table}[!h]
\centering
\caption{Descriptive Statistics: Exporters vs Non-Exporters}
\label{tab:descriptive}
\begin{tabular*}{\textwidth}{@{\extracolsep{\fill}} lcccc}
\toprule
\toprule
            &   Exporters&Non-Exporters&  Difference&          SE\\
\midrule
Revenue per Worker*&        1.49&        0.84&        0.66&      (0.02)\\
Capital Stock*&      114.10&       17.68&       96.43&      (2.70)\\
Material Expenditure*&      200.83&       38.52&      162.31&      (3.22)\\
Number of Workers&      212.46&       50.05&      162.41&      (2.22)\\
Skill Ratios&        0.57&        0.46&        0.11&      (0.01)\\
Skill Premia&        1.90&        1.57&        0.33&      (0.01)\\
\midrule
N. of Observations &     6,340 &    43,989 & & \\
\bottomrule
\end{tabular*}
\note{The table displays pooled sample averages over 1981--1991 by exporting status and the difference in means between groups, for the estimation sample described in Section~\ref{s:data}. Variables marked with an asterisk (*) are expressed in deflated millions of Colombian Pesos. \textit{Number of Workers} is expressed in headcounts. \textit{Skill Ratios} represent the ratio of skilled to unskilled workforce. \textit{Skill Premia} denote the ratio of average annual wage rates for skilled versus unskilled workers. Standard errors of the difference in means appear in parentheses.}
\end{table}

We specify the outcome variable of interest $y_{jt}$ as a function of years spent in the export market, $a$, along with cohort of export market entrants $c$, sector $s$, and year $t$ fixed effects:
\begin{align}
   y_{jt} =  \sum_{a = 0}^{A} \lambda_{a} D_{ja} + \nu_{c} + \nu_{s}+ \nu_{t} + u_{jt}. \label{eq:lifecycle}
\end{align}
Here, $D_{ja}$ denotes an indicator variable equal to one if plant $j$ has spent $a$ years in the export market, and $u_{jt}$ represents the residual term. We bin $A$ at five or more years of export market tenure ($A \geq 5$). Our interest centers on the estimated coefficients $\lambda_{a}$, which capture the dynamics of the outcome variable relative to its level in the year preceding the plant's initial export entry.

\begin{figure}[!h]
\centering
\begin{subfigure}[t]{0.46\textwidth}
\centering
\includegraphics[width=\linewidth]{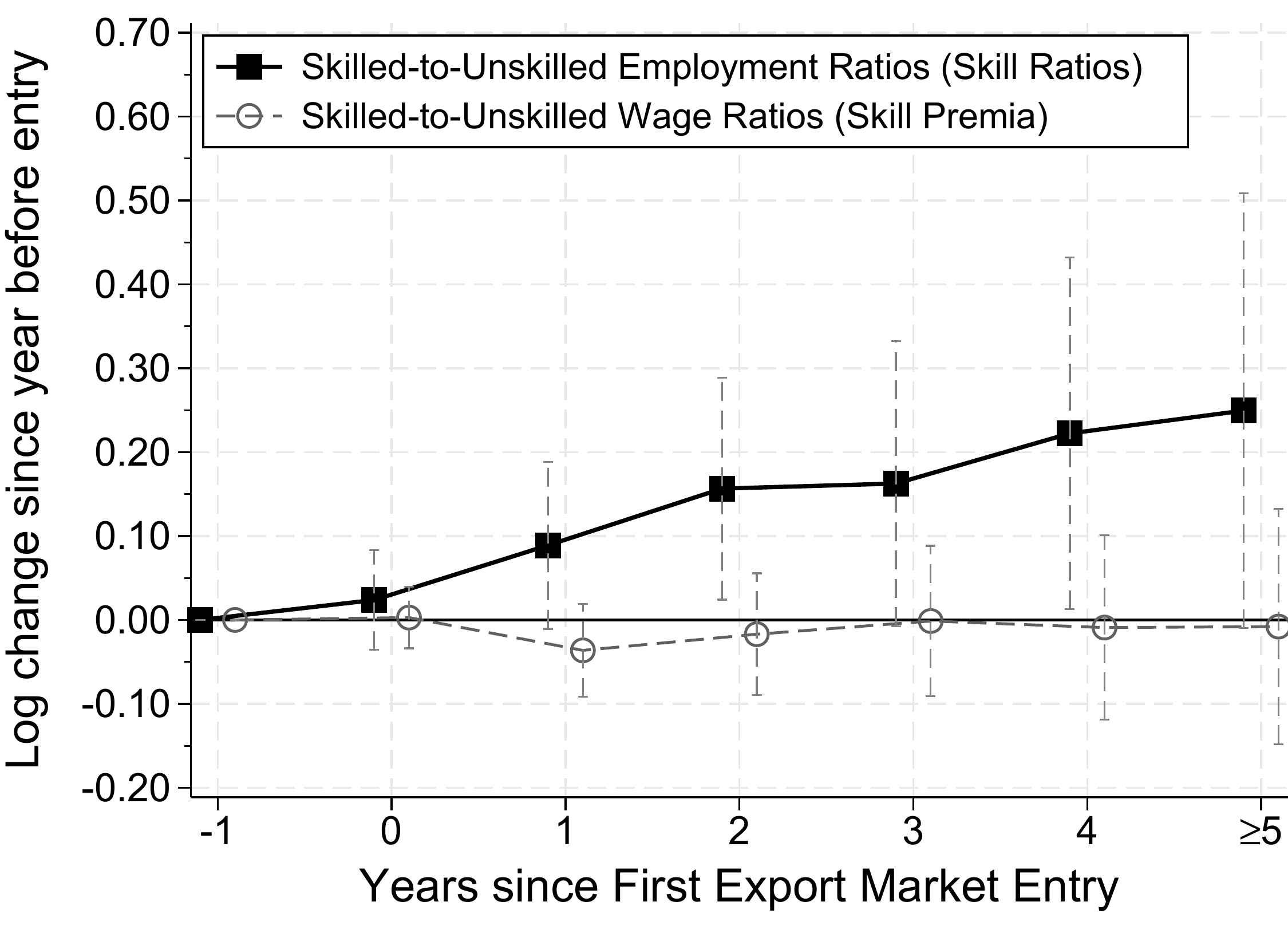}
\subcaption{Skill Ratios and Skill Premia}
\label{fig:skill_rat}
\end{subfigure}
\hfill
\begin{subfigure}[t]{0.46\textwidth}
\centering
\includegraphics[width=\linewidth]{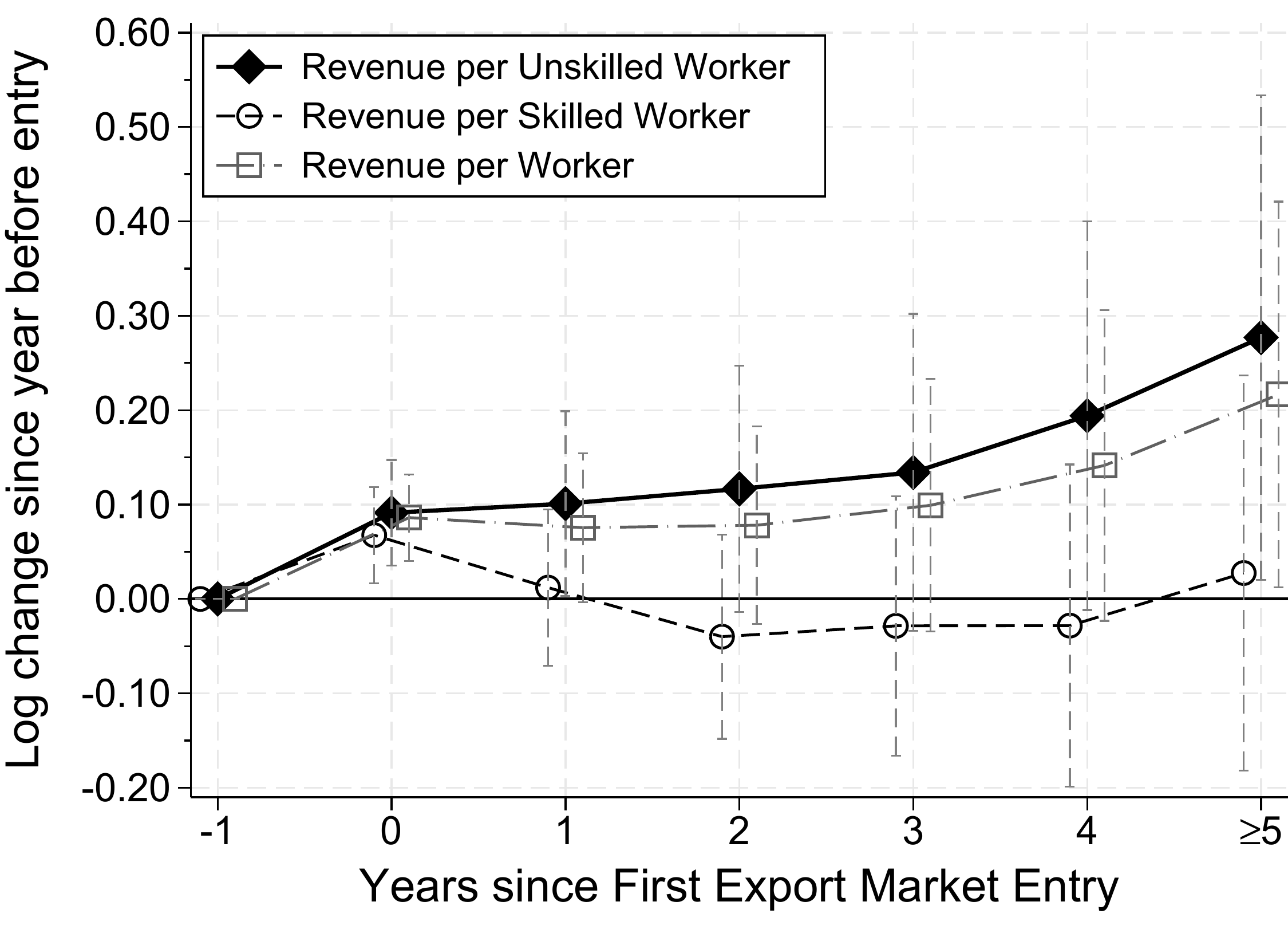}
\subcaption{Labor Productivity by Skill Category}
\label{fig:skill_vs_unskill}
\end{subfigure}
\caption{Skill Composition and Labor Productivity Trajectories}
\note{This figure displays the coefficients from estimating equation \eqref{eq:lifecycle} on establishment-level (log) skill ratios and (log) skill premia (Figure \ref{fig:skill_rat}) and on (log) revenue per worker by skill category (Figure \ref{fig:skill_vs_unskill}), along with the corresponding 90\% confidence intervals. Standard errors are clustered at the plant level. The sample comprises new exporters that entered the export market by 1989, restricted to those with complete panels over their observed years. Source: Colombian Annual Manufacturing Survey (DANE), 1981--1991.}
\end{figure}

Figure \ref{fig:skill_rat} shows that post-entry dynamics in skill premia cannot account for the observed trajectory of skill ratios, consistent with a skill-biased pattern associated with exporting. We observe that skill ratios increase by approximately 5\% annually following export entry, relative to one year prior to exporting. In contrast, skill premia exhibit no statistically significant dynamics after export entry. This pattern suggests that, although exporters become more skill-intensive post-entry, the relative compensation for skilled workers shows no statistically detectable change.

Furthermore, Figure \ref{fig:skill_vs_unskill} shows that labor productivity dynamics differ sharply across worker categories, where we measure productivity as revenue per worker. Unskilled worker productivity grows by approximately 5.5\% annually following export entry, whereas skilled worker productivity remains statistically unchanged. Overall labor productivity rises by approximately 4\% annually. This differential growth pattern is consistent with organizational or technological changes that disproportionately enhance unskilled operations relative to skilled operations, rather than uniformly elevating productive efficiency across the workforce.

\begin{figure}[!h]
\centering
\begin{subfigure}[t]{0.46\textwidth}
\centering
\includegraphics[width=\linewidth]{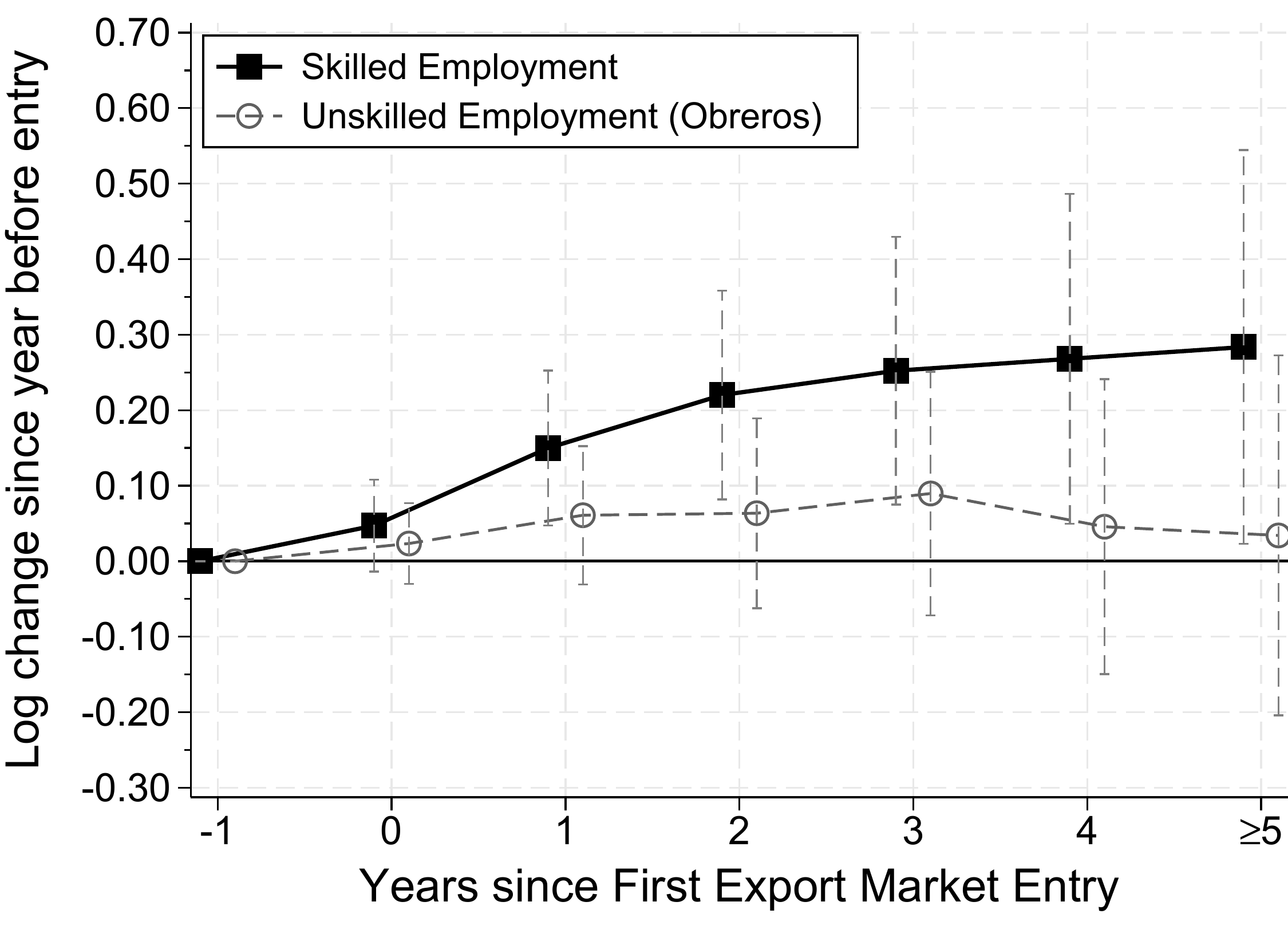}
\subcaption{Skilled and Unskilled Employment}
\label{fig:emp}
\end{subfigure}
\hfill
\begin{subfigure}[t]{0.46\textwidth}
\centering
\includegraphics[width=\linewidth]{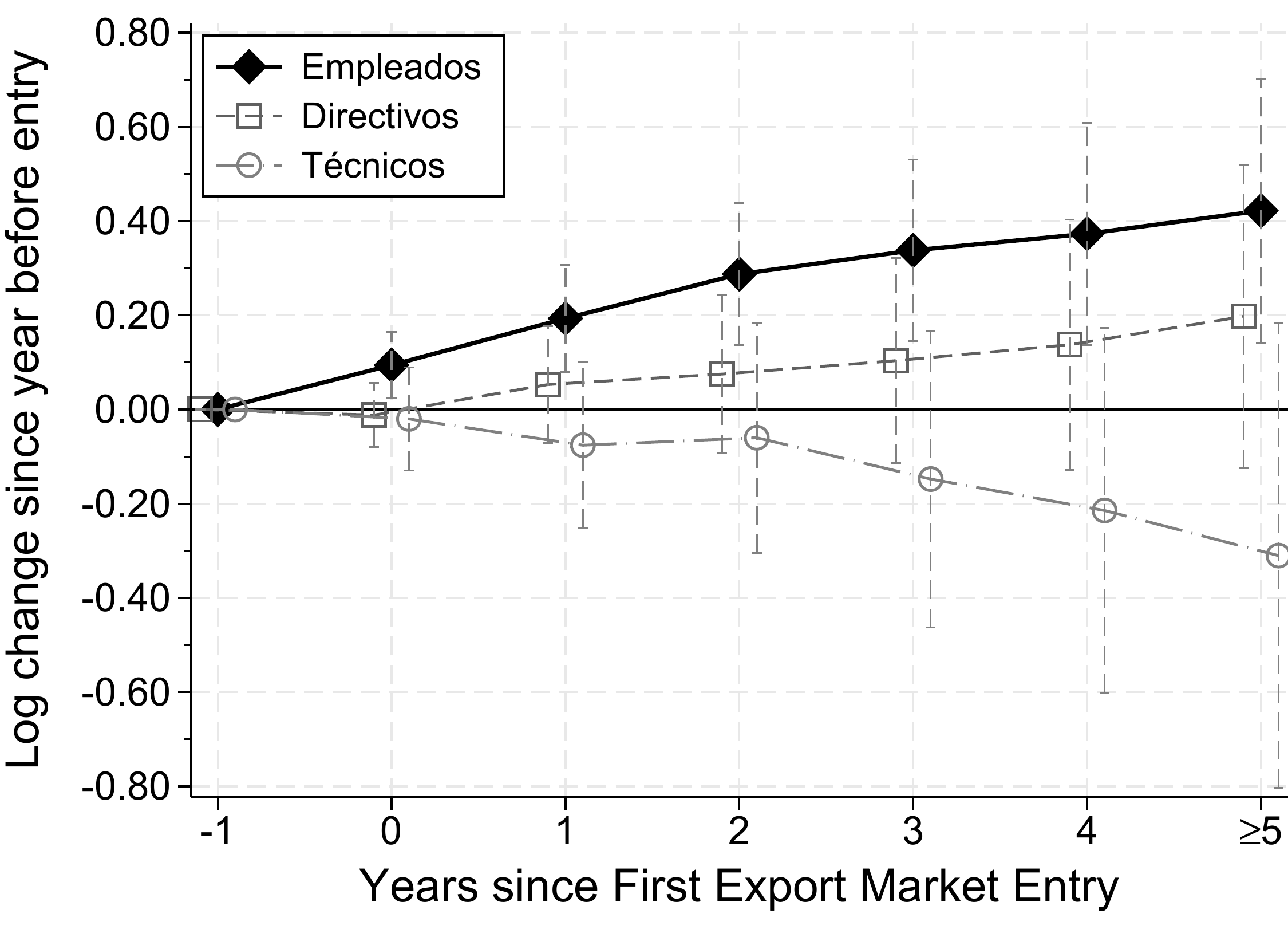}
\subcaption{Skilled Employment by Occupation}
\label{fig:labor_div}
\end{subfigure}
\caption{Employment Trajectories}
\note{This figure displays the coefficients from estimating equation \eqref{eq:lifecycle} on establishment-level (log) skilled and unskilled (\textit{obreros}) employment (Figure \ref{fig:emp}) and on (log) employment across skilled worker occupational categories: employees (\textit{empleados}), managers (\textit{directivos}), and technicians (\textit{técnicos}) (Figure \ref{fig:labor_div}), along with the corresponding 90\% confidence intervals. Standard errors are clustered at the plant level. The sample comprises new exporters that entered the export market by 1989, restricted to those with complete panels over their observed years. Source: Colombian Annual Manufacturing Survey (DANE), 1981--1991.}
\end{figure}

Figure \ref{fig:emp} shows that rising skill ratios coincide with disproportionate expansion of skilled employment. We find that skilled labor grows by 5.6\% annually following export entry, whereas unskilled labor remains flat. This pattern indicates that export market participation is associated with selective workforce expansion concentrated among skilled workers.

The pattern extends to occupational composition: Figure \ref{fig:labor_div} shows that skilled employment growth concentrates almost entirely within the \textit{empleados} (employees) category, which expands by 8\% annually. In contrast, \textit{directivos} (managers) grow by roughly 4\% annually, whereas \textit{técnicos} (technicians) decline, though neither coefficient attains statistical significance. This compositional pattern shows that export market participation coincides with selective workforce expansion concentrated among employees rather than managerial or specialized technical personnel.

Overall, these descriptive trajectories show new exporters becoming more skill-intensive: skill ratios increase by 5\% annually, whereas skill premia remain stable, employment expands selectively among skilled workers, and productivity gains concentrate among unskilled workers. This asymmetric pattern is central to the factor-biased efficiency gains we quantify in this article.

We now turn to a structural production function framework in Section \ref{s:theory}. This framework provides two key advantages. First, it allows us to separately estimate skilled and unskilled labor-augmenting productivities alongside identifying the elasticity of substitution between these labor types, thereby decomposing aggregate productivity growth into its factor-specific components. Second, it identifies the state variables governing each plant's export-entry decision, which guide the conditioning variables we use to match new exporters with otherwise comparable non-exporters, allowing us to assess factor-biased productivity gains from exporting.

\section{Empirical Model}\label{s:theory}

We integrate a production framework with factor-augmenting productivities into a dynamic model of exporting and investment. Each period, plants maximize short-run profits by optimally allocating skilled labor, unskilled labor, and materials, then weigh expected benefits of export market participation against entry or continuation costs. We specify productivity processes that permit exporting to be associated with factor-biased efficiency gains. Plants subsequently choose physical capital investment based on current state variables and anticipated export status. We model capital accumulation through standard investment dynamics, establishing intertemporal linkages between current decisions and future production capabilities.

\subsection{Environment}

\textit{Technology and Productivity Dynamics}: We model Colombian manufacturing plants producing output ($Q_{jt}$) through a nested constant elasticity of substitution (CES) production function that combines physical capital ($K_{jt}$), intermediates ($M_{jt}$), skilled labor ($S_{jt}$), and unskilled labor ($U_{jt}$):
\begin{equation}
Q_{jt} = \left[\tilde{\alpha}_K  K_{jt}^{\rho} + \tilde{\alpha}_M M_{jt}^{\rho} + \tilde{\alpha}_L (\exp(\omega_{L,jt}) L_{jt})^{\rho} \right]^{\frac{1}{\rho}}\exp(\omega_{H,jt}) \label{eq:prod}
\end{equation}
\begin{equation}
L_{jt} = \left[\tilde{\alpha}_S S_{jt}^{\theta} + \tilde{\alpha}_U(\exp(\omega_{R,jt}) U_{jt})^{\theta}\right]^{\frac{1}{\theta}} \label{eq:labor}
\end{equation}
The CES share parameters $\tilde{\alpha}_K$, $\tilde{\alpha}_M$, $\tilde{\alpha}_L$, $\tilde{\alpha}_S$, and $\tilde{\alpha}_U$ govern factor intensity. The parameter $\rho$ controls substitution elasticity among capital, materials, and the labor composite, whereas $\theta$ governs substitution between skilled and unskilled labor.\footnote{The nested structure implies $\sigma_{KS} = s_S/(1-\rho) + (1-s_S)/(1-\theta)$ and $\sigma_{KU} = (1-s_S)/(1-\rho) + s_S/(1-\theta)$, where $s_S$ is the skilled labor cost share of payroll. Capital-skill complementarity, the condition that capital complements skilled labor more strongly than unskilled labor \citep{griliches1969capital, krusell2000capital}, holds if and only if $\sigma_{KS} < \sigma_{KU}$. See Appendices \ref{app:skill-capital} and \ref{a:cap_skill_empirical} for the derivation and empirical verification.} We incorporate three heterogeneous productivity components: $\omega_{L,jt}$ augments the labor composite, $\omega_{R,jt}$ denotes the unskilled-to-skilled relative labor-augmenting efficiency, and $\omega_{H,jt}$ represents Hicks-neutral productivity.\footnote{Several empirical studies explore the role of input efficiencies in various empirical contexts. \citet{raval2019micro} estimates the capital-labor elasticity of substitution and documents that labor-augmenting productivity is persistent and correlated with exports, firm size, and growth.}

We specify productivity evolution as independent first-order Markov processes, where each component responds to the plant's lagged export status and investment decisions \citep{aw2011r, de2013detecting, doraszelski2013r}:
\begin{equation}
    \omega_{X,jt} = \iota_t + \iota_s + \rho_X \omega_{X,jt-1} + \beta_X^e e_{jt-1} + \beta_X^i i_{jt-1} + \beta_X^{ei} \left( e_{jt-1} \cdot i_{jt-1} \right) + \xi_{X,jt}, \quad\text{for } X\in\{H,L,R\}\label{eq:omega}
\end{equation}
The terms $\iota_{t}$ and $\iota_{s}$ are year and sector fixed effects. The binary indicator $e_{jt-1}$ equals one if plant $j$ exported in period $t-1$, and $i_{jt-1}$ equals one if the plant invested in equipment during that period. The interaction term $e_{jt-1} \cdot i_{jt-1}$ allows for potential complementarities between exporting and investment in productivity dynamics.

The innovation terms $\xi_{H,jt}$, $\xi_{L,jt}$, and $\xi_{R,jt}$ are mean-zero idiosyncratic shocks, independent of past productivity, export decisions, and investment choices. We assume the Hicks-neutral innovation $\xi_{H,jt}$ follows a distribution with variance $\sigma^2_H$. Plants observe all three current productivity draws, $\omega_{H,jt}$, $\omega_{L,jt}$, and $\omega_{R,jt}$, before allocating labor and intermediate inputs within each period.

\par\medskip
\noindent\textit{Output and Input Markets}: We model each plant as a monopolistic competitor facing a Dixit-Stiglitz inverse demand curve combining domestic and foreign markets:
\begin{equation}
P_{jt}=P_{It}\left(\frac{Q_{jt}}{Q_{It}}\right)^{\frac{1}{\eta}}, \label{eq:output_demand}
\end{equation}
where $Q_{jt}$ denotes plant $j$'s aggregate quantity demanded at time $t$, $P_{jt}$ represents the corresponding price, $Q_{It}$ and $P_{It}$ capture industry-level quantity and price indices, and $\eta$ measures the aggregate demand elasticity.

We treat skilled and unskilled wage rates, $W_{S,jt}$ and $W_{U,jt}$, as exogenous state variables, following \citet{doraszelski2018measuring}. We do not model the wage-generating processes explicitly, allowing wages to vary across plants and over time through plant-specific characteristics and temporal factors. We assume competitive pricing in the intermediate materials market at price $P_{M,jt}$.

\subsection{Static Input Allocation Decision}

We collect the plant's state variables in the vector
\begin{equation}
    \Omega_{jt} = \left(K_{jt}, W_{S,jt}, W_{U,jt}, P_{M,jt}, P_{It}, Q_{It}, \exp(\omega_{L,jt}), \exp(\omega_{R,jt}), \exp(\omega_{H,jt})\right).
\end{equation}
Conditional on $\Omega_{jt}$ and the export decision $e_{jt}$, plant $j$ chooses materials $M_{jt}$, skilled labor $S_{jt}$, and unskilled labor $U_{jt}$ to maximize short-run profits:
\begin{equation}
\begin{aligned}
\label{eq:profit_max}
\pi(\Omega_{jt}, e_{jt}) = & \max_{M_{jt}, S_{jt}, U_{jt}} \quad P_{jt}Q_{jt} - W_{S,jt}S_{jt}-W_{U,jt}U_{jt}-P_{M,jt}M_{jt} \\
\textrm{s.t.} \quad & Q_{jt} \geq (1 - e_{jt}){Q}^{0} + e_{jt}{Q}^{1} \\
& Q_{jt} = \left[\tilde{\alpha}_K K_{jt}^{\rho} + \tilde{\alpha}_M M_{jt}^{\rho} + \tilde{\alpha}_L (\exp(\omega_{L,jt})L_{jt})^{\rho} \right]^{\frac{1}{\rho}} \exp(\omega_{H,jt}) \\
& L_{jt} = \left[\tilde{\alpha}_S S_{jt}^{\theta} + \tilde{\alpha}_U (\exp(\omega_{R,jt}) U_{jt})^{\theta} \right]^{\frac{1}{\theta}} \\
& P_{jt}=P_{It}\left(\frac{Q_{jt}}{Q_{It}}\right)^{\frac{1}{\eta}}.
\end{aligned}
\end{equation}
The output constraint requires production to meet a regime-specific minimum scale, $Q^{0}$ when the plant sells only domestically ($e_{jt}=0$) and $Q^{1}$ when it also serves the export market ($e_{jt}=1$).\footnote{We assume the constraint does not bind at the optimum, so input choices satisfy the unconstrained first-order conditions.}

\subsection{Dynamic Export and Investment Decisions, and End of Period}
After input allocation, plant $j$ decides whether to enter the export market in period $t+1$. Following \citet*{das2007market, aw2011r, rho2016firm}; \citet*{cooper2026exporting}, we model this choice as involving non-convex costs $\gamma_{jt}$, which the plant observes but the econometrician does not. We specify these costs as independent draws from distribution $G(\cdot|e_{jt})$, conditional on current export status. Non-exporters ($e_{jt}=0$) face start-up costs that differ from the continuation costs borne by active exporters ($e_{jt}=1$). After resolving the export decision, the plant chooses investment level $I_{jt}$, incurring convex adjustment costs $C(I_{jt}, K_{jt})$.

We represent the ex-ante expected value of plant $j$ through the Bellman equation:
\begin{equation}
    \begin{aligned}
     EV(\Omega_{jt}, e_{jt}) = \pi(\Omega_{jt}, e_{jt}) +\int_{\gamma_{jt}}  \max_{e_{jt+1}} \left\{EV^{1}(\Omega_{jt}, e_{jt}) - \gamma_{jt}, EV^{0}(\Omega_{jt}, e_{jt}) \right\} \text{d} G(\gamma_{jt}|e_{jt})\label{eq:exporting}
\end{aligned}
\end{equation}
Current-period short-run profits $\pi(\Omega_{jt}, e_{jt})$ constitute the flow value, whereas the integral captures expected continuation value. The plant compares the value of exporting $EV^{1}(\Omega_{jt}, e_{jt})$ net of costs $\gamma_{jt}$ against the value of remaining a domestic producer $EV^{0}(\Omega_{jt}, e_{jt})$.

We define continuation values $EV^{1}$ and $EV^{0}$ by optimally choosing investment $I_{jt}$ conditional on next period's export status. For $x \in\{0,1\}$:
\begin{equation}
\begin{aligned}
EV^{x}(\Omega_{jt}, e_{jt}) = \max_{I_{jt}} &\left\{-C(I_{jt}, K_{jt}) + \beta \int_{\Omega_{jt+1}} EV(\Omega_{jt+1}, e_{jt+1}=x) \text{d} F(\Omega_{jt+1}|\Omega_{jt}, e_{jt}, I_{jt})\right\}\\
\textrm{s.t.} \quad & K_{jt+1}=(1-\delta)K_{jt}+I_{jt}
\label{eq:invest_e}
\end{aligned}
\end{equation}
The plant trades off investment costs $C(I_{jt}, K_{jt})$ against discounted future value conditional on export status. The transition density $F(\Omega_{jt+1}|\Omega_{jt}, e_{jt}, I_{jt})$ describes state evolution, whereas capital accumulates through standard depreciation at rate $\delta$ and investment $I_{jt}$. The plant discounts future payoffs at rate $\beta$.

\par\medskip
\noindent\textit{Exporting and Investment Policy Functions}: We derive plant $j$'s conditional choice probability of exporting in period $t+1$ from Equation \eqref{eq:exporting}:
\begin{equation}
    \Pr(e_{jt+1}=1|\Omega_{jt}, e_{jt}) = \int_{\gamma_{jt}} \mathbb{1}\left\{ EV^{1}(\Omega_{jt}, e_{jt}) - EV^{0}(\Omega_{jt}, e_{jt}) \geq \gamma_{jt} \right\} \text{d} G(\gamma_{jt}|e_{jt}). \label{eq:CCP}
\end{equation}
Equation \eqref{eq:invest_e} likewise yields the investment policy function, which depends on current states and anticipated export status:
\begin{equation}
\label{eq:investment}
    I_{jt} = \mathcal{I}(\Omega_{jt}, e_{jt}, e_{jt+1})
\end{equation}
An important implication of these choice probabilities is that conditioning on the state variables $(\Omega_{jt}, e_{jt})$ does not generate degenerate export probabilities across plants, as the costs $\gamma_{jt}$ enter independently. Consider non-exporters sharing identical state $(\Omega_{jt}, e_{jt}=0)$ at time $t$. Their entry probabilities in period $t+1$ depend solely on idiosyncratic cost draws. Under the i.i.d. sunk-cost draws, export entry is as good as random conditional on $\Omega_{jt}$.

\par\medskip
\noindent\textit{End of Period}: We define planned revenue as the product of price and quantity:
\begin{equation}
    R_{jt}\equiv P_{jt}Q_{jt}. \label{eq:revenue}
\end{equation}
Observed revenue is subject to end-of-period measurement error $\zeta_{jt}$, distributed i.i.d. with mean zero and variance $\sigma^2_\zeta$,\footnote{$\sigma^2_H$ and $\sigma^2_\zeta$ are the only variance parameters that require specification for separating measurement error from Hicks-neutral productivity.} independent of state variables, the export decision, and the productivity innovations $\xi_{X,jt}$:
\begin{equation}
   \tilde{R}_{jt} = R_{jt} \exp(\zeta_{jt}) \label{eq:quantity}
\end{equation}

\par\medskip
\noindent\textit{Total Factor Productivity}: We recast our production function specification into a framework with separate productivity terms for each labor type. Skilled labor-augmenting productivity equals overall labor-augmenting productivity, $\omega_{S,jt}=\omega_{L,jt}$, and unskilled labor-augmenting productivity is recoverable from labor-augmenting productivity and the relative productivity of unskilled to skilled workers, $\omega_{U,jt}=\omega_{R,jt}+\omega_{S,jt}$. We then compute log TFP as a weighted sum of the Hicks-neutral and factor-augmenting productivities\footnote{This aggregation is exact in differential form and, in levels, when the output elasticities are constant. Appendix \ref{app:tfp_derivation} provides the derivation.}:
\begin{equation}
\label{TFP}
\log TFP_{jt} \approx \omega_{H,jt}+\varepsilon_{S,jt}\omega_{S,jt}+\varepsilon_{U,jt}\omega_{U,jt}
\end{equation}
where $\varepsilon_{S,jt}$ and $\varepsilon_{U,jt}$ denote the output elasticities of skilled and unskilled labor, respectively.

\section{Estimation Strategy}\label{s:empirics}
This section develops our empirical strategy in five steps. We first estimate the production function via GMM (Steps 1--3), embedding equilibrium conditions and Markov assumptions governing the productivity process. We then apply a univariate Kalman filter to decompose estimated production residuals into a Hicks-neutral productivity component and measurement error (Step 4). Finally, we apply a matched difference-in-differences design (Step 5) to estimate the dynamic productivity gains from export entry, comparing new exporters with never-exporting controls matched on the model's state variables.

We normalize the CES production function using geometric means, following \citet{grieco2016production}.\footnote{This normalization ensures the normalized factor share parameters capture average marginal returns to inputs. It also simplifies computation by allowing these parameters to depend solely on the data and, at most, a single unknown common parameter. Appendix \ref{a:geo_means} provides the theoretical foundations and implementation details.} For each variable $X_{jt}$, we define its normalized counterpart $\ddot{X}_{jt} = X_{jt}/\bar{X}$, where $\bar{X} = \left(\prod_{n=1}^N X_n\right)^{1/N}$ is the geometric mean across plants. We denote the normalized CES factor share parameters by $\alpha_X$ for $X \in \{K, M, L, S, U\}$.

\par\medskip
\noindent\textit{Step 1. Estimating the Relative Demand for Skilled versus Unskilled Labor} -- We denote plant-level expenditures on materials, skilled labor, and unskilled labor by $E_{M,jt}$, $E_{S,jt}$, and $E_{U,jt}$, with total payroll $E_{L,jt} = E_{S,jt} + E_{U,jt}$. The first-order conditions from the normalized profit-maximization problem (Appendix \ref{a:geo_means}) yield relative demand for skilled versus unskilled labor:
\begin{equation}
\label{eq:omega_US_char}
\log\left(\frac{\ddot{S}_{jt}}{\ddot{U}_{jt}}\right)= -{\sigma_{\theta}}\log\left(\frac{\ddot{W}_{S,jt}}{\ddot{W}_{U,jt}}\right) + \widetilde{\ddot\omega}_{R,jt},
\end{equation}
where $\sigma_{\theta} = \frac{1}{1-\theta}$ is the elasticity of substitution between skilled and unskilled workers, and the residual $\widetilde{\ddot\omega}_{R,jt} = (1-\sigma_{\theta})\ddot{\omega}_{R,jt}$ is a rescaled measure of the relative efficiency of unskilled labor, which is endogenous to the skill premium. We estimate $\sigma_{\theta}$ by combining equation \eqref{eq:omega_US_char} with the normalized Markov process for $\ddot\omega_{R,jt}$ via GMM. Following \citet{blundell2000gmm}, we take $\rho_{R}$-differences of equation \eqref{eq:omega_US_char} and rearrange to construct the moment condition:
\begin{equation}
\label{eq:bb_R}
\begin{aligned}
\mathbb{E}\Bigg[
&\log\frac{\ddot{S}_{jt}}{\ddot{U}_{jt}} - \rho_{R}\log\frac{\ddot{S}_{jt-1}}{\ddot{U}_{jt-1}}
+ \sigma_{\theta}\left\{\log\frac{\ddot{W}_{S,jt}}{\ddot{W}_{U,jt}} - \rho_{R}\log\frac{\ddot{W}_{S,jt-1}}{\ddot{W}_{U,jt-1}}\right\} \\
&\quad - \beta_{R}^{e} e_{jt-1} - \beta_{R}^{i} i_{jt-1} - \beta_{R}^{ei} (e_{jt-1} \cdot i_{jt-1})
\,\Bigg|\, \iota_{t}, \iota_{s}, Z_{R,jt}
\Bigg] = 0,
\end{aligned}
\end{equation}
where $Z_{R,jt}$ denotes our instrument set.

We identify the Step 1 parameters using the exporting and investment dummies, their interactions, the lagged relative allocation $\log\left(\frac{\ddot{S}_{jt-1}}{\ddot{U}_{jt-1}}\right)$, the lagged log skill premium $\log\left(\frac{\ddot{W}_{S,jt-1}}{\ddot{W}_{U,jt-1}}\right)$, and a shift-share instrument for the skill premium in the spirit of \citet{raval2019micro}. The two lagged instruments are uncorrelated with the period-$t$ innovation under the first-order Markov assumption and remain informative about current relative demand because productivity is persistent. The shift-share instrument provides the supply-side variation in the skill premium needed to identify $\sigma_\theta$, which we estimate from the relative demand for skilled labor. It exploits regional exposure to skill-biased labor-demand shocks originating in non-sampled industries, which shift the premium sampled plants take as given.

Specifically, for region $r$ and period $t$,\footnote{Here $r$ indexes Colombian administrative departments (see Figure~\ref{fig:map_Colombia} in Appendix~\ref{a:map_Colombia}). So that $Z^{R1}_{rt}$ is well defined at the plant level, the estimation sample keeps only plants with stable department assignments and excludes the isolated San Andrés archipelago. Both restrictions apply sample-wide.} we define the instrument $Z_{rt}^{R1}$ as:
\begin{equation}
      Z_{rt}^{R1} = \sum_{I \notin \mathcal{F}} \left( s_{rI,t-4}^{S} - s_{rI,t-4}^{U} \right) \left( g_{It}^{S} - g_{It}^{U} \right),
\end{equation}
where $\mathcal{F}$ is the set of 19 sampled industries, $s_{rI,t-4}^{S} = S_{rI,t-4} / S_{r,t-4}$ denotes the share of region $r$'s skilled employment concentrated in industry $I$ (lagged four years), $s_{rI,t-4}^{U}$ is defined analogously for unskilled workers, and
\begin{equation}
      g_{It}^{X} = \frac{\log X_{It} - \log X_{I,t-4}}{4}, \quad X \in \{S, U\},
\end{equation}
is the annualized national-industry growth rate. Summing over non-sampled industries is intended to reduce contamination from plants' own-industry shocks. The four-year lag on the exposure shares makes them predetermined with respect to contemporaneous plant-level decisions. It also leaves the instrument undefined before 1985, so we estimate the Step-1 moment conditions on the subsample of plant-years for which it is available.

\par\medskip
\noindent\textit{Step 2. Estimating the Relative Demand for Materials versus Labor} -- Substituting equation \eqref{eq:omega_US_char} into the normalized labor aggregator\footnote{See equation \eqref{eq:lnorm1} in Appendix \ref{a:geo_means}.} and evaluating at $\hat{\sigma}_\theta$, we obtain the closed-form expression for aggregate labor:
\begin{equation}
\label{eq:L_char}
    \hat{\ddot{L}}_{jt} = \ddot{S}_{jt}\left(\frac{\ddot{E}_{L,jt}}{\ddot{E}_{S,jt}}\right)^{\frac{\hat{\sigma}_\theta}{\hat{\sigma}_\theta-1}},
\end{equation}
where $\ddot{E}_{L,jt}=\frac{E_{S,jt}+E_{U,jt}}{\bar{E}_{S}+\bar{E}_{U}}$. Substituting this expression into the first-order conditions and taking the ratio of materials to aggregate labor yields:
\begin{equation}
\begin{aligned}
\label{eq:omega_S_char}
\log\left(\frac{\ddot{M}_{jt}}{\ddot{L}_{jt}}\right)= -\sigma_{\rho} \log\left( \frac{\ddot{P}_{M,jt}}{\ddot{W}_{L,jt}}\right)+ \widetilde{\ddot{\omega}}_{L,jt},
\end{aligned}
\end{equation}
where $\ddot{W}_{L,jt} = \ddot{E}_{L,jt}/\ddot{L}_{jt}$ is the aggregate labor wage rate, $\sigma_{\rho} = \frac{1}{1-\rho}$ is the elasticity of substitution between labor and materials, and $\widetilde{\ddot\omega}_{L,jt} = (1-\sigma_{\rho}) \ddot{\omega}_{L,jt}$ is a rescaled measure of labor-augmenting productivity, endogenous to the ratio of wages to material prices. We estimate $\sigma_{\rho}$ and the Markov process parameters for $\ddot\omega_{L,jt}$ via GMM, combining equations \eqref{eq:omega_S_char} and \eqref{eq:L_char} with the Markov process for $\ddot{\omega}_{L,jt}$. As in Step 1, we take $\rho_{L}$-differences of equation \eqref{eq:omega_S_char} and rearrange to construct the moment condition:
\begin{equation}
\label{eq:bb_L}
\begin{aligned}
\mathbb{E}\Bigg[
&\log\frac{\ddot{M}_{jt}}{\ddot{L}_{jt}} - \rho_{L}\log\frac{\ddot{M}_{jt-1}}{\ddot{L}_{jt-1}}
+ \sigma_{\rho}\left\{\log\frac{\ddot{P}_{M,jt}}{\ddot{W}_{L,jt}} - \rho_{L}\log\frac{\ddot{P}_{M,jt-1}}{\ddot{W}_{L,jt-1}}\right\} \\
&\quad - \beta_{L}^{e} e_{jt-1} - \beta_{L}^{i} i_{jt-1} - \beta_{L}^{ei} (e_{jt-1} \cdot i_{jt-1})
\,\Bigg|\, \iota_{t}, \iota_{s}, Z_{L,jt}
\Bigg] = 0,
\end{aligned}
\end{equation}
where the bracketed expression is the innovation $\xi_{L,jt}$ to labor-augmenting productivity and $Z_{L,jt}$ denotes our instrument set.

As in Step 1, we identify the Step 2 parameters using the exporting and investment dummies, their interactions, the lagged material-to-labor ratio $\log\!\left(\frac{\ddot{M}_{jt-1}}{\ddot{L}_{jt-1}}\right)$, and the lagged expenditure ratio of materials to skilled labor $\log\!\left(\frac{E_{M,jt-1}}{E_{S,jt-1}}\right)$. Both lagged instruments are uncorrelated with the period-$t$ innovation $\xi_{L,jt}$ under the first-order Markov assumption. The lagged material-to-labor ratio remains informative about current relative input demand because productivity is persistent, whereas the lagged expenditure ratio supplies the plant-level variation needed to identify $\sigma_{\rho}$, which the industry-year material price index cannot provide.\footnote{Appendix \ref{a:rel_inputs} sets out the inner-nest identity behind this argument and the predeterminedness assumption it requires.}

\par\medskip
\noindent\textit{Step 3. Estimating $(\alpha_K,\alpha_L,\alpha_M)$ and the Markov process for $\ddot{\omega}_{H,jt}$} -- We define the CES component of production as
\begin{equation}
    f(\boldsymbol{\alpha}, \sigma_\rho, \ddot{X}_{jt}) =
    \left[
    \alpha_K \ddot{K}_{jt}^{\frac{\sigma_\rho-1}{\sigma_\rho}}
    + \alpha_M \ddot{M}_{jt}^{\frac{\sigma_\rho-1}{\sigma_\rho}}
    + \alpha_L \left( \exp(\ddot{\omega}_{L,jt}) \ddot{L}_{jt} \right)^{\frac{\sigma_\rho-1}{\sigma_\rho}}
    \right]^{\frac{\sigma_\rho}{\sigma_\rho-1}},
\end{equation}
where $\boldsymbol{\alpha} = (\alpha_K,\alpha_L,\alpha_M)$ and $\ddot{X}_{jt} = (\ddot{K}_{jt}, \ddot{M}_{jt}, \ddot{L}_{jt}, \ddot{\omega}_{L,jt})$. We either observe $\sigma_\rho$ and $\ddot{X}_{jt}$ directly in the data or recover them from the previous steps. Substituting the inverse expression for $\ddot{\omega}_{L,jt}$ into the CES aggregator, and imposing the functional forms for the normalized factor share parameters described in Appendix \ref{a:geo_means}, yields
\begin{align}
    \log f(\tau, \sigma_\rho, \ddot{X}_{jt})
    &\equiv \frac{\sigma_\rho}{\sigma_\rho - 1}\log \alpha_M(\tau) + \log \ddot{M}_{jt}
    + \frac{\sigma_\rho}{\sigma_\rho - 1}
    \log \left[
    \tau \left( \frac{\ddot{K}_{jt}}{\ddot{M}_{jt}} \right)^{\frac{\sigma_\rho-1}{\sigma_\rho}}
    + \frac{TVC_{jt}}{E_{M,jt}}
    \right],
    \label{eq:normalized_production}
\end{align}
where $TVC_{jt} = E_{M,jt} + E_{L,jt}$, and $\tau$ is a capital-share parameter whose interpretation and calibration we develop below.

Combining \eqref{eq:normalized_production} with the demand equation \eqref{eq:output_demand} yields the following revenue equation:
\begin{align}
    \log \frac{\ddot{\tilde{R}}_{jt}}{\ddot{P}_{It}}
    =
    - \frac{1}{\eta} \log \frac{\ddot{R}_{It}}{\ddot{P}_{It}}
    + \frac{\eta + 1}{\eta} \log f(\tau, \widehat{\sigma}_{\rho}, \ddot{X}_{jt})
    + \frac{\eta + 1}{\eta}{\ddot{\omega}}_{H,jt}
    + \zeta_{jt},
    \label{eq:revenue_generating_eq}
\end{align}
where $\ddot{\tilde{R}}_{jt}$ denotes normalized plant revenue, and $\ddot{P}_{It}$ and $\ddot{R}_{It}$ denote the industry-level price and revenue indices, respectively. As in the previous steps, Hicks-neutral productivity is endogenous to current input choices, in particular materials and total variable costs, $(\ddot{M}_{jt},TVC_{jt})$. The term $\zeta_{jt}$ captures end-of-period measurement error in revenue.

We estimate $(\tau,\eta)$ jointly with the Markov process for $\ddot{\omega}_{H,jt}$ using a combination of calibration and GMM. Following \citet{blundell2000gmm}, we quasi-difference \eqref{eq:revenue_generating_eq} by $\rho_H$ and rearrange to construct the moment condition:
\begin{equation}
\label{eq:bb_H}
\begin{aligned}
\mathbb{E}\Bigg[
&\log \frac{\ddot{\tilde{R}}_{jt}}{\ddot{P}_{It}} - \rho_H \log \frac{\ddot{\tilde{R}}_{jt-1}}{\ddot{P}_{It-1}}
+ \frac{1}{\eta}\left\{ \log \frac{\ddot{R}_{It}}{\ddot{P}_{It}} - \rho_H \log \frac{\ddot{R}_{It-1}}{\ddot{P}_{It-1}} \right\} \\
&\quad - \frac{\eta + 1}{\eta}\left\{ \log f(\tau, \widehat{\sigma}_{\rho}, \ddot{X}_{jt}) - \rho_H \log f(\tau, \widehat{\sigma}_{\rho}, \ddot{X}_{jt-1}) \right\} \\
&\quad - \beta_H^e e_{jt-1}
    - \beta_H^i i_{jt-1}
    - \beta_H^{ei}(e_{jt-1}\cdot i_{jt-1})
\,\Bigg|\, \iota_t, \iota_s, Z_{H,jt}
\Bigg] = 0.
\end{aligned}
\end{equation}
The error inside this moment condition is the composite $\chi_{jt} = \tilde{\xi}_{H,jt} + \zeta_{jt} - \rho_H \zeta_{jt-1}$, combining the innovation to Hicks-neutral productivity with revenue measurement error, and $Z_{H,jt}$ denotes the instrument set. Specifically, we first exploit our normalization and calibrate $\tau$ directly as the ratio of the geometric mean of capital service expenditures (Appendix \ref{a:data_construction}) to the geometric mean of material expenditures:
\begin{align}
    \tau=\frac{\bar{E}_K}{\bar{E}_M}. \label{eq:tau_calibration}
\end{align}

The parameter $\tau$ is the wedge between capital's static first-order condition and its observed expenditure ratio (Appendix \ref{a:geo_means}). Capital is chosen through the dynamic investment problem \eqref{eq:invest_e}, so its static condition need not hold at the sample mean, and $\tau$ measures the mean deviation of the capital stock from its static optimum \citep{grieco2016production}. Setting $\tau=\bar{E}_K/\bar{E}_M$ imputes capital's rental at its user cost and recovers the familiar cost-share form for the capital weight.\footnote{We calibrate rather than estimate $\tau$ because the revenue moments identify it only weakly, as is common when capital variation is limited. Appendix \ref{a:tau_sensitivity} reports the variant in which $\tau$ is estimated jointly.}

To estimate $\eta$ and the Markov process of $\ddot{\omega}_{H,jt}$, we employ the lagged export and investment dummies, their interaction, and lagged materials, $\ddot{M}_{jt-1}$. Lagged materials are predetermined with respect to innovations in Hicks-neutral productivity and uncorrelated with $\zeta_{jt}$, the contemporaneous revenue measurement error, yet remain informative about current production because productivity is persistent. We include the lagged capital-to-material ratio $\frac{\ddot{K}_{jt-1}}{\ddot{M}_{jt-1}}$, lagged in the denominator because $\ddot{M}_{jt}$ is endogenous to $\xi_{H,jt}$, as an additional instrument that provides predetermined variation in $\log f(\cdot)$, helping identify the Markov persistence $\rho_H$. Identification of $\eta$ then relies on lagged aggregate industry deflated revenue, $\log\!\left(\frac{\ddot{R}_{It-1}}{\ddot{P}_{It-1}}\right)$, under the exclusion restriction that it is uncorrelated with the period-$t$ composite error $\chi_{jt}$ conditional on sector and year fixed effects; because each plant is a negligible share of its industry aggregate, the plant's own lagged revenue measurement error does not contaminate the instrument.

At this point, we recover the composite term
\[
\widetilde{\omega}_{jt}
=
\widehat{\left(
\ddot{\omega}_{H,jt}
+
\left(\frac{\eta}{1+\eta}\right)\zeta_{jt}
\right)},
\]
which combines Hicks-neutral productivity with measurement error.

\par\medskip
\noindent{\textit{Step 4. Filtering $\ddot{\omega}_{H,jt}$ and $\zeta_{jt}$}} -- Step 3 yields $\widetilde{\omega}_{jt}$, a composite measure conflating Hicks-neutral productivity and measurement error. To disentangle these components, we apply Kalman filtering \citep{hamilton1994filter}, exploiting their differing persistence.

We cast the quasi-differenced revenue equation in state-space form using parameter estimates from Steps 1--3. Define the state vector collecting the unobserved components, $X_{jt} = [\xi_{H,jt} ~ \zeta_{jt} ~ \zeta_{jt-1}]'$. The quasi-differenced equation is then represented by the measurement equation:
\begin{equation}
\begin{aligned}
\label{eq:state_space_measurement}
       \widetilde{\omega}_{jt} &= \hat{\rho}_{H} \widetilde{\omega}_{jt-1} + \begin{bmatrix} 1 & \frac{\hat{\eta}}{1+\hat{\eta}} & -\frac{\hat{\eta}}{1+\hat{\eta}}\hat{\rho}_H \end{bmatrix} X_{jt} \\ &\quad + \hat{\beta}_{H}^{e} e_{jt-1} + \hat{\beta}_{H}^{i} i_{jt-1} + \hat{\beta}_{H}^{ei} (e_{jt-1} \cdot i_{jt-1}) + \hat{\iota}_{H,t} + \hat{\iota}_{H,s}.
       \end{aligned}
\end{equation}
The state vector $X_{jt}$ evolves according to the transition equation:
\begin{equation}
\label{eq:state_space_transition}
  \begin{bmatrix} \xi_{H,jt+1} \\ \zeta_{jt+1} \\ \zeta_{jt}\end{bmatrix}=\begin{bmatrix} 0 & 0 & 0 \\ 0 & 0 & 0 \\ 0 & 1 & 0 \end{bmatrix}\begin{bmatrix} \xi_{H,jt} \\ \zeta_{jt} \\ \zeta_{jt-1}\end{bmatrix}+\begin{bmatrix} 1 & 0 \\ 0 & 1 \\ 0 & 0 \end{bmatrix}\begin{bmatrix} \xi_{H,jt+1} \\ \zeta_{jt+1}\end{bmatrix}.
\end{equation}
The innovation vector $[\xi_{H,jt} ~ \zeta_{jt}]'$ is white noise with diagonal covariance $\text{diag}(\sigma_H^2, \sigma_\zeta^2)$ and is orthogonal to the entire history of past states.

We estimate the variance parameters $(\sigma_H^2, \sigma_\zeta^2)$ by quasi-maximum likelihood, with the likelihood constructed by the Kalman filter under Gaussian approximations for $[\xi_{H,jt} ~ \zeta_{jt}]'$.\footnote{Distributional assumptions on the innovation vector are unnecessary: the pseudo-likelihood delivers consistent and asymptotically Normal estimates under regularity conditions \citep{hamilton1994filter}.} With variance estimates $\hat{\sigma}_H$ and $\hat{\sigma}_\zeta$ in hand, we run the smoothing recursions \citep{rauch1965maximum, hamilton1994smooth} to extract $\hat{\xi}_{H,jt}$ and $\hat{\zeta}_{jt}$ across all observations, thereby recovering $\hat{\ddot{\omega}}_{H,jt}$. Appendix \ref{a:kalman} details the filtering and smoothing algorithms along with the likelihood function construction.

\par\medskip
\noindent {\textit{Step 5. Matched Difference-in-Differences to Estimate Productivity Gains from Exporting}} -- We analyze productivity dynamics around export entry. The productivity Markov-process coefficients \citep{de2013detecting} control for self-selection through prior productivity but do not capture how export effects evolve over time, for example by building gradually after entry.

We therefore implement a matched difference-in-differences approach \citep{heckman1997matching, de2007exports, garcia2019exporting} to estimate the average treatment effect on plants that entered export markets during the sample period and persisted thereafter.\footnote{We exclude plants that exported continuously or exited export markets during the period.} The estimator combines two steps: a propensity-score match between each new exporter and observationally equivalent never-exporting controls, followed by an export entry cohort-stacked difference-in-differences on the matched panel \citep{cengiz2019effect} to address the staggered nature of export entry.\footnote{See \citet{de2020two, callaway2021difference, goodman2021difference, sun2021estimating} for related advances in the staggered difference-in-differences literature.}

According to the exporting choice probability in Equation \eqref{eq:CCP}, conditional on the state vector $(\ddot{\Omega}_{jt}, e_{jt})$, export entry depends only on the i.i.d. sunk-cost draw $\gamma_{jt}$, so the state variables constitute the relevant conditioning set for matching \citep{rosenbaum1983central}. We therefore match each new exporter, in the year preceding export entry, with the three nearest never-exporting controls within the same entry cohort and 3-digit SIC industry. This stratified matching fixes the industry-year states and pre-entry export status by construction. Hence, we can match on a logit propensity score built from the other plant-level state variables within each stratum.\footnote{We impose a caliper of 0.1 on the propensity score.} As we match with replacement, a never-exporter can serve as a control for more than one new exporter. Beyond this ignorability condition, identification requires parallel trends in potential outcomes between new exporters and their matched controls within each cohort window.

We then estimate dynamic treatment effects on the matched panel. Each cohort $c$, indexed by its entry year $T_c$, contributes an analysis sample of the new exporters entering in $T_c$ and the never-exporting controls matched to any of them, restricted to the four years before and three years after entry. Stacking all cohort samples and indexing each observation by plant $j$, cohort $c$, and calendar year $t$, we estimate:
\begin{align}
    y_{jct} = \alpha_{jc} + \delta_{tc} + \sum_{\substack{h = -4 \\ h \neq -1}}^{3} \beta_h \cdot \mathbb{1}\{t - T_c = h\} \cdot D_{jc} + u_{jct}, \label{eq:diff_in_diff}
\end{align}
where $D_{jc}$ equals one if plant $j$ is a new exporter in cohort $c$ and zero otherwise. The cohort-specific plant fixed effects $\alpha_{jc}$ allow the same control plant to receive a separate intercept in each cohort it serves, and the cohort-specific year fixed effects $\delta_{tc}$ absorb aggregate shocks within each cohort's event window. The coefficients $\beta_h$ estimate the average treatment effect on the treated (ATT) for new exporters that continue to export, at event time $h$ relative to the omitted reference $h = -1$. For the total-factor-productivity outcome, we use $\log TFP_{jt}$ from \eqref{TFP}, freezing the output elasticities at the cohort baseline $T_c - 2$ to isolate productivity dynamics from contemporaneous elasticity changes.

For notational simplicity, we suppress the normalization symbol ($\ddot{\phantom{x}}$) hereafter; all subsequent results pertain to the normalized model.

\section{Results and Discussion}\label{s:results}
\subsection{Estimates of the Production Function Model Parameters}
Table \ref{tab:structural} presents the baseline estimates of the production function parameters. Skilled and unskilled labor act as gross complements in Colombian manufacturing sectors during the sample period. The estimated elasticity of substitution, \( \sigma_{\theta} \), of 0.344 implies that, for a cost-minimizing plant facing constant relative wages, an increase (decrease) in the relative productivity of unskilled workers translates into a smaller (larger) relative use of unskilled labor.

This estimate deviates from the consensus elasticity of about 1.5 reported by \citet{katz1992changes} and \citet{autor2008trends}, and from the average in the meta-analysis by \citet{Havranek2024pub}.\footnote{We re-estimate \( \sigma_{\theta} \) by two-stage least squares in Section~\ref{s:robustness}, instrumenting relative wages with local skill-premium variation across departments and industries. The four instruments with strong first stages yield estimates between 0.47 and 0.76, all below one and none significantly different from the baseline, reinforcing gross complementarity (Table~\ref{tab:2SLS}).} We conjecture that the gap reflects our occupational, rather than educational, definition of skill: the skilled and unskilled categories defined in Section~\ref{s:data} map to distinct plant functions that complement one another in production rather than competing for the same work. Such functions may substitute less readily than coarser educational classifications that pool distinct occupations together. This complementarity opens the possibility that the observed increase in skilled labor intensity among exporters reflects productivity gains that specifically enhance the efficiency of unskilled labor, a hypothesis we evaluate in the next subsection.

\begin{table}[!ht]
\caption{Estimates of Structural Parameters}
\resizebox{\textwidth}{!}{%
\begin{tabular}{ccccccccccccc}
\toprule\toprule
 $\mu$ & & $\sigma_{\rho}$ & $\sigma_{\theta}$ & & $\alpha_L$ & $\alpha_M$ & $\alpha_K$ & $\alpha_S$ & $\alpha_U$ & & $\sigma_H$ & $\sigma_\zeta$ \\
\midrule
 1.091 & & 0.399 & 0.344 & & 0.233 & 0.702 & 0.065 & 0.317 & 0.683 & & 0.263 & 0.109 \\
 (0.025) & & (0.105) & (0.022) & & (0.002) & (0.002) & (0.001) & (0.002) & (0.002) & & (0.026) & (0.025) \\
\bottomrule
\end{tabular}

}
\note{This table reports estimates of the model's structural parameters for the baseline calibrated specification. The markup is defined as \( \mu = \eta/(\eta+1) \), with an implied demand elasticity \( \hat{\eta} = -11.97 \) (bootstrap SE 7.36). The share parameters \( \alpha_L \), \( \alpha_M \), and \( \alpha_K \) are derived from the calibrated ratio \( \tau = \bar{E}_K / \bar{E}_M \) and the input-share geometric means; their small bootstrap standard errors reflect the stability of the input-expenditure means across resamples. Bootstrap standard errors appear in parentheses. We draw 1,000 bootstrap samples with replacement, clustered at the establishment level, and re-estimate the structural model for each sample; 998 replications converged successfully. Appendix \ref{a:tau_sensitivity} reports the corresponding estimates under the alternative specification in which \( \tau \) is estimated jointly with the other Step-3 parameters. \( \sigma_\theta \) and \( \sigma_\rho \) denote the elasticities of substitution between skilled and unskilled labor and between the labor composite, capital, and materials, respectively. Estimates use the structural panel of 10,023 plants (50,329 plant-year observations).}
\label{tab:structural}
\end{table}
We estimate the outer-nest elasticity of input substitution, \( \sigma_{\rho} \), at 0.399, which indicates strong complementarity between capital, materials, and composite labor.\footnote{This estimate closely aligns with the findings of \citet{raval2019micro} and the broader insights from the meta-analysis by \citet{gechert2022measuring}.} The estimated markup of 1.091 corresponds to an implied \( \hat{\eta} = -11.97 \).\footnote{We report the markup \( \mu = \eta/(\eta+1) \) rather than the demand elasticity \( \eta \) in levels because \( \eta \) is bounded above by \( -1 \) and has a right-skewed bootstrap distribution, whereas \( \mu \) is unconstrained and near-normal.} We also recover the CES share parameters, which, under our normalization, represent the average marginal returns of each input. They are 0.233 for composite labor, 0.702 for materials, and 0.065 for capital, as well as 0.317 for skilled labor and 0.683 for unskilled labor. Finally, we estimate the standard deviation of the innovation to Hicks-neutral productivity, \(\sigma_H\), at 0.263 and the standard deviation of measurement error, \(\sigma_\zeta\), at 0.109, which yields a signal-to-noise ratio of 2.41 (the ratio of standard deviations \(\sigma_H/\sigma_\zeta\)).\footnote{The Hansen-J overidentification test fails to reject the moment conditions at Steps 1 and 3, with plant-clustered \(p\)-values of 0.754 and 0.523. Step 2 is just-identified. The excluded instruments are strong at both steps. In the first stages implied by each step's moment conditions, the plant-clustered Kleibergen--Paap \(rk\) Wald \(F\) is 584.4 at Step 1 (the shift-share instrument \( Z^{R1} \) and the lagged skill premium) and 95.4 at Step 2 (the lagged materials-to-skilled-labor expenditure ratio).}

In Table \ref{tab:lbe}, we present the baseline estimates for the coefficients of the Markov productivity processes, each modeled as an AR(1). Each productivity component exhibits high persistence, with autoregressive coefficients of 0.780 for Hicks-neutral productivity, 0.864 for labor-augmenting productivity, and 0.855 for unskilled labor relative productivity. Past exporting is associated with a 6.0\% higher conditional mean of unskilled labor relative productivity, but we find no statistically significant conditional association between past exporting and either Hicks-neutral or labor-augmenting productivity. Past investment in machinery and equipment is associated with a 3.1\% higher conditional mean of labor-augmenting productivity and a 6.2\% higher conditional mean of unskilled labor relative productivity; its conditional association with Hicks-neutral productivity is not statistically significant.

\begin{table}[!h]
\caption{Estimates of the Productivity Markov Process Coefficients}
\begin{tabular}{lcccc}
\toprule\toprule
 & $\ddot{\omega}_{-1}$ & $e_{-1}$ & $i_{-1}$ & $e_{-1}i_{-1}$ \\
\midrule
$\ddot{\omega}_H$ & 0.780 & 0.028 & 0.002 & 0.006 \\
 & (0.032) & (0.018) & (0.010) & (0.017) \\
$\ddot{\omega}_L$ & 0.864 & 0.039 & 0.031 & 0.023 \\
 & (0.004) & (0.040) & (0.011) & (0.041) \\
$\ddot{\omega}_R$ & 0.855 & 0.060 & 0.062 & 0.001 \\
 & (0.004) & (0.028) & (0.009) & (0.028) \\
\bottomrule
\end{tabular}

\note{This table presents the coefficients for the productivity Markov process in the baseline calibrated specification. All regressions include industry and year fixed effects. Standard errors are computed using a nonparametric bootstrap clustered at the establishment level. We draw 1,000 bootstrap samples with replacement and re-estimate the structural model for each sample; 998 replications converged successfully. Appendix \ref{a:tau_sensitivity} reports the corresponding estimates under the alternative specification in which \( \tau \) is estimated. The regressors \( e_{-1} \) and \( i_{-1} \) are one-year-lagged indicators for export status and equipment investment, and \( e_{-1} i_{-1} \) is their interaction. Estimates use the structural panel of 10,023 plants (50,329 plant-year observations).}
\label{tab:lbe}
\end{table}

These average effects, however, mask compositional heterogeneity by combining new exporters, who likely experience stronger effects, with continuing exporters. Additionally, although controlling for prior productivity levels mitigates selection bias \citep{de2013detecting}, the control group includes a mix of never-exporters, future exporters, and firms that exited the export market. For these reasons, we now turn to our preferred analysis of local, factor-biased productivity gains from exporting specifically for new exporters.

\subsection{Local Estimates of Factor-Biased Productivity Dynamics around Export Entry}

The PSM procedure matches 305 new exporters with 655 never-exporting controls. Before estimating these dynamics, we verify that this match constructs a suitable control group. The matched sample satisfies the two standard pre-treatment diagnostics, common support and balance on the matching covariates, which we report in Appendix \ref{a:appendix_matching}.

Panel A of Figure \ref{fig:event-study-combined} reports the matched DiD estimates from equation \eqref{eq:diff_in_diff} for the model's factor-augmenting productivity components over the four years before and three years after export entry. Against flat pre-trends, unskilled-augmenting productivity ($\ddot\omega_U$) rises by 22\% in the year of entry and by 35\% on average over the post-entry window. Skilled-augmenting productivity ($\ddot\omega_S$), by contrast, shows no detectable change, so the gain is driven almost entirely by the relative-unskilled component ($\ddot\omega_R$), which captures the efficiency of unskilled labor relative to skilled.\footnote{$\ddot\omega_U = \ddot\omega_S + \ddot\omega_R$ by construction, so unskilled-augmenting productivity decomposes into the skilled-augmenting and relative-unskilled components.} We also detect no Hicks-neutral ($\ddot\omega_H$) effect. TFP rises modestly, by 8\% on average over the post-entry window (Panel B of Figure~\ref{fig:event-study-combined}).

In contrast to the null finding of \citet{clerides1998learning},\footnote{\citet{clerides1998learning} study the same Colombian plant panel over 1983--1991, a slightly shorter window than our 1981--1991 sample, and find that distributed lags of export experience contribute little to plant marginal costs once plant fixed effects and capital stocks are conditioned on.} we detect a modest post-entry rise in aggregate productivity. The factor-augmenting decomposition identifies this gain as factor-biased rather than Hicks-neutral: it originates in rising unskilled-labor efficiency, not in a neutral improvement common to all inputs. Such bias is invisible to the conventional Hicks-neutral treatment of export-induced productivity gains, which would attribute the modest aggregate movement to a uniform productivity shift and miss the underlying reallocation of efficiency toward unskilled labor. The factor-augmenting model recovers it by separating the input-specific margins a single neutral term collapses into one.

\begin{figure}[!ht]
\caption{Estimated Local Effects of Exporting on Productivity}
\label{fig:event-study-combined}
\centering
\begin{subfigure}[t]{0.46\textwidth}
\centering
\includegraphics[width=\linewidth]{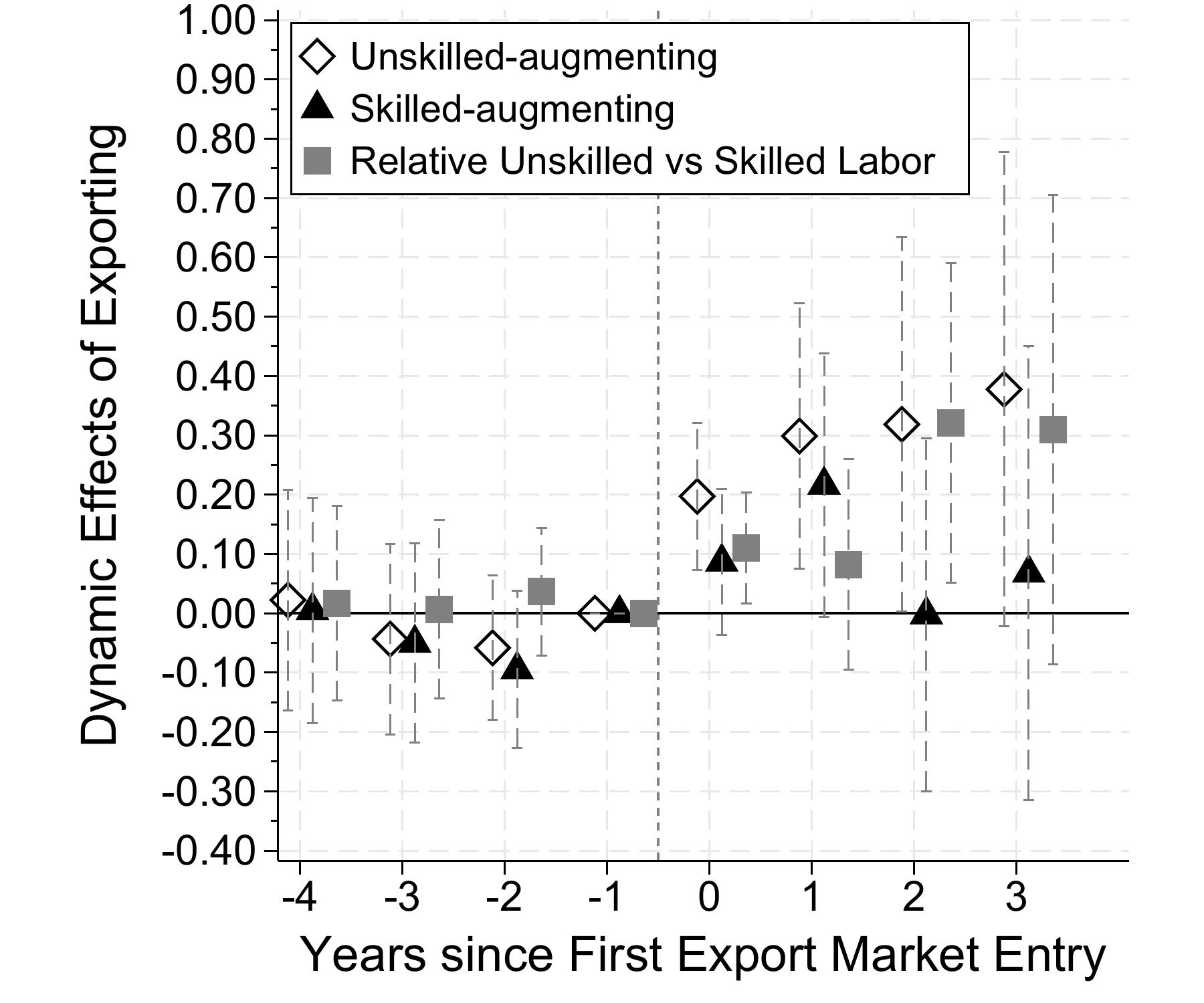}
\subcaption{Factor-Augmenting Productivity}\label{fig:event-study-factor}
\end{subfigure}\hfill
\begin{subfigure}[t]{0.46\textwidth}
\centering
\includegraphics[width=\linewidth]{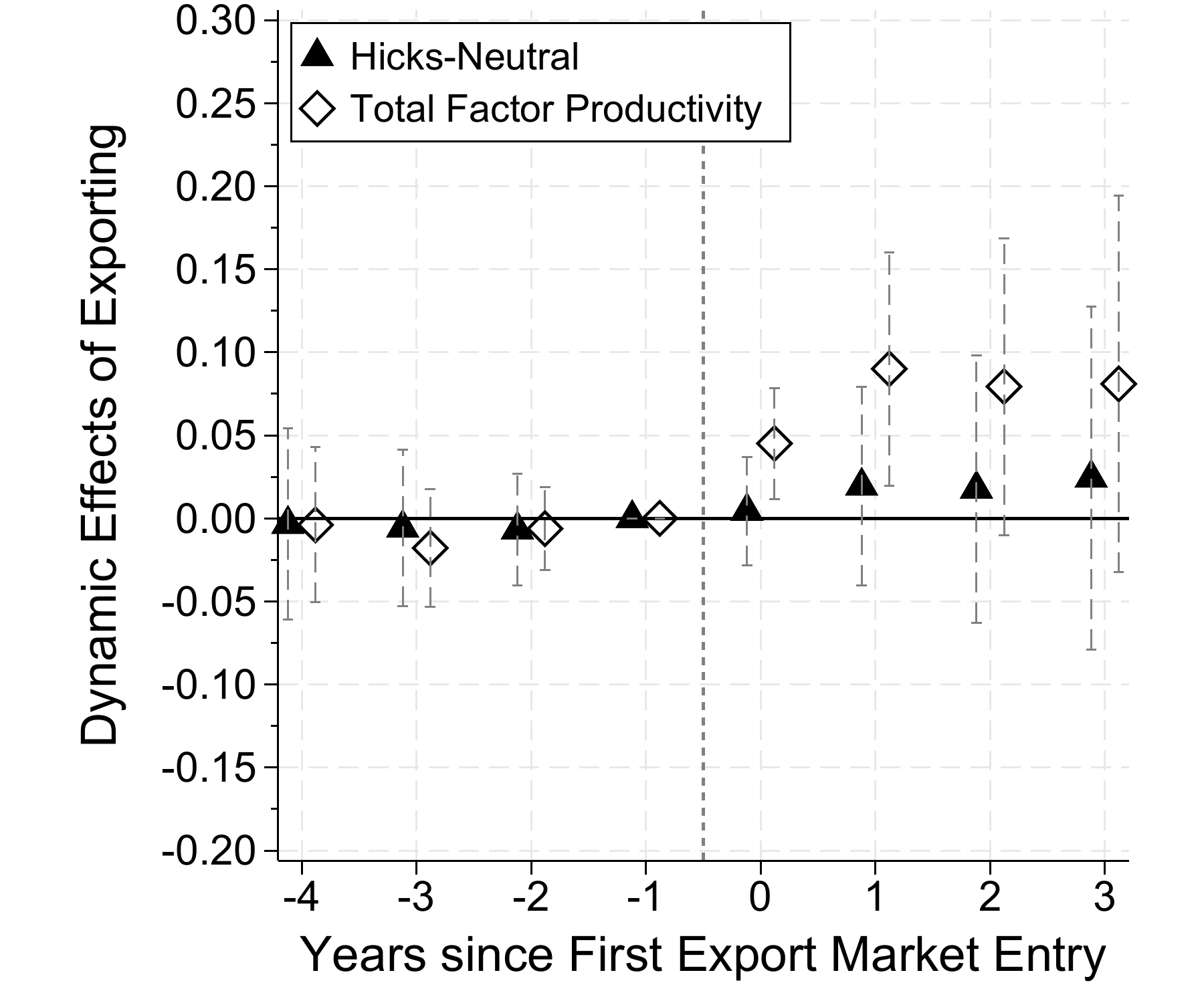}
\subcaption{Hicks-Neutral and Total Factor Productivity}\label{fig:event-study-output}
\end{subfigure}
\note{Panel A: dynamic effects of exporting on skilled-augmenting productivity ($\ddot\omega_S$), unskilled-augmenting productivity ($\ddot\omega_U$), and the relative unskilled-vs-skilled productivity component ($\ddot\omega_R$). Panel B: dynamic effects on Hicks-neutral productivity ($\ddot\omega_H$) and total factor productivity (TFP). Estimates derive from a matched cohort-stacked event-study specification, with 90\% confidence intervals constructed from 998 nonparametric bootstrap replications. The matched sample comprises 305 new exporters and 655 never-exporting controls. The omitted reference period is $h=-1$.}
\end{figure}

\begin{figure}[!ht]
\caption{Estimated Local Effects of Exporting on Capital Assets}
\label{fig:event-study2}
\centering
\includegraphics[width=\linewidth]{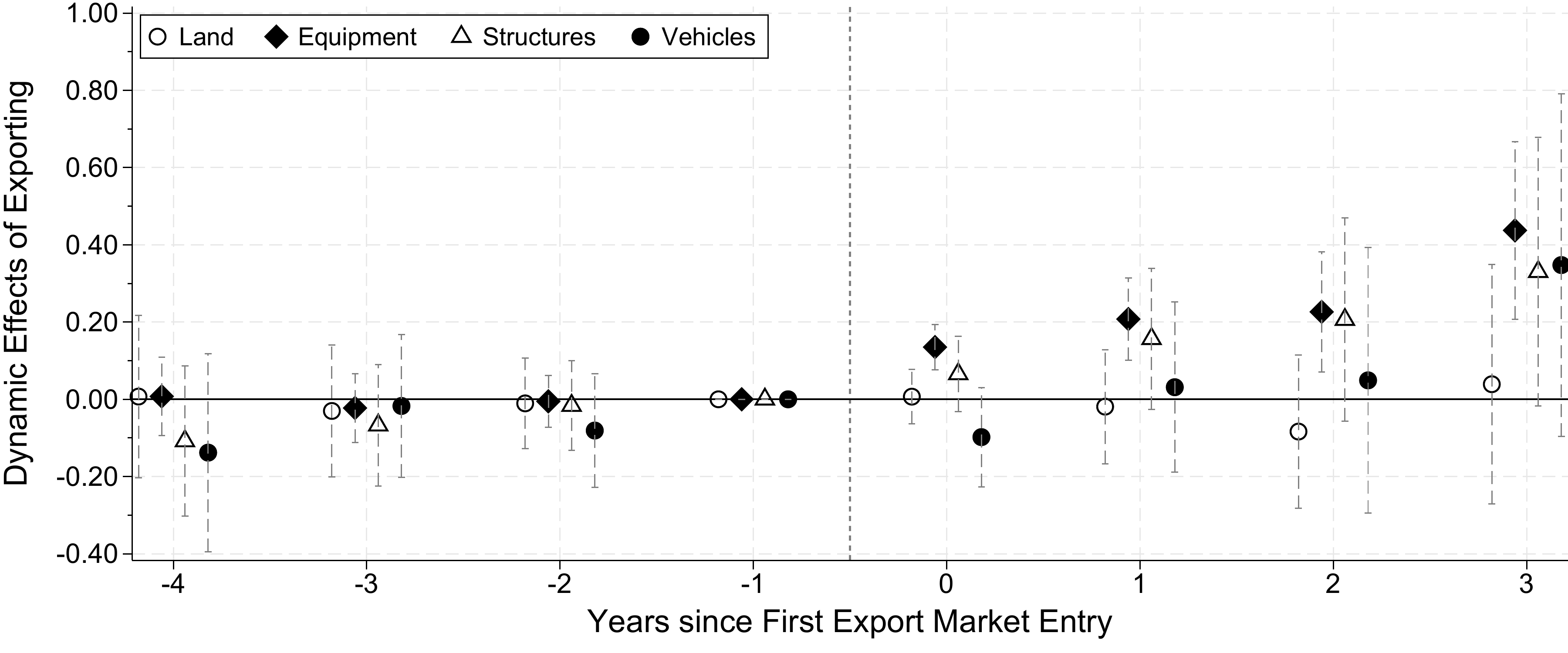}
\note{Dynamic effects of exporting on the four capital types separately observed in the plant panel: land, structures, equipment, and vehicles. Estimates derive from a matched cohort-stacked event-study specification, with 90\% confidence intervals constructed from 998 nonparametric bootstrap replications. The matched sample comprises 305 new exporters and 655 never-exporting controls. The omitted reference period is $h=-1$.}
\end{figure}

These factor-biased productivity dynamics coincide with a shift in plant capital toward equipment. Figure \ref{fig:event-study2} reports event-study estimates for the four capital types separately observed in the plant panel: land, structures, equipment, and vehicles. Equipment rises sharply after entry, by 29\% on average over the post-entry window and 55\% three years after entry; structures rise more modestly, whereas vehicles and land are imprecisely estimated. The joint rise of equipment capital and unskilled-augmenting productivity around export entry is consistent with accounts of technology upgrading upon entering export markets \citep{bustos2011trade, garcia2019exporting}.

\begin{figure}[!ht]
\caption{Dynamic Effects of Exporting on Observed Skill Ratios, Premia, and Employment}
\label{fig:rawdid}
\centering
\includegraphics[width=0.85\textwidth]{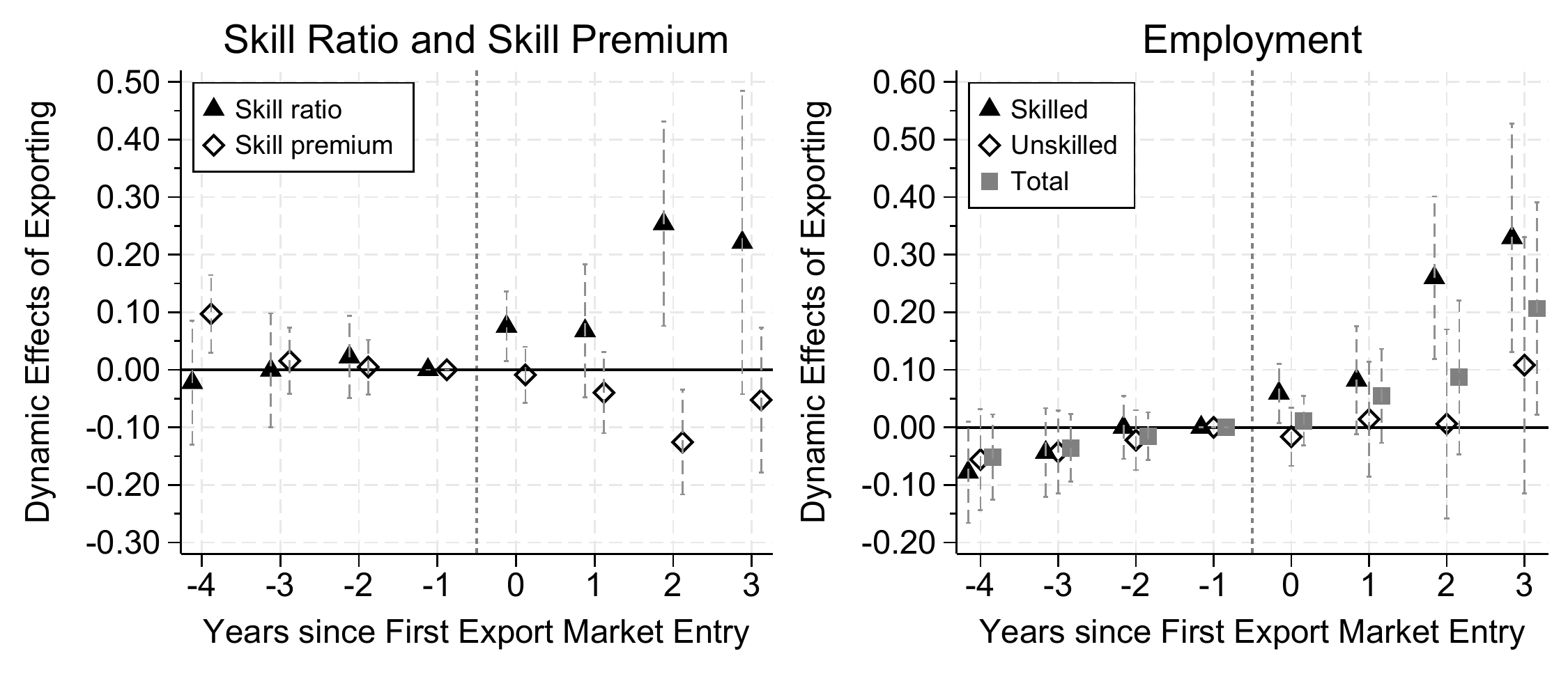}
\note{This figure reports cohort-stacked event-study estimates of the dynamic effects of exporting on observed outcomes: the log skill ratio, the log skill premium, log employment by skill group, and log total employment. Estimates use the baseline propensity-score match on the model-implied covariates; the omitted reference period is $h=-1$. Bars are 90\% confidence intervals from 998 nonparametric bootstrap replications.}
\end{figure}

The productivity series behind these event studies are estimated objects, so we verify that the effects are visible in the observed data. We re-run the baseline matched event study replacing the estimated outcomes with observed ones: the skill ratio, the skill premium, and employment by skill group. We show the estimates in Figure~\ref{fig:rawdid}. Averaged over the post-entry window, the skill ratio of new exporters is about 17\% higher than that of their matched controls, whereas the skill premium is about 5\% lower.\footnote{The premium's joint pre-entry test rejects at the 5\% level ($p = 0.011$); the skill ratio's does not ($p = 0.78$).} The levels show a directed expansion: skilled employment is about 20\% higher, unskilled employment is essentially unchanged, and total employment is about 9\% higher. These movements reproduce the structural estimate: the effect on relative-unskilled productivity combines the skill-ratio and premium effects through the estimated elasticity of substitution $\sigma_{\theta}$.\footnote{Equation \eqref{eq:omega_US_char} inverts to $\omega_{R} = \left[\log(S/U) + \sigma_{\theta}\log(W_{S}/W_{U})\right]/(1-\sigma_{\theta})$, up to demeaning. The event study is linear on a common matched sample, so the identity carries to the estimated effects: $\beta_{R}(h) = \left[\beta_{S/U}(h) + \sigma_{\theta}\,\beta_{W}(h)\right]/(1-\sigma_{\theta})$, with $\beta_{S/U}(h)$ and $\beta_{W}(h)$ the skill-ratio and premium effects. Over the post-entry window, the right-hand side reproduces $\beta_{R}$ estimated directly from the structural series to the third decimal.}

\subsection{Aggregate Implications for Skilled Labor Intensity}\label{ss:decomposition}

In this subsection, we construct a sector-wide counterfactual aggregate skilled labor share $\bar{s}_t^{cf} = S_t^{cf}/(S_t^{cf} + U_t^{cf})$, where a superscript $cf$ denotes a counterfactual quantity and the unsuperscripted symbol its observed value. The counterfactual subtracts the estimated post-entry effects from each new-exporter plant-year at post-entry horizon $h_{jt}$ (zero otherwise), setting $\mathrm{d}\omega_v = -\beta_v(h_{jt})$ for $v \in \{H, L, R\}$,\footnote{Beyond the estimation window we carry the terminal effect $\beta_v(+3)$ forward.} where $\beta_v(h)$ is the matched stacked-DiD event-study coefficient on $\omega_v$ at horizon $h$ from equation~\eqref{eq:diff_in_diff} (with $\beta_S = \beta_L$ and $\beta_U = \beta_L + \beta_R$). The counterfactual is a partial-equilibrium experiment: we hold capital, factor prices, and the industry price and quantity indices at their observed values, and let each plant's output adjust along its demand curve. 

We linearize the plant's static first-order conditions around the observed allocation.\footnote{The first-order construction parallels the productivity-shutdown counterfactual of \citet{zhang2019non}; the approximation error is second order in the post-entry effects $\beta_v(h)$, which are small over the four-year window.} The resulting loading matrix, derived in Appendix~\ref{a:cf_inversion}, translates the productivity changes $(\mathrm{d}\omega_H, \mathrm{d}\omega_L, \mathrm{d}\omega_R)$ into log-changes in the three static inputs, which we denote $(\hat S_{jt}, \hat U_{jt}, \hat M_{jt})$. The counterfactual input levels are then
\begin{equation}
S_{jt}^{cf} = S_{jt}\,e^{\hat S_{jt}}, \qquad U_{jt}^{cf} = U_{jt}\,e^{\hat U_{jt}}, \qquad M_{jt}^{cf} = M_{jt}\,e^{\hat M_{jt}}. \label{eq:cf_inputs}
\end{equation} The inner-nest relative first-order condition makes the skill ratio respond only to the relative-unskilled shift $\beta_R \equiv \beta_U - \beta_S$, with $\log(S_{jt}^{cf}/U_{jt}^{cf}) - \log(S_{jt}/U_{jt}) = -(1-\sigma_\theta)\,\beta_R(h_{jt})$, whereas the Hicks-neutral and labor-augmenting channels are neutral with respect to the skill mix, moving skilled and unskilled labor in common proportion. 

To aggregate, we write the observed skill share as the employment-weighted mean $\bar{s}_t = \sum_j \theta_{jt}\, s_{jt}$, where $s_{jt} = S_{jt}/N_{jt}$ is the plant skill share, $N_{jt} = S_{jt}+U_{jt}$ its employment, and $\theta_{jt} = N_{jt}/\sum_k N_{kt}$ its employment weight. The counterfactual share takes the same form,
\begin{equation}
\bar{s}_t^{cf} = \sum_j \theta_{jt}^{cf}\, s_{jt}^{cf}, \qquad \theta_{jt}^{cf} = \frac{N_{jt}^{cf}}{\sum_k N_{kt}^{cf}}, \label{eq:cf_share}
\end{equation}
with each plant's intensity and employment replaced by their counterfactual counterparts. The intensity $s_{jt}^{cf}$ follows the exact inner-nest ratio shift above, so it depends only on $\sigma_\theta$ and $\beta_R$, whereas the employment $N_{jt}^{cf}$ follows the level responses in equation~\eqref{eq:cf_inputs}, so it carries all three channels. The Hicks-neutral and labor-augmenting gains therefore reach the aggregate share only through the weights, by reallocating employment across plants.

\begin{figure}[!ht]
\caption{Aggregate Skill Composition: Counterfactual and Decomposition}
\label{fig:aggregate_cf}
\centering
\begin{subfigure}[t]{0.46\textwidth}
\centering
\subcaption{Observed vs.\ counterfactual aggregate skill share}\label{fig:aggregate_cf_a}
\includegraphics[width=\linewidth]{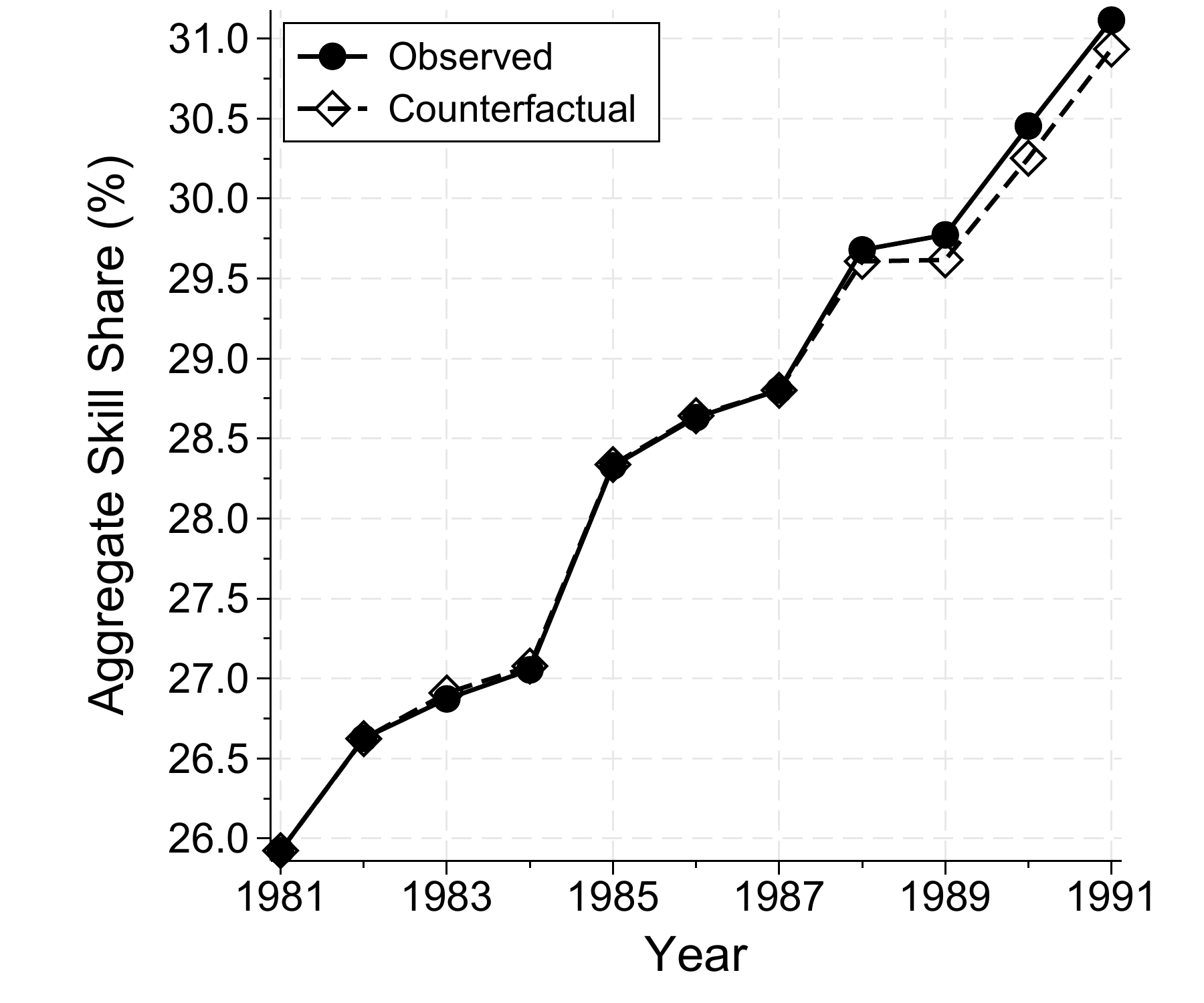}
\end{subfigure}\hfill
\begin{subfigure}[t]{0.46\textwidth}
\centering
\subcaption{Decomposition of the cumulative export effect}\label{fig:aggregate_cf_b}
\includegraphics[width=\linewidth]{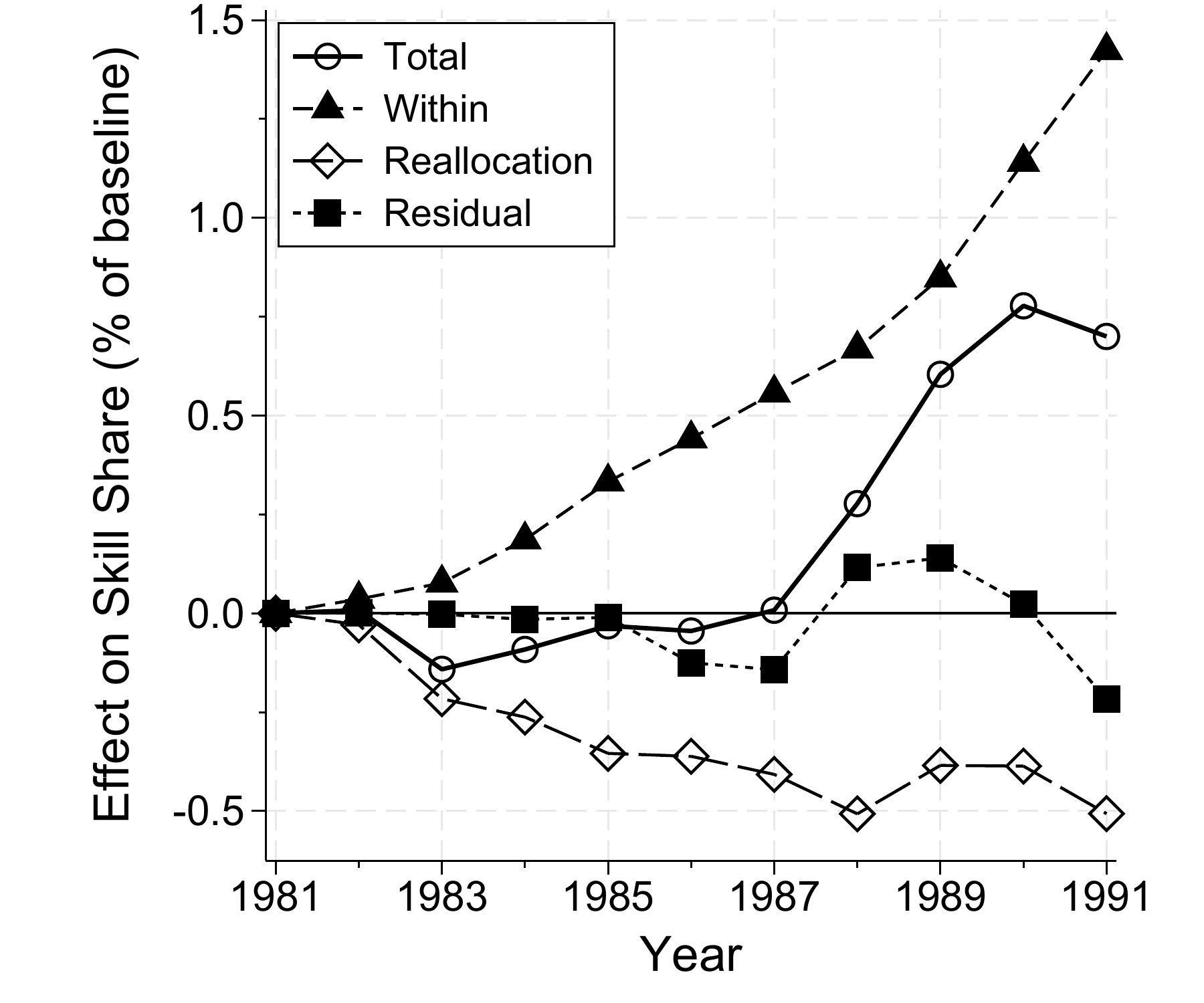}
\end{subfigure}
\note{Panel A: observed aggregate skill share $\bar{s}_t = S_t/(S_t+U_t)$ over 1981--1991, against the counterfactual that switches off the export-induced gains in all three productivity channels $\omega_H$, $\omega_L$, and $\omega_R$. The counterfactual subtracts the matched stacked-DiD post-entry effects from each new-exporter plant-year, holds the industry indices, capital, and factor prices fixed, lets output adjust along the demand curve, and re-aggregates. Panel B: cumulative export effect on $\bar{s}_t$ ($= \bar{s}_t - \bar{s}_t^{cf}$ for the full counterfactual), expressed as a percent of the 1981 counterfactual baseline and decomposed per equation~\eqref{eq:fhk_subdecomp}. The ``Within'' line plots $\Phi^{\Delta s}$ (the within-plant skill-share channel), ``Reallocation'' plots $\Phi^{\Delta\theta}$ (the cross-plant employment-weight channel), and ``Residual'' plots $\Phi^{\mathrm{res}}$ (the FHK interaction sub-pieces plus the plant entry-exit contribution). The plotted lines are point estimates.}
\end{figure}

Panel A of Figure~\ref{fig:aggregate_cf} plots the observed aggregate skill share against the counterfactual that switches off all three export channels. The observed skilled labor share rises from 25.9\% in 1981 to 31.1\% in 1991. The counterfactual reaches 30.9\% in 1991, a cumulative gap of $0.18$ percentage points, equivalently 0.70\% of the 1981 baseline aggregate skill share. New exporters during their post-entry window constitute about 2\% of all plant-years in the panel, yet their productivity gains deliver a measurable shift in aggregate skill intensity.

The skill-share gap originates within plants. Panel B decomposes the cumulative export effect on $\bar{s}_t$ following the year-by-year accounting of \citet{foster2001aggregate} (FHK), applied separately to the observed and counterfactual worlds and subtracted. Each world forms these weights from its own employment, so each captures the cross-plant employment-weight reallocation native to its own world. 

We further rearrange the standard FHK Within, Between, and Cross terms algebraically to distinguish between within-plant skill-share changes and cross-plant employment-weight changes:
\begin{equation}
\bar{s}_t - \bar{s}_t^{cf} \;=\; \sum_{\tau=2}^{t} \Big[\Phi^{\Delta s}_\tau + \Phi^{\Delta\theta}_\tau + \Phi^{\mathrm{res}}_\tau\Big],
\label{eq:fhk_subdecomp}
\end{equation}
where $\Phi^{\Delta s}_\tau$ is the year-$\tau$ contribution of the within-plant skill-share channel (the observed-minus-counterfactual gap in $\Delta s_{jt}$, at observed employment weights), $\Phi^{\Delta\theta}_\tau$ that of the cross-plant reallocation channel (the same gap in $\Delta\theta_{jt}$, at counterfactual lagged shares), and $\Phi^{\mathrm{res}}_\tau$ a residual collecting the FHK interaction sub-pieces and the plant entry-exit channel. Explicit formulas are derived in Appendix~\ref{a:fhk_subdecomp}. 

By 1991 the within-plant channel contributes 1.42\% of the 1981 baseline ($0.37$ percentage points), more than the total 0.70\% aggregate effect (the $0.18$-percentage-point gap of Panel~A); the cross-plant reallocation channel subtracts 0.51\% ($0.13$ percentage points) and the residual 0.22\% ($0.06$ percentage points). The aggregate skill-share rise thus originates within plants: new exporters upgrade their own skill intensity, whereas the scale-driven reallocation of employment toward less skill-intensive new exporters partially offsets the rise rather than reinforcing it.

In head-count terms, the same gains expanded skilled employment by approximately $5{,}596$ workers by 1991 (7.3\% of the 1981 baseline skilled stock) and unskilled employment by approximately $10{,}458$ workers (4.7\% of the baseline unskilled stock). The expansion is \emph{directed} rather than uniform: skilled labor grows faster than unskilled, consistent with unskilled-labor-saving technical change under the estimated complementarity.\footnote{Because the level responses are governed by the small denominator $\Delta_{jt}$ of Appendix~\ref{a:cf_inversion}, the head-count magnitudes are sensitive to the demand elasticity $\eta$. The within-plant skill intensity depends only on $\sigma_\theta$ and $\beta_R$, but the aggregate skill share also weights plants by their employment, so its cross-plant reallocation component inherits the sensitivity to $\eta$.}

Because output is free to adjust, the construction captures both margins of the plant's response to export-induced technical change: the \emph{composition} of inputs, which reshapes the organization of production, and the \emph{scale} of production, as a more productive plant expands along its demand curve. Composition and scale leave distinct footprints: the compositional margin is the within-plant upgrading of Panel~B, whereas the scale margin shows up directly in the head-count levels above and enters the skill share through the cross-plant reallocation channel, which partially offsets the within-plant rise. The corresponding movement in aggregate TFP is reported in Appendix~\ref{a:tfp_cf}.

\section{Robustness}\label{s:robustness}
\noindent{\textit{Skilled-Unskilled Workers Substitution}} -- We re-estimate \(\sigma_\theta\) using alternative instruments that exploit variation in the skill premium across time, industries, and Colombian administrative departments. Combining the characterization equation \eqref{eq:omega_US_char} with the normalized Markov process \eqref{eq:omega} yields a regression equation with $\sigma_\theta$ as the coefficient $b_1$:
\begin{equation}
\begin{aligned}
\log\left(\frac{\ddot{S}_{jt}}{\ddot{U}_{jt}}\right)& = b_1\log\left(\frac{\ddot{W}_{U,jt}}{\ddot{W}_{S,jt}}\right) + b_2\log\left(\frac{\ddot{S}_{jt-1}}{\ddot{U}_{jt-1}}\right) + b_3\log\left(\frac{\ddot{W}_{U,jt-1}}{\ddot{W}_{S,jt-1}}\right) \\
&+ b_4 e_{jt-1} + b_5 i_{jt-1} +b_6 \left(e_{jt-1} \cdot i_{jt-1}\right) + \iota_t + \iota_s + u_{jt}. \label{eq:omega_R}
\end{aligned}
\end{equation}

To address the endogeneity of contemporaneous relative wages, we build five instruments from the log relative wage of unskilled to skilled workers: (i) the lagged output-weighted average across same-SIC3 plants in administrative departments contiguous to plant $j$'s; (ii) the output-weighted mean across same-SIC3, same-department plants other than plant $j$; (iii) the lagged version of (ii); (iv) the output-weighted average across same-department plants in a different SIC3 industry; and (v) the lagged version of (iv).

The instruments are relevant because local labor market conditions are key determinants of the wages a plant faces. Exclusion requires that, conditional on year and industry fixed effects and the Markov controls in equation \eqref{eq:omega_R}, other plants' wage premia affect plant $j$'s skill ratio only through its relative wages. Instruments (i), (iv), and (v) draw on variation outside plant $j$'s department-by-SIC3 cell, a different department in (i) and a different SIC3 within the same department in (iv) and (v); they are less exposed to within-cell productivity shocks but more weakly correlated with plant $j$'s own wages. Instruments (ii) and (iii), drawn from within plant $j$'s department-by-SIC3 cell, deliver stronger first stages but rely more heavily on the fixed effects and Markov controls to absorb shocks common to the cell.

Table \ref{tab:2SLS} presents the two-stage least squares (2SLS) estimates of $\sigma_\theta$. Four of the five instruments, (ii)--(v), clear the conventional first-stage threshold of $F>10$, with Kleibergen--Paap rank $F$-statistics of 30.8, 19.5, 16.3, and 15.2 and point estimates of 0.47, 0.60, 0.76, and 0.65; all lie below one and reinforce the complementarity finding from Table \ref{tab:structural}. None is statistically different from the baseline value of 0.344 at conventional levels. The remaining instrument~(i) yields a point estimate of 1.61 but rests on a weak first stage ($F=3.2$) and is reported for completeness.

\begin{table}[!h]
\caption{Robustness: 2SLS Results}
\label{tab:2SLS}
{\footnotesize\renewcommand{\arraystretch}{0.5}%
\begin{tabular}{lcccccc}
\toprule  \toprule
               & OLS   & IV (i)  & IV (ii)  & IV (iii) & IV (iv) & IV (v)  \\
\midrule
  $\log\left({\ddot{W}_{U,jt}}/{\ddot{W}_{S,jt}}\right)$
               & 0.379 & 1.607 & 0.466 & 0.598 & 0.760 & 0.654  \\
               & (0.010) & (1.026) & (0.229) & (0.295) & (0.361) & (0.380) \\
\midrule
\multicolumn{7}{l}{\textbf{First Stage}}  \\
               &       &       &       &       &       &        \\
Instrument~1
               &       & -0.011 &       &       &       &        \\
               &       & (0.006) &       &       &       &        \\
Instrument~2
               &       &       & -0.053 &       &       &        \\
               &       &       & (0.010) &       &       &        \\
Instrument~3
               &       &       &       & -0.041 &       &        \\
               &       &       &       & (0.009) &       &        \\
Instrument~4
               &       &       &       &       & -0.089 &        \\
               &       &       &       &       & (0.022) &        \\
Instrument~5
               &       &       &       &       &       & -0.077  \\
               &       &       &       &       &       & (0.020) \\
\midrule
KP rk Wald $F$ &       & 3.22 & 30.79 & 19.49 & 16.32 & 15.19 \\
\midrule
Industry FE    & YES   & YES    & YES    & YES   & YES   & YES    \\
Year FE        & YES   & YES    & YES    & YES   & YES   & YES    \\
\midrule
Observations   & 39,506 & 37,668 & 39,126 & 39,120 & 39,506 & 39,506 \\
\bottomrule
\end{tabular}
}
\note{This table presents the 2SLS estimation results for equation \eqref{eq:omega_R}, whose dependent variable is the log skilled-to-unskilled employment ratio $\log(\ddot{S}_{jt}/\ddot{U}_{jt})$, alongside OLS estimates from the same specification (without instrumentation) for comparison. Instruments (i)--(v) are defined in the text. Standard errors clustered at the plant level are reported in parentheses. The table also reports the first-stage coefficient for each instrument and the Kleibergen--Paap rank $F$-statistic. Sample sizes differ across columns because the instruments draw on lagged and neighboring-cell wage information unavailable for some plant-years.}
\end{table}

\begin{figure}[h!]
    \centering
    \includegraphics[width=0.85\textwidth]{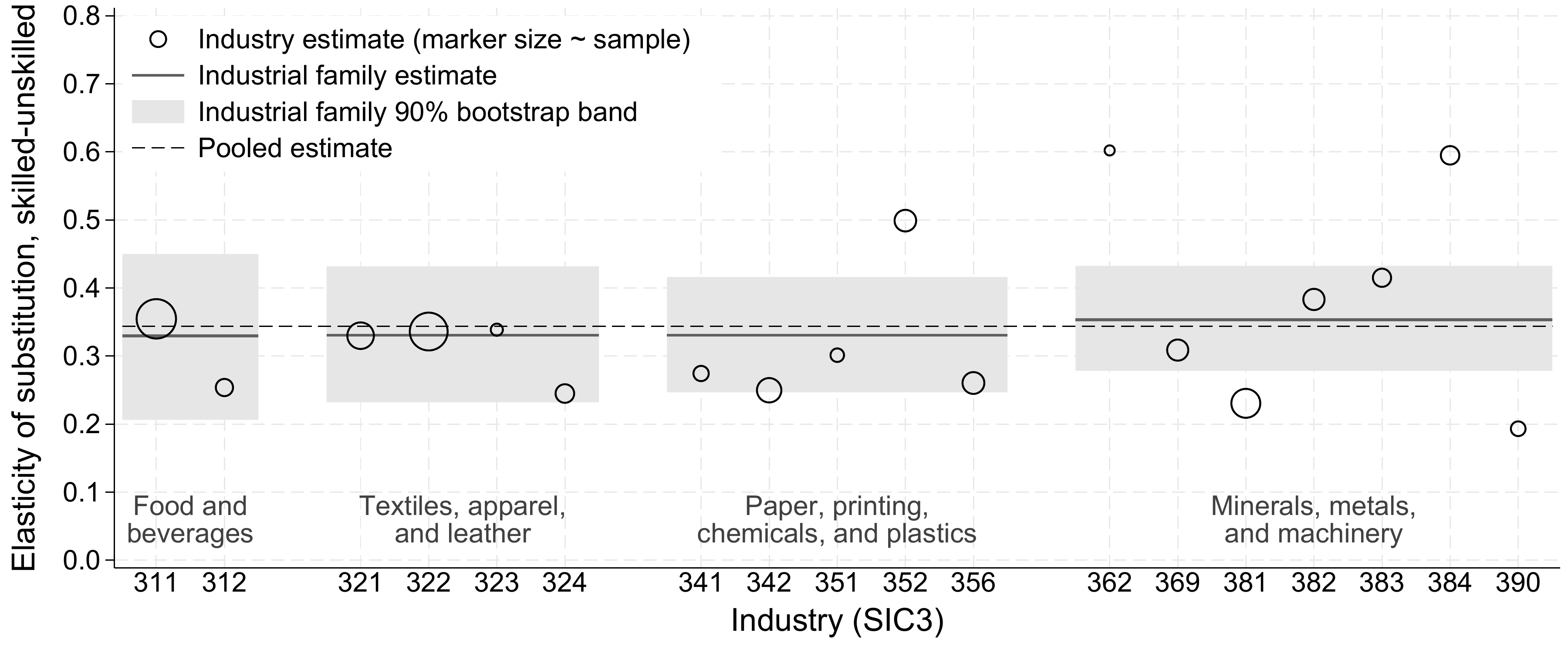}
    \caption{Skilled-Unskilled Elasticity of Substitution by Industry}
    \label{fig:sigma_by_industry}
\note{This figure plots the Step~1 GMM estimates of $\sigma_\theta$ for the 18 SIC3 industries in which the estimator converges (circles, with marker size proportional to the estimation sample) and for the four industry families (horizontal lines, with shaded 90\% confidence bands from 998 bootstrap replications). The dashed line marks the pooled estimate of 0.344. All estimates lie below one, the value above which skilled and unskilled labor would be gross substitutes. Iron and steel (SIC3 371) is omitted: its Step~1 GMM does not converge.}
\end{figure}

The complementarity finding also holds within industries. We re-estimate $\sigma_\theta$ separately for four industry families: food and beverages (SIC3 311--312), textiles, apparel, and leather (321--324), paper, printing, chemicals, and plastics (341--356), and minerals, metals, and machinery (362--390). The family-level Step~1 estimates range from 0.330 to 0.353, tightly clustered around the pooled 0.344. We show the industry-level and family-level estimates in Figure~\ref{fig:sigma_by_industry}.

\par\medskip

\noindent{\textit{Cross-Validation Through an Alternative ACF Estimator}} -- We cross-validate the baseline estimate of Hicks-neutral productivity by replacing the GMM of Step~3 and the Kalman filter of Step~4 with a proxy-variable inversion in the spirit of \citet{olley1996dynamics}, \citet{levinsohn2003estimating}, and \citet{ackerberg2015identification} (see Appendix~\ref{a:acf} for details). In the alternative estimator, Steps~1 and 2 are unchanged from the baseline, so the proxy-variable inversion isolates the role of Steps~3 and~4 in the estimates of $\eta$, $\sigma_H$, $\sigma_\zeta$, and the $\ddot{\omega}_{H}$ Markov coefficients.

The implied markup is $\widehat{\mu} = 1.075$, and the standard deviations of the Hicks-neutral innovation and the measurement error are $\widehat{\sigma}_H = 0.235$ and $\widehat{\sigma}_\zeta = 0.156$ (Table~\ref{tab:structural_acf}); all three are close to their baseline values in Table~\ref{tab:structural}. The ACF and baseline estimates of Hicks-neutral productivity are highly correlated, both pointwise and within plants, on the subsample of plant-years with positive investment, and their distributions agree closely (Figures~\ref{fig:comparison_density} and~\ref{fig:comparison} in the Appendix). We retain the baseline estimation strategy because it is well-defined on the full panel, whereas the proxy-variable inversion is restricted to plant-years with positive investment.

\begin{table}[!h]
\caption{Estimates of Production Function Parameters: Alternative ACF Specification}
\centering
{\small\renewcommand{\arraystretch}{0.9}%
\begin{tabular}{cccccccccc}
\toprule\toprule
 $\mu$ & & $\alpha_L$ & $\alpha_M$ & $\alpha_K$ & $\alpha_S$ & $\alpha_U$ & & $\sigma_H$ & $\sigma_\zeta$ \\
\midrule
 1.075 & & 0.217 & 0.714 & 0.069 & 0.341 & 0.659 & & 0.235 & 0.156 \\
 (0.020) & & (0.002) & (0.002) & (0.001) & (0.002) & (0.002) & & (0.021) & (0.001) \\
\bottomrule
\end{tabular}
}
\note{This table reports structural-parameter estimates from the alternative ACF specification of Appendix~\ref{a:acf}. $\sigma_\theta$ and $\sigma_\rho$ are unchanged from Steps~1--2 of the body's baseline (Table~\ref{tab:structural}) and are therefore omitted. The markup is $\mu = \eta/(\eta+1)$. The factor share parameters $\alpha_K$, $\alpha_S$, and $\alpha_U$ are computed on the $\ddot{I}_{jt}>0$ subsample using subsample geometric means and do not directly compare to those in Table~\ref{tab:structural}. Bootstrap standard errors appear in parentheses, drawn from 1,000 bootstrap samples with replacement, clustered at the establishment level, with each sample re-estimating the full structural model. Of these, 998 replications converged across the full pipeline, and all 998 are retained for $\widehat{\mu}$, $\widehat{\sigma}_H$, and the $\ddot{\omega}_H$ Markov coefficients (Table~\ref{tab:lbe_acf}). Standard errors on $\widehat{\sigma}_\zeta$ and the share parameters likewise use all 998 replications.}
\label{tab:structural_acf}
\end{table}
\par\medskip

\noindent{\textit{Unfiltered $\ddot{\omega}_{H,jt}$}} -- The local effects of exporting hold whether the analysis uses the filtered or unfiltered Hicks-neutral productivity, both as the outcome and as a state in the matching step (Figure \ref{fig:unfiltered} in the Appendix). Using the unfiltered measure as the outcome widens the bootstrap confidence intervals, more so for the Hicks-neutral effects than for the TFP effects. Matching plants on the unfiltered productivity widens the intervals no further, and the Hicks-neutral estimates stay near zero across the three scenarios.\footnote{The matching process results in a balanced sample in terms of observable characteristics and satisfies the common support assumption. See Figure \ref{fig:matching_descriptive_noisy}, panels \subref{fig:combined_noisy} and \subref{tab:matching_unfiltered}, in the Appendix.} The TFP effects stay positive throughout, running modestly lower at the later horizons when the matching uses the unfiltered measure but remaining within the bootstrap intervals across all three scenarios.

\par\medskip

\noindent{\textit{Matching Design}} -- We probe the matched stacked event study along its two design choices, the conditioning set and the control group, holding the stacking, fixed effects, and event window fixed. The baseline matches new exporters to never-exporters on the model-implied covariates: the three productivity states, capital, both skill-group wages, and the industry-year aggregates. We compare it against three variants. The \textit{observables-only} variant changes the conditioning set, building the propensity score from the same covariates but excluding the three estimated productivity states. The \textit{unmatched} variant skips the propensity-score step and retains the entire never-exporter pool, so that both the treated group and the control group change. The \textit{matched treated, full pool} variant keeps the baseline-matched new exporters and sets them against that same full pool, so that only the control group changes. We show the results in Figure~\ref{fig:robustness_matching} in the Appendix. Across the four designs the effect of export entry on relative-unskilled productivity is positive and builds over the post-entry window, the Hicks-neutral effect is indistinguishable from zero, and the TFP effect is positive at the early horizons. The estimates agree in sign and timing, and the differences in magnitude are small relative to the bootstrap intervals, leaving the baseline reading intact.
\par\medskip

\noindent{\textit{Placebo}} -- We test whether the matched stacked event study could manufacture the estimated effects in the absence of any true export entry. In each of 500 placebo draws, we remove all new exporters from the sample, assign pseudo entry years to never-exporters that replicate the true cohort-by-industry composition of entry, and re-run the full procedure, from the propensity-score matching to the stacked regressions. We show the placebo distributions in Figure~\ref{fig:placebo_hist} in the Appendix. The placebo estimates center on zero for every outcome. For relative-unskilled productivity, unskilled-augmenting productivity, and equipment, no draw produces an effect as large in magnitude as the one we estimate in the data. The effect of export entry on TFP is also unlikely to arise by chance, with only a few draws reaching its magnitude.

\par\medskip

\noindent{\textit{The 1990--91 Import Liberalization}} -- Colombia sharply liberalized imports at the end of the sample: the \textit{apertura} of 1990--91 was the abrupt, largely unexpected final phase of reforms that between 1985 and 1991 removed most nontariff barriers and cut average manufacturing tariffs from 32\% to 12\% \citep{fieler2018trade, attanasio2004trade, bussolo2003globalisation}. The import liberalization overlaps the post-entry window of the late entry cohorts, so cheaper imported machinery and intermediates could in principle produce the same unskilled-labor-saving pattern we attribute to export entry.

To remove the overlap between the event window and the import liberalization, we re-estimate the baseline matched event study on early entry cohorts, at two cutoffs. The first drops the 1990 and 1991 cohorts, so that no plant enters export markets during the liberalization (117 new exporters remain); the second keeps only plants that entered by 1986, whose full event window ends by 1989, so that every observation predates the \textit{apertura} (47 plants). We show the event studies for the two cohort restrictions alongside the baseline in Figure~\ref{fig:apertura_compare} in the Appendix. The effect of exporting on relative-unskilled productivity is larger than the baseline at the first cutoff and larger still at the second. At both cutoffs, the effect of exporting on Hicks-neutral productivity stays indistinguishable from zero and the effect on TFP stays positive in the first years after entry. The effect of exporting on relative-unskilled productivity thus grows as the sample's overlap with the import liberalization shrinks, whereas an effect driven by cheaper imports would instead fade. The pattern is consistent with export entry, not the import liberalization, driving the results.

\section{Summary and Concluding Remarks}\label{s:conclusion}
To conclude, our findings indicate that the productivity gains from exporting are factor-biased rather than Hicks-neutral. In Colombian manufacturing between 1981 and 1991, export entry raises unskilled-labor productivity by about 22\% on entry, with statistically insignificant changes in skilled-labor and Hicks-neutral productivity, whereas total factor productivity rises by only about 8\% on average after entry. Because skilled and unskilled labor are complements, these gains are unskilled-labor-saving: they raise the skill intensity of production and move together with the post-entry expansion of machinery and equipment.

Our findings speak to export-led industrialization: export market entry can drive unskilled-labor-saving technology adoption, shifting relative labor demand toward skilled workers. These distributional consequences underscore the value of policies that help workers adapt as exporting reshapes the demand for skill.

To better understand the allocation of skilled and unskilled labor, future research could endogenize wages by adding their supply curves. This would clarify how shifts in plants' skill composition feed back into equilibrium wages and welfare. Embedding labor supply in our production function framework would also support counterfactual simulations of the welfare consequences of exporting for both skilled and unskilled workers.

\clearpage
\setcounter{secnumdepth}{3}
\setcounter{subsection}{0}
\renewcommand\thesubsection{A.\arabic{subsection}}
\renewcommand\thefigure{A\arabic{figure}}
\renewcommand\thetable{A\arabic{table}}
\renewcommand\theequation{A.\arabic{equation}}
\let\appsubsection\subsection
\renewcommand{\subsection}{\FloatBarrier\appsubsection}
\renewcommand{\topfraction}{0.9}
\renewcommand{\textfraction}{0.05}
\section*{Appendix}

\subsection{Data Construction}\label{a:data_construction}

This appendix describes the construction of the estimation sample from the Colombian Annual Manufacturing Survey and the measurement of the variables used in the article.

We drop plants that switched 3-digit industry across observed years. We restrict the sample to establishments employing both skilled and unskilled workers, dropping 7.3\% of observations. We retain only plants operating for at least two consecutive years and remove 910 outlier observations with extreme changes in labor employment or output.\footnote{We follow the outlier removal procedure outlined in \citet{ruhl2017new}. Plants with year-on-year changes exceeding 150\% in either employment or output are dropped as outliers.} To address potential measurement error, we further trim observations at the 1st and 99th percentiles of the skilled-to-unskilled payroll expenditure ratio, the payroll-to-material expenditure ratio, and the variable-cost-to-revenue ratio, dropping 3,059 observations. The final dataset comprises 10,023 plants spanning 50,329 plant-year observations.

We measure plant-level revenue as aggregate sales across domestic and export markets. We construct capital stocks for four asset categories (land, structures, equipment, and transportation equipment) using the perpetual inventory method. We measure capital service expenditure as $E_{K,jt} = \sum_i (r_t + \delta_i) K^i_{jt}$, where $r_t$ denotes Colombia's real interest rate, obtained from the World Bank WDI database (for years prior to 1986, we use the 1986 value because earlier observations are unavailable), $\delta_i$ denotes the asset-specific depreciation rate, taken from U.S. Bureau of Economic Analysis industry estimates following \citet{raval2023testing} (land does not depreciate), and $K^i_{jt}$ is the deflated capital stock of asset type $i$. We measure skilled and unskilled labor by workforce headcounts and their corresponding payroll expenditures (comprising wages, salaries, and non-salary benefits). We deflate revenue, worker payrolls, and intermediate-input expenditures (which comprise raw materials, electricity, and fuels) using the GDP deflator, and capital stocks (except land, which is deflated by the structures deflator) using investment deflators, both sourced from \citet{raval2023testing}. We obtain the annual SIC-3 industry-level output price index directly from the survey.\footnote{We construct this index by dividing plant-level sales by the reported real output value, rescaled by the GDP deflator.} We construct aggregate industry revenue as the geometric mean of plant revenue, weighted by each plant's share of sectoral sales, following \citet{klette1996inconsistency}.
To recover real intermediate quantities, we deflate each component of intermediate inputs by its own price index and sum the deflated components. The three indices are a 3-digit industry intermediate-goods price index for raw materials, an electricity unit-value index for electricity, and a petroleum-and-coal output-price index for fuels.

\subsection{Elasticities of Substitution in Nested CES Production}\label{app:skill-capital}

This appendix derives the elasticities of substitution between capital and each labor type in the nested CES production function from Section \ref{s:theory}. We suppress firm $(j)$ and time $(t)$ subscripts throughout. The derivation is invariant to the geometric-mean normalization presented in Appendix \ref{a:geo_means}: the normalization constants and productivity shifters cancel in every log-ratio derivative.

The two-level nested CES technology is:
\begin{align}
Q &= \left[\tilde{\alpha}_K K^{\rho} + \tilde{\alpha}_M M^{\rho} + \tilde{\alpha}_L \left(\exp(\omega_L) L\right)^{\rho}\right]^{1/\rho}\exp(\omega_H), \label{eq:top} \\
L &= \left[\tilde{\alpha}_S S^{\theta} + \tilde{\alpha}_U \left(\exp(\omega_R) U\right)^{\theta}\right]^{1/\theta}, \label{eq:bottom}
\end{align}
where $\rho, \theta \in (-\infty, 1]$ govern within-nest substitutability. Let $w_K$, $w_M$, $w_S$, and $w_U$ denote the prices of capital, materials, skilled labor, and unskilled labor, respectively. We use the direct partial elasticity of substitution, $\sigma_{ij} \equiv d\ln(i/j) / d\ln(w_j/w_i)$. For within-nest pairs, standard CES algebra gives $\sigma_{KM} = \sigma_{KL} = \sigma_{ML} = \sigma_\rho \equiv 1/(1-\rho)$ and $\sigma_{SU} = \sigma_\theta \equiv 1/(1-\theta)$.

\subsubsection{Cross-Nest Elasticities}

We derive $\sigma_{KS}$, the elasticity of substitution between capital and skilled labor. The two inputs sit in different nests, so their substitutability runs through the labor aggregate $L$. Decomposing $\ln(K/S) = \ln(K/L) + \ln(L/S)$ and differentiating with respect to $\ln w_S$ (with $w_K$ held fixed, so $d\ln(w_S/w_K) = d\ln w_S$):
\begin{equation}
\sigma_{KS} = \frac{\partial \ln(K/L)}{\partial \ln w_S} + \frac{\partial \ln(L/S)}{\partial \ln w_S}.
\label{eq:sigma_KS_decomp}
\end{equation}
The outer-nest first-order conditions (FOCs) for $K$ and $L$ give $\partial \ln(K/L)/\partial \ln w_L = \sigma_\rho$. Shephard's lemma applied to the inner-nest unit cost function $w_L(w_S, w_U)$ yields $\partial \ln w_L / \partial \ln w_S = s_S \equiv w_S S / (w_S S + w_U U)$, the skilled labor cost share. Applying the chain rule,
\begin{equation*}
\frac{\partial \ln(K/L)}{\partial \ln w_S} = \frac{\partial \ln(K/L)}{\partial \ln w_L} \cdot \frac{\partial \ln w_L}{\partial \ln w_S} = \sigma_\rho \, s_S.
\end{equation*}

By cost minimization within the labor nest, the conditional (cost-minimizing) demand for $S$ given the aggregate $L$ is
\begin{equation*}
S = \tilde{\alpha}_S^{\sigma_\theta}\, L \left(\frac{w_S}{w_L}\right)^{-\sigma_\theta},
\end{equation*}
where $w_L = w_L(w_S, w_U)$ is the inner-nest unit cost (the CES price index of $L$) introduced above. Taking logs,
\begin{equation*}
\ln(L/S) = -\sigma_\theta \ln \tilde{\alpha}_S + \sigma_\theta \ln(w_S/w_L),
\end{equation*}
whose first term is independent of $w_S$.
Differentiating with respect to $\ln w_S$ and substituting $\partial \ln w_L / \partial \ln w_S = s_S$,
\begin{equation*}
\frac{\partial \ln(L/S)}{\partial \ln w_S} = \sigma_\theta \left(1 - \frac{\partial \ln w_L}{\partial \ln w_S}\right) = \sigma_\theta(1 - s_S).
\end{equation*}
Substituting into \eqref{eq:sigma_KS_decomp}:
\begin{equation}
\sigma_{KS} = s_S \sigma_\rho + (1 - s_S)\sigma_\theta.
\label{eq:sigma_KS_final}
\end{equation}
By the symmetry between $S$ and $U$ in the inner nest, repeating the derivation with their roles swapped replaces $s_S$ with $1 - s_S$ and yields:
\begin{equation}
\sigma_{KU} = (1 - s_S) \sigma_\rho + s_S \sigma_\theta.
\label{eq:sigma_KU}
\end{equation}

\subsubsection{Capital-Skill Complementarity}

Capital-skill complementarity ($\sigma_{KS} < \sigma_{KU}$) holds if and only if
\begin{equation}
\sigma_{KU} - \sigma_{KS} = (1 - 2 s_S)(\sigma_\rho - \sigma_\theta) > 0.
\label{eq:sigma_diff}
\end{equation}
When unskilled labor dominates the wage bill ($s_S < 1/2$), this reduces to the condition $\sigma_\rho > \sigma_\theta$: the outer nest must be more substitutable than the inner nest. Appendix \ref{a:cap_skill_empirical} verifies that both conditions hold in our sample.

\subsection{Capital-Skill Complementarity in the Sample}\label{a:cap_skill_empirical}

\begin{figure}[ht]
    \centering
    \includegraphics[width=0.8\textwidth]{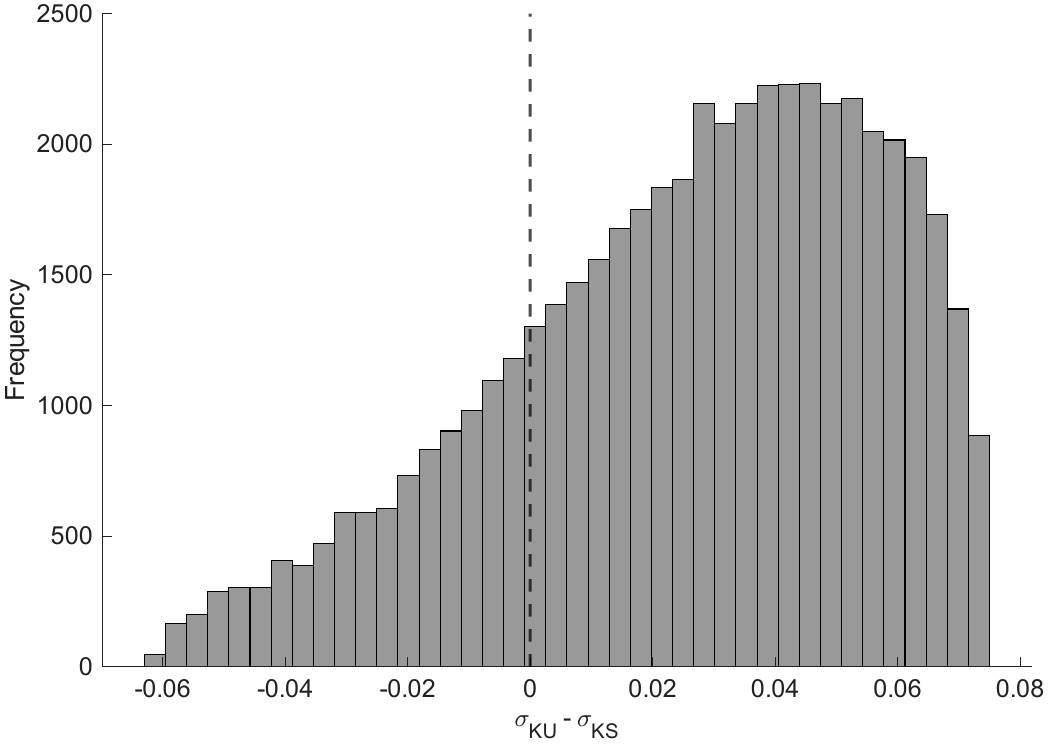}
\caption{Distribution of the Capital-Skill Complementarity Gap $\sigma_{KU} - \sigma_{KS}$ Across Plant-Year Observations}
\label{fig:cap_skill_comp}
\note{This figure displays the pooled distribution of the difference between the capital-unskilled and capital-skilled elasticities of substitution, $\sigma_{KU} - \sigma_{KS} = (1 - 2 s_S)(\sigma_\rho - \sigma_\theta)$, across all plant-year observations in the post-cleaning sample (50,329 plant-years), using the baseline production-function estimates of $\sigma_\rho$ and $\sigma_\theta$ from Table \ref{tab:structural} and the plant-year skilled labor cost share $s_S = W_S S / (W_S S + W_U U)$. Positive values indicate capital-skill complementarity. The dashed vertical line marks zero.}
\end{figure}

In our sample, unskilled workers account for the majority of labor costs ($\bar{s}_U = 1 - \bar{s}_S = 0.68$), placing us in the $s_S < 1/2$ regime of equation \eqref{eq:sigma_diff}. Figure \ref{fig:cap_skill_comp} plots, across plant-year observations, the degree of capital-skill complementarity $\sigma_{KU} - \sigma_{KS}$ (a deterministic transformation of $s_S$ given $\hat\sigma_\rho$ and $\hat\sigma_\theta$); a positive value indicates that capital complements skilled labor more strongly than unskilled labor ($\sigma_{KS} < \sigma_{KU}$). The mass of the distribution lies to the right of zero, so capital-skill complementarity holds for most plants, consistent with the findings of \citet{griliches1969capital} and \citet{krusell2000capital}. It reverses only for a minority of plants with high skilled labor cost shares ($s_S$ above $1/2$, where capital instead complements unskilled labor more, $\sigma_{KS} > \sigma_{KU}$), consistent with the heterogeneity permitted by the nested CES specification.

\subsection{Derivation of TFP Aggregation}
\label{app:tfp_derivation}

This appendix derives the aggregation of total factor productivity (TFP) from the nested CES production structure we specify in Section \ref{s:theory}, adapting the derivation of \citet{doraszelski2018measuring} to our three-productivity parameterization. We show that log TFP decomposes into a weighted sum of Hicks-neutral and factor-augmenting productivity terms, where output elasticities provide the weights.

Per the relabeling at the end of Section \ref{s:theory}, skilled and unskilled labor-augmenting productivities satisfy $\omega_{S,jt} = \omega_{L,jt}$ and $\omega_{U,jt} = \omega_{L,jt} + \omega_{R,jt}$. By CES homogeneity of degree one, the outer labor composite in \eqref{eq:prod} rewrites as $\exp(\omega_{L,jt}) L_{jt} = \mathcal{L}_{jt}$, where
\begin{equation*}
\mathcal{L}_{jt} = \left[\tilde{\alpha}_S (\exp(\omega_{S,jt}) S_{jt})^\theta + \tilde{\alpha}_U (\exp(\omega_{U,jt}) U_{jt})^\theta\right]^{1/\theta},
\end{equation*}
so the production function reads $Q_{jt} = [\tilde{\alpha}_K K_{jt}^\rho + \tilde{\alpha}_M M_{jt}^\rho + \tilde{\alpha}_L \mathcal{L}_{jt}^\rho]^{1/\rho} \exp(\omega_{H,jt})$. We work with this reparameterized form throughout.

For the reparameterized production function above, we define the output elasticity with respect to input $i \in \{K, M, \mathcal{L}\}$ as:
\begin{equation}
\varepsilon_{i,jt} \equiv \frac{\partial \log Q_{jt}}{\partial \log X_{i,jt}} = \frac{\tilde{\alpha}_i X_{i,jt}^{\rho}}{\sum_{k} \tilde{\alpha}_k X_{k,jt}^{\rho}}
\label{eq:top_elasticity}
\end{equation}
Constant returns to scale imply $\sum_i \varepsilon_{i,jt} = 1$. Within the labor nest, we define conditional elasticities for skilled and unskilled workers:
\begin{equation}
\varepsilon_{S|L,jt} = \frac{\tilde{\alpha}_S (\exp(\omega_{S,jt}) S_{jt})^{\theta}}{\tilde{\alpha}_S (\exp(\omega_{S,jt}) S_{jt})^{\theta} + \tilde{\alpha}_U (\exp(\omega_{U,jt}) U_{jt})^{\theta}}, \qquad \varepsilon_{U|L,jt} = 1 - \varepsilon_{S|L,jt}
\label{eq:cond_elas}
\end{equation}
Applying the chain rule yields the unconditional output elasticities for each labor type:
\begin{equation}
\varepsilon_{S,jt} = \varepsilon_{L,jt} \cdot \varepsilon_{S|L,jt}, \qquad \varepsilon_{U,jt} = \varepsilon_{L,jt} \cdot \varepsilon_{U|L,jt}
\label{eq:unconditional}
\end{equation}
We totally differentiate the log quantity production function above to obtain:
\begin{equation}
d\log Q_{jt} = \varepsilon_{K,jt} \, d\log K_{jt} + \varepsilon_{M,jt} \, d\log M_{jt} + \varepsilon_{L,jt} \, d\log \mathcal{L}_{jt} + d\omega_{H,jt}
\label{eq:diff_top}
\end{equation}
Similarly, we differentiate the log labor aggregate:
\begin{equation}
d\log \mathcal{L}_{jt} = \varepsilon_{S|L,jt} \, d\log(\exp(\omega_{S,jt}) S_{jt}) + \varepsilon_{U|L,jt} \, d\log(\exp(\omega_{U,jt}) U_{jt})
\label{eq:diff_labor}
\end{equation}
Substituting \eqref{eq:diff_labor} into \eqref{eq:diff_top} and applying \eqref{eq:unconditional} yields:
\begin{equation}
d\log Q_{jt} = \sum_{i \in \{K,M,S,U\}} \varepsilon_{i,jt} \, d\log X_{i,jt} + d\omega_{H,jt} + \varepsilon_{S,jt} \, d\omega_{S,jt} + \varepsilon_{U,jt} \, d\omega_{U,jt}
\label{eq:combined}
\end{equation}
We measure TFP growth as the Solow residual:
\begin{equation}
d\log \text{TFP}_{jt} \equiv d\log Q_{jt} - \sum_{i \in \{K,M,S,U\}} \varepsilon_{i,jt} \, d\log X_{i,jt}
\label{eq:tfp_def_app}
\end{equation}
Substituting \eqref{eq:combined} into \eqref{eq:tfp_def_app} yields the exact differential relationship:
\begin{equation}
d\log \text{TFP}_{jt} = d\omega_{H,jt} + \varepsilon_{S,jt} \, d\omega_{S,jt} + \varepsilon_{U,jt} \, d\omega_{U,jt}
\label{eq:tfp_diff}
\end{equation}
We integrate \eqref{eq:tfp_diff} from the origin (where all productivity terms equal zero):
\begin{equation}
\log \text{TFP}_{jt} = \omega_{H,jt} + \int_0^{\omega_{S,jt}} \varepsilon_{S}(\omega_S, \omega_U) \, d\omega_S + \int_0^{\omega_{U,jt}} \varepsilon_{U}(\omega_S, \omega_U) \, d\omega_U
\label{eq:tfp_integral}
\end{equation}
Because the elasticities vary with the productivity levels, we approximate the integrals by evaluating the elasticities at their current-period values; the approximation is exact when the elasticities are constant:
\begin{equation}
\log \text{TFP}_{jt} \approx \omega_{H,jt} + \varepsilon_{S,jt} \, \omega_{S,jt} + \varepsilon_{U,jt} \, \omega_{U,jt}.
\label{eq:tfp_final}
\end{equation}

\subsection{Normalization via Geometric Means}\label{a:geo_means}

We normalize the CES production function using geometric means, following \citet{grieco2016production}.\footnote{The geometric-mean normalization was introduced by \citet{de1989quest} and developed through \citet{klump2000ces, klump2000economic, de2006conjecture, leon2010identifying}.} This normalization achieves three objectives. First, it eliminates the influence of measurement units by scaling each variable to its sample mean. Second, it renders the normalized factor share parameters interpretable as baseline expenditure shares (equal to output elasticities under cost minimization and constant returns to scale). Third, it permits all factor share parameters to be expressed as functions of observable data and a single unknown common parameter. We define the geometric mean of any variable $X$ as:
\[
\bar{X} = \left( \prod_{n=1}^N X_n \right)^{\frac{1}{N}}.
\]
\par\medskip
\noindent\textit{Technology and Productivity Dynamics} -- We define the baseline point as the configuration in which all variables equal their geometric means. Evaluating equations (\ref{eq:prod}) and (\ref{eq:labor}) at this baseline, treating intermediates, skilled labor, and unskilled labor as static inputs, yields:
\begin{align}
& \bar Q = \left[ \tilde{\alpha}_K \bar K^{\rho} + \tilde{\alpha}_M \bar M^{\rho} + \tilde{\alpha}_L(\exp(\bar\omega_L)\bar{L})^{\rho}\right]^{\frac{1}{\rho}}\exp(\bar\omega_H) \label{eq:baseprod}\\
& \bar L =\left[\tilde{\alpha}_S\bar  S^{\theta} + \tilde{\alpha}_U(\exp(\bar \omega_R) \bar U)^{\theta}\right]^{\frac{1}{\theta}}\label{eq:baseL}\\
& \tilde{\alpha}_K + \tilde{\alpha}_M + \tilde{\alpha}_L = 1 \label{eq:basealpha1}\\
& \tilde{\alpha}_S + \tilde{\alpha}_U = 1 \label{eq:basealpha2}\\
& \left(\frac{MP_U}{MP_S}\right)_{\bar Z} = \frac{\tilde{\alpha}_U(\exp{(\bar\omega_R)}\bar U)^{\theta}\bar S}{\tilde{\alpha}_S\bar S^{\theta}\bar U} = \bar \mu_{US} \label{eq:baseSU}\\
& \left(\frac{MP_L}{MP_M}\right)_{\bar Z} = \frac{\tilde{\alpha}_L(\exp(\bar\omega_L)\bar L)^{\rho}\bar M}{\tilde{\alpha}_M \bar M ^{\rho}\bar L} = \bar \mu_{LM} \label{eq:baseML}\\
& \left(\frac{MP_M}{MP_K}\right)_{\bar Z} = \frac{\tilde{\alpha}_M\bar M^{\rho}\bar K}{\tilde{\alpha}_K\bar K^{\rho}\bar M} = \left(\frac{\bar E_K}{\tau\bar E_M}\right)\bar \mu_{MK}\label{eq:baseMK}
\end{align}
Here $\bar\omega_H$, $\bar\omega_L$, and $\bar\omega_R$ denote Hicks-neutral, labor-augmenting, and unskilled-labor-augmenting productivity, respectively. The quantities $\bar\mu_{US}$, $\bar\mu_{LM}$, and $\bar\mu_{MK}$ denote the corresponding baseline relative input prices.

Under the profit-maximization problem \eqref{eq:profit_max}, the marginal rates of substitution among the static inputs equal these relative prices, $\bar\mu_{US} = \frac{\bar W_U}{\bar W_S}$ and $\bar\mu_{LM} = \frac{\bar W_L}{\bar P_M}$. Here $\bar W_S$ and $\bar W_U$ denote geometric mean wages of skilled and unskilled workers, $\bar P_M$ and $\bar P_K$ denote geometric mean input prices for materials and capital, and $\bar W_L$ denotes the exact CES price index of the labor aggregate at the baseline point:
\begin{equation}
\bar{W}_L=\left(\tilde{\alpha}_S^{\frac{1}{1-\theta}} (\bar{W}_S)^{\frac{\theta}{\theta-1}} + \tilde{\alpha}_U^{\frac{1}{1-\theta}} \left(\frac{\bar{W}_U}{\exp(\bar\omega_R)}\right)^{\frac{\theta}{\theta-1}}\right)^{\frac{\theta-1}{\theta}}.
\end{equation}
Capital requires separate treatment. Because it evolves through the dynamic investment problem \eqref{eq:invest_e}, its static first-order condition need not hold at the baseline, so the relative price $\bar\mu_{MK} = \frac{\bar P_M}{\bar P_K}$ cannot be equated with the corresponding marginal rate of substitution. Equation (\ref{eq:baseMK}) therefore introduces $\tau$, the wedge between capital's static first-order condition and its observed expenditure ratio. Under the setup of \citet{grieco2016production} (their Online Appendix 4), $\tau$ admits the interpretation of the mean deviation of the capital stock from its static optimum. We denote geometric mean expenditures on materials, capital, skilled labor, and unskilled labor by $\bar E_M$, $\bar E_K$, $\bar E_S$, and $\bar E_U$, respectively. In the baseline specification we calibrate $\tau = \bar E_K / \bar E_M$.\footnote{Section \ref{s:empirics} (Step 3) develops this calibration and the robustness variant in which $\tau$ is estimated jointly with the remaining Step 3 parameters via GMM.}

The two adding-up constraints plus three MRS conditions close the five-share system, with $\tau$ carrying the dynamic-capital degree of freedom. We solve equations (\ref{eq:basealpha1}), (\ref{eq:basealpha2}), (\ref{eq:baseML}), (\ref{eq:baseMK}), and (\ref{eq:baseSU}) to recover the five CES factor share parameters\footnote{Under cost minimization, the MRS equals the observed relative price, and $\bar E_X = \bar W_X \bar X$ for $X \in \{S, U\}$ (analogously for materials and capital, using $\bar P$). For example, (\ref{eq:baseSU}) becomes $\tilde\alpha_U(\exp(\bar\omega_R)\bar U)^\theta / (\tilde\alpha_S \bar S^\theta) = \bar E_U/\bar E_S$, which combined with $\tilde\alpha_S + \tilde\alpha_U = 1$ yields (\ref{alpha_U}). Analogous reductions apply to (\ref{eq:baseML}) and (\ref{eq:baseMK}).}:
\begin{align}
& \tilde{\alpha}_U = \frac{\frac{\bar E_U}{(\exp(\bar \omega_R)\bar U)^\theta}}{\frac{\bar E_U}{(\exp(\bar \omega_R)\bar U)^\theta} +\frac{\bar E_S}{\bar S^\theta}} \label{alpha_U}\\
& \tilde{\alpha}_S = 1 - \tilde{\alpha}_U \\
& \tilde{\alpha}_M = \frac{\frac{\bar E_M}{\bar M^\rho}}{\frac{\bar E_M}{\bar M^\rho} + \frac{\bar E_L}{(\exp(\bar \omega_L)\bar L)^\rho} + \frac{\tau\bar E_M}{\bar K^\rho}}\\
& \tilde{\alpha}_L =\frac{\frac{\bar E_L}{(\exp(\bar \omega_L)\bar L)^\rho}}{\frac{\bar E_M}{\bar M^\rho} + \frac{\bar E_L}{(\exp(\bar \omega_L)\bar L)^\rho} + \frac{\tau\bar E_M}{\bar K^\rho}}\\
& \tilde{\alpha}_K = 1 - \tilde{\alpha}_M - \tilde{\alpha}_L \label{alpha_K}
\end{align}

Substituting these solutions, equations \eqref{alpha_U}--\eqref{alpha_K}, into the production function equations \eqref{eq:prod} and \eqref{eq:labor} yields the normalized production function:
\begin{equation}
 Q_{jt} = \bar Q\left[ \alpha_K\left( \frac{K_{jt}}{\bar K}\right)^{\rho} + \alpha_M\left(\frac{M_{jt}}{\bar M}\right)^{\rho} +  \alpha_L\left(\exp(\ddot\omega_{L,jt})\frac{L_{jt}}{\bar L}\right)^\rho \right]^{\frac{1}{\rho}}\exp(\ddot\omega_{H,jt}),  \label{eq:qnorm1}
\end{equation}
where
\begin{equation}
     \frac{L_{jt}}{\bar L} = \left[\alpha_S\left(\frac{S_{jt}}{\bar S}\right)^{\theta} + \alpha_U \left(\exp(\ddot\omega_{R,jt}) \frac{U_{jt}}{\bar U}\right)^{\theta}\right]^{\frac{1}{\theta}}, \label{eq:lnorm1}
\end{equation}
and transformed productivity is defined as
\begin{equation}
    \ddot\omega_{X,jt} = \omega_{X,jt}-\bar\omega_X, \quad X \in\{H,L,R\}.
\end{equation}
For all other variables, the double-dot diacritic denotes division by the geometric mean: $\ddot X_{jt} \equiv X_{jt}/\bar X$ for plant-level inputs and prices, and $\ddot X_{It} \equiv X_{It}/\bar X_I$ for industry-level indices.
Rescaling the outer CES aggregator by $\bar Q_*^\rho$, where $\bar Q_* = \bar Q/\exp(\bar\omega_H)$, gives $\alpha_X = \tilde\alpha_X \bar X^\rho / \bar Q_*^\rho$ for $X \in \{K, M\}$ and $\alpha_L = \tilde\alpha_L(\exp(\bar\omega_L)\bar L)^\rho / \bar Q_*^\rho$. Rescaling the labor nest by $\bar L^\theta$ gives $\alpha_S = \tilde\alpha_S \bar S^\theta / \bar L^\theta$ and $\alpha_U = \tilde\alpha_U(\exp(\bar\omega_R)\bar U)^\theta / \bar L^\theta$. Substituting the tilde-shares from (\ref{alpha_U})--(\ref{alpha_K}) and simplifying yields:
\begin{align}
& \alpha_S = \frac{\bar E_S}{\bar E_S + \bar E_U} \label{alphaS}\\
& \alpha_U = 1 - \alpha_S \label{alphaU}\\
& \alpha_M = \frac{\bar E_M}{\bar E_M + \bar E_L + \tau\bar E_M}=\frac{\bar E_M}{\bar E_M + \bar E_S +\bar E_U  + \tau\bar E_M} \label{alphaM}\\
& \alpha_L = \frac{\bar E_L}{\bar E_M + \bar E_L + \tau\bar E_M}=\frac{\bar E_S + \bar E_U}{\bar E_M + \bar E_S +\bar E_U  + \tau\bar E_M} \label{alphaL}\\
& \alpha_K = \frac{\tau\bar E_M}{\bar E_M + \bar E_L + \tau\bar E_M}=\frac{\tau\bar E_M}{\bar E_M + \bar E_S +\bar E_U  + \tau\bar E_M} \label{alphaK}
\end{align}
Substituting the baseline calibration $\tau = \bar E_K/\bar E_M$ gives the familiar cost-share form $\alpha_X = \bar E_X/(\bar E_M + \bar E_L + \bar E_K)$ for $X \in \{M, L, K\}$, with capital rental imputed as $\bar E_K$. At the baseline point, constant returns to scale of the labor aggregator and cost minimization give $\bar W_L \bar L = \bar W_S \bar S + \bar W_U \bar U$. Because geometric means are multiplicative, $\bar W_S \bar S = \bar E_S$ and $\bar W_U \bar U = \bar E_U$, so baseline labor expenditure is $\bar E_L = \bar E_S + \bar E_U$.

Because the normalized productivity terms $\ddot{\omega}_{X,jt}$ differ from $\omega_{X,jt}$ only by a constant, they inherit the same Markov structure. For $X\in \{H,L,R\}$:
\begin{equation}
\ddot{\omega}_{X,jt} = \ddot{\iota}_t + \ddot{\iota}_s + \rho_X \ddot{\omega}_{X,jt-1} + \beta_X^e e_{jt-1} + \beta_X^i i_{jt-1} + \beta_X^{ei} \left(e_{jt-1} \cdot i_{jt-1}\right) + \xi_{X,jt}. \label{eq:omega_norm}
\end{equation}
Throughout the article we refer to $\ddot\omega_{X,jt}$ simply as ``productivity''.

\par\medskip
\noindent\textit{Output and Input Markets} -- The Dixit-Stiglitz inverse demand depends only on price and quantity ratios, which are invariant under the geometric-mean normalization. We therefore write:
\begin{equation}
\ddot{P}_{jt}=\ddot{P}_{It}\left(\frac{\ddot{Q}_{jt}}{\ddot{Q}_{It}}\right)^{\frac{1}{\eta}}. \label{eq:normalized_demand}
\end{equation}

\subsubsection{Static Input Allocation Decision}
Plant $j$ chooses variable inputs to maximize short-run profits. The production function and labor aggregator are expressed in normalized form, whereas prices, wages, and industry indices remain in levels as observed in the data:
\begin{equation}
\label{eq:profit_max_norm}
    \begin{aligned}
\pi(\ddot{\Omega}_{jt}, e_{jt}) = & \max_{M_{jt}, S_{jt}, U_{jt}} \quad P_{jt}Q_{jt} - W_{S,jt}S_{jt}-W_{U,jt}U_{jt}-P_{M,jt}M_{jt} \\
\textrm{s.t.} \quad & Q_{jt} \geq (1-e_{jt}) Q^{0} + e_{jt} Q^{1} \\
& Q_{jt} = \bar Q\left[ \alpha_K\ddot{K}_{jt}^{\rho} + \alpha_M\ddot{M}_{jt}^{\rho} +  \alpha_L(\exp(\ddot\omega_{L,jt})\ddot{L}_{jt})^\rho \right]^{\frac{1}{\rho}}\exp(\ddot\omega_{H,jt}) \\
& \ddot{L}_{jt} = \left[\alpha_S\ddot{S}_{jt}^{\theta} + \alpha_U \left(\exp(\ddot\omega_{R,jt}) \ddot{U}_{jt}\right)^{\theta}\right]^{\frac{1}{\theta}}\\
& {P}_{jt}={P}_{It}\left(\frac{{Q}_{jt}}{{Q}_{It}}\right)^{\frac{1}{\eta}},
\end{aligned}
\end{equation}
where $\ddot{\Omega}_{jt}$ collects the relevant state variables:
\begin{equation}
\ddot{\Omega}_{jt}\equiv (\ddot{K}_{jt}, W_{S,jt}, W_{U,jt}, P_{M,jt}, P_{It}, Q_{It}, \exp(\ddot{\omega}_{H,jt}), \exp(\ddot{\omega}_{L,jt}), \exp(\ddot{\omega}_{R,jt})).
\end{equation}

The first-order conditions for the three variable inputs, at an allocation where the output constraint does not bind, are:
\begin{align*}
    E_{S,jt} &= \left(\frac{1+\eta}{\eta}\right) {R}_{jt} \left(\frac{\alpha_L\left(\exp(\ddot\omega_{L,jt}) \ddot{L}_{jt}\right)^\rho}{{\alpha_K} \ddot{K}_{jt}^{\rho} + \alpha_M \ddot{M}_{jt}^{\rho} + \alpha_L \left(\exp(\ddot\omega_{L,jt}) \ddot{L}_{jt}\right)^\rho}\right) \left(\frac{\alpha_S \ddot{S}_{jt}^{\theta}}{\ddot{L}_{jt}^{\theta}}\right), \\
    E_{U,jt} &= \left(\frac{1+\eta}{\eta}\right) {R}_{jt} \left(\frac{\alpha_L\left(\exp(\ddot\omega_{L,jt}) \ddot{L}_{jt}\right)^\rho}{{\alpha_K}\ddot{K}_{jt}^{\rho} + \alpha_M \ddot{M}_{jt}^{\rho} + \alpha_L \left(\exp(\ddot\omega_{L,jt}) \ddot{L}_{jt}\right)^\rho}\right) \left(\frac{\alpha_U \left(\exp(\ddot\omega_{R,jt}) \ddot{U}_{jt}\right)^{\theta}}{\ddot{L}_{jt}^{\theta}}\right), \\
    E_{M,jt} &= \left(\frac{1+\eta}{\eta}\right) {R}_{jt} \left(\frac{\alpha_M\ddot{M}_{jt}^{\rho}}{{\alpha_K} \ddot{K}_{jt}^{\rho} + \alpha_M \ddot{M}_{jt}^{\rho} + \alpha_L \left(\exp(\ddot\omega_{L,jt}) \ddot{L}_{jt}\right)^\rho}\right),
\end{align*}
where $E_{S,jt}$, $E_{U,jt}$, and $E_{M,jt}$ denote expenditures on skilled labor, unskilled labor, and materials. Taking ratios of these conditions yields the relative-demand equations of Steps 1 and 2 in Section \ref{s:empirics}.

\subsubsection{Dynamic Export and Investment Decisions, and End of Period}
\noindent\textit{Exporting and Investment Policy Functions} -- Because prices and expenditures scale homogeneously under the normalization, the flow profits $\pi$ and value functions $EV^x$ in \eqref{eq:exporting}--\eqref{eq:invest_e} transform into their normalized counterparts up to plant-independent multiplicative constants, leaving the argmax invariant. The investment policy on the normalized state space therefore satisfies:
\begin{equation}
\label{eq:investment_norm}
    \ddot{I}_{jt} = \mathcal{I}(\ddot{\Omega}_{jt},e_{jt},e_{jt+1}).
\end{equation}
Equation \eqref{eq:exporting} yields plant $j$'s conditional choice probability of exporting in period $t+1$:
\begin{equation}
    \Pr(e_{jt+1}=1|\ddot{\Omega}_{jt}, e_{jt}) = \int_{\gamma_{jt}} \mathbb{1}\left\{ EV^{1}(\ddot{\Omega}_{jt}, e_{jt}) - EV^{0}(\ddot{\Omega}_{jt}, e_{jt}) \geq \gamma_{jt} \right\} \text{d} G(\gamma_{jt}|e_{jt}). \label{eq:CCP_norm}
\end{equation}
\par\medskip
\noindent\textit{End of Period} -- We define normalized planned revenue as:
\begin{equation}
   \ddot{R}_{jt}\equiv  \ddot{P}_{jt}\ddot{Q}_{jt}. \label{eq:normalized_output}
\end{equation}
Because measurement error $\zeta_{jt}$ has zero mean, the geometric mean of observed revenue $\tilde{R}_{jt}$ (defined in \eqref{eq:quantity}) coincides with that of planned revenue $R_{jt}$, both equal to $\bar{R}$. Normalized observed revenue therefore satisfies:
\begin{equation}
      \ddot{\tilde{R}}_{jt} = \ddot{R}_{jt} \exp(\zeta_{jt}). \label{eq:normalized_output_observed}
\end{equation}

\subsection{Relative Input Use and Relative Input Prices}\label{a:rel_inputs}

The instrument set of Step 2 uses the lagged expenditure ratio of materials to skilled labor. Relative material use and relative input prices share a common component in the inner-nest expenditure ratio:
\begin{align}
      \frac{\ddot{M}_{jt}}{\ddot{L}_{jt}}
      &= \frac{\ddot{M}_{jt}}{\ddot{S}_{jt}}
      \left( \frac{\ddot{E}_{S,jt}}{\ddot{E}_{L,jt}} \right)^{\frac{\sigma_{\theta}}{\sigma_{\theta}-1}}, \label{eq:rel_use} \\
      \frac{\ddot{P}_{M,jt}}{\ddot{W}_{L,jt}}
      &= \left( \frac{\ddot{P}_{M,jt}}{\ddot{W}_{S,jt}} \right)
      \left( \frac{\ddot{E}_{S,jt}}{\ddot{E}_{L,jt}} \right)^{\frac{1}{1-\sigma_{\theta}}}. \label{eq:rel_price}
\end{align}
Because the lagged material-to-labor ratio already identifies the persistence of $\ddot{\omega}_{L,jt}$, identification of $\sigma_{\rho}$ rests on additional variation in relative input prices. Material prices vary only at the industry-year level, so the lagged material-to-skilled-labor price ratio contributes little within-industry cross-sectional variation. Under the assumption that the lagged expenditure ratio is predetermined with respect to $\xi_{L,jt}$, it supplies the plant-level variation that identifies $\sigma_{\rho}$.

\subsection{Colombia Administrative Departments}\label{a:map_Colombia}
\begin{figure}[h!]
    \centering
    \includegraphics[width=0.7\textwidth]{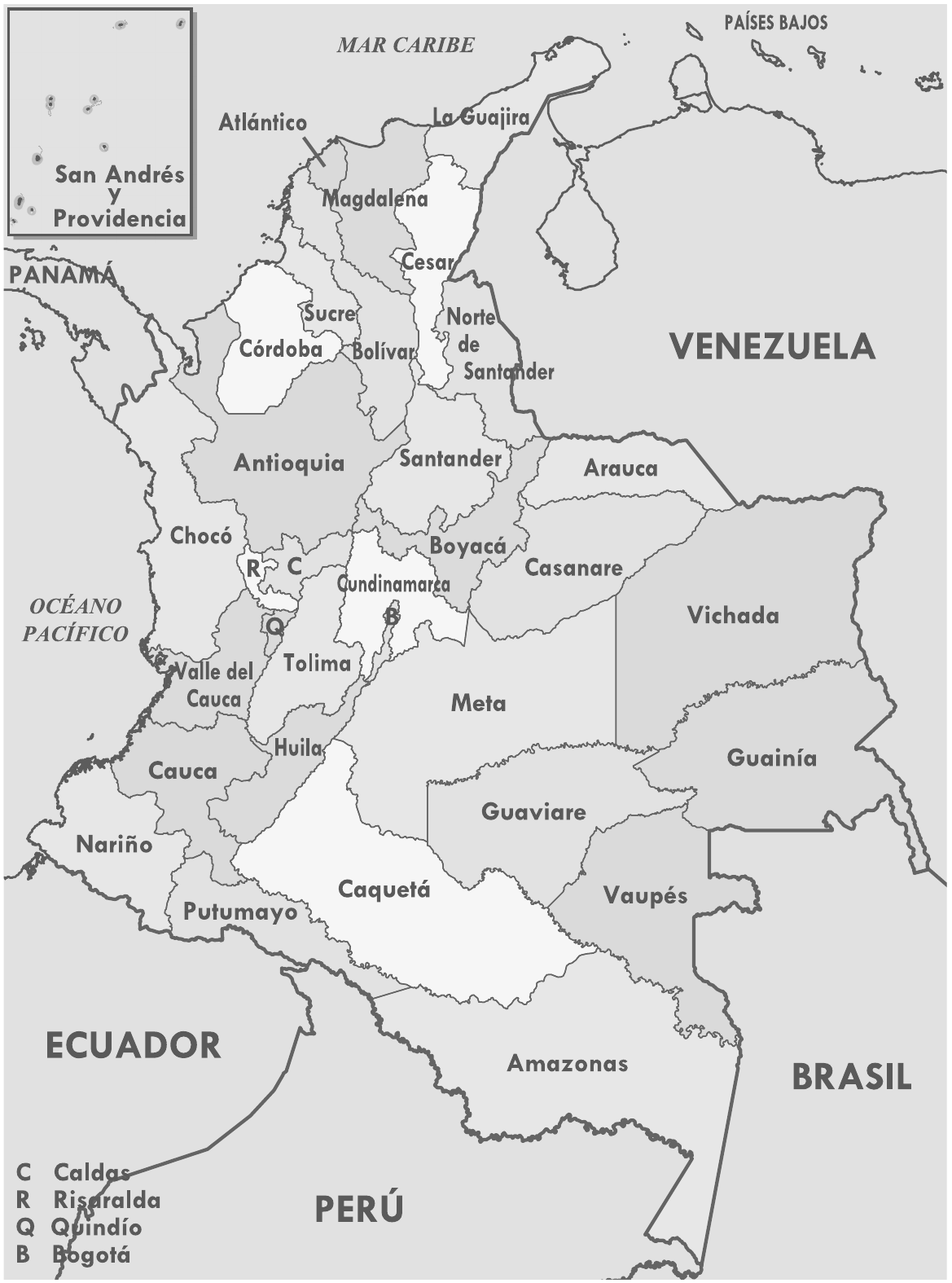}
    \caption{Map of Colombia's Administrative Departments}
\label{fig:map_Colombia}
\note{Sourced from Wikipedia.\\
By Milenioscuro, own work, CC BY-SA 4.0, \url{https://commons.wikimedia.org/w/index.php?curid=44931327}.
}
\end{figure}

\subsection{Kalman Procedures}\label{a:kalman}
This appendix details the Kalman filtering procedure used to compute the likelihood function for the state-space model in Section \ref{s:empirics}. Specifically, we reframe the dynamic panel residual $\widetilde{\omega}_{jt}$ as a linear state-space model:
\begin{align*}
    \widetilde{\omega}_{jt} &= H \pi_{jt} + A Y_{jt} + \ddot{\iota}_t + \ddot{\iota}_s, \\
    \pi_{jt} &= F \pi_{jt-1} + G u_{jt},
\end{align*}
where the measurement equation includes:
\begin{align*}
    H &= \begin{bmatrix} 1 & \frac{\hat{\eta}}{1+\hat{\eta}} & -\frac{\hat{\eta}}{1+\hat{\eta}}\hat{\rho}_{H} \end{bmatrix}, \quad
    A = \begin{bmatrix} \hat{\rho}_{H} & \hat{\beta}_{H}^{e} & \hat{\beta}_{H}^{i} & \hat{\beta}_{H}^{ei} \end{bmatrix},\\
    Y_{jt} &= \begin{bmatrix} \widetilde{\omega}_{jt-1} & e_{jt-1} & i_{jt-1} & e_{jt-1}\cdot i_{jt-1} \end{bmatrix}'.
\end{align*}
Equivalently, year and SIC dummies can be absorbed into an extended $Y_{jt}$, with $A$ extended by the corresponding dummy coefficients. The transition equation specifies:
\begin{align*}
    \pi_{jt} &= \begin{bmatrix} \xi_{H,jt} \\ \zeta_{jt} \\ \zeta_{jt-1} \end{bmatrix}, \quad 
    u_{jt} = \begin{bmatrix} \xi_{H,jt} \\ \zeta_{jt} \end{bmatrix}, \quad
    F = \begin{bmatrix} 0 & 0 & 0 \\ 0 & 0 & 0 \\ 0 & 1 & 0 \end{bmatrix}, \quad 
    G = \begin{bmatrix} 1 & 0 \\ 0 & 1 \\ 0 & 0 \end{bmatrix}, \\
    \text{var}(u_{jt}) &= \Sigma = \begin{bmatrix} \sigma_{H}^{2} & 0 \\ 0 & \sigma_{\zeta}^{2} \end{bmatrix}.
\end{align*}
We parameterize $\sigma_{H}^{2} = \exp(\theta_{1})$ and $\sigma_{\zeta}^{2} = \exp(\theta_{2})$ with $\theta_{1}, \theta_{2} \in \mathbb{R}$ to enforce positivity under unconstrained optimization. For each plant \( j \), the Kalman filter is initialized with the stationary mean and covariance of $\pi_{jt}$ under $(F, G, \Sigma)$ as the pre-sample prior at $t=0$; the first observation enters at $t=1$. The recursion proceeds as follows, with $k=3$ the state dimension \citep{hamilton1994filter}:
\begin{align*}
    & \pi_{j,0|0} = \mathbb{E}(\pi_{j0}) = \vec{0}_{3 \times 1}, \\
    & P_{j,0|0} = \mathrm{reshape}\!\left((I_{k^2} - F \otimes F)^{-1}\,\mathrm{vec}(G \Sigma G'),\,k,\,k\right), \\
    & \pi_{j,t+1|t} = F \pi_{j, t|t}, \\
    & P_{j,t+1|t} = F P_{j, t|t} F'  + G \Sigma G', \\
    & \nu_{jt+1} = \widetilde{\omega}_{jt+1} - H \pi_{j,t+1|t} - A Y_{jt+1} - \ddot{\iota}_{t+1} - \ddot{\iota}_s, \\
    & N_{j,t+1} = H P_{j,t+1|t} H', \\
    & O_{j,t+1} = P_{j,t+1|t} H' N_{j,t+1}^{-1}, \\
    & \pi_{j,t+1|t+1} =  \pi_{j,t+1|t}  +  O_{j,t+1}  \nu_{jt+1}, \\
    & P_{j,t+1|t+1} = (I - O_{j,t+1} H) P_{j,t+1|t}.
\end{align*}
At each iteration, we compute the log-likelihood of observing \(\widetilde{\omega}_{jt+1}\) using \(\nu_{jt+1}\) and \(N_{j,t+1}\):
\begin{align*}
    \log{f(\widetilde{\omega}_{jt+1} | \mathcal{H}_{jt})} = - \tfrac{1}{2}\log{(\text{det}(N_{j,t+1}))} - \tfrac{1}{2}\nu_{jt+1}' N_{j,t+1}^{-1} \nu_{jt+1} - \tfrac{1}{2}\log(2\pi).
\end{align*}
The contribution of plant \( j \) to the overall likelihood is then:
\begin{align*}
    L_{j}((\sigma_{H}^{2}, \sigma_{\zeta}^{2})|D) = - \tfrac{1}{2}\sum_{t=1}^{T_{j}} \log{(\text{det}(N_{j,t}))} - \tfrac{1}{2}\sum_{t=1}^{T_{j}} \nu_{jt}' N_{j,t}^{-1} \nu_{jt} - \tfrac{T_j}{2}\log(2\pi),
\end{align*}
and the total likelihood function is
\begin{align*}
    \mathcal{L}((\sigma_{H}^{2}, \sigma_{\zeta}^{2})|D) = \sum_{j=1}^{J} L_{j}((\sigma_{H}^{2}, \sigma_{\zeta}^{2})|D).
\end{align*}
Upon estimating \((\sigma_{H}^{2}, \sigma_{\zeta}^{2})\) via MLE, we recover the estimates of \(\xi_{H,jt}\) and \(\zeta_{jt}\) using the Kalman smoother \citep{rauch1965maximum}, based on all available information over the sample period. The smoother yields more precise estimates of \(\pi_{jt}\) by conditioning on the full sample \(T_{j}\), denoted \(\pi_{j,t|T_{j}}\). Specifically, given the Kalman-filtered variables \((\pi_{j,t|t}, P_{j,t|t},\allowbreak \pi_{j,t+1|t},\allowbreak P_{j,t+1|t})\), we recursively apply the Kalman smoother from \(T_{j}\) back to \(t=1\):
\begin{align*}
    & \widehat{\pi}_{j,T_{j}|T_{j}} = \pi_{j,T_{j}|T_{j}}, \\
    & \widehat{P}_{j,T_{j}|T_{j}} = P_{j,T_{j}|T_{j}}, \\
    & C_{jt} = P_{j,t|t} F' P_{j,t+1|t}^{-1}, \\
    & \widehat{\pi}_{j,t|T_{j}} = \pi_{j,t|t} + C_{jt}(\widehat{\pi}_{j,t+1|T_{j}} - \pi_{j,t+1|t}), \\
    & \widehat{P}_{j,t|T_{j}} = P_{j,t|t} + C_{jt}(\widehat{P}_{j,t+1|T_{j}} - P_{j,t+1|t}) C_{jt}'.
\end{align*}
The smoother returns $\hat{\zeta}_{jt}$ as the second element of $\widehat{\pi}_{j,t|T_{j}}$. We then recover Hicks-neutral productivity via $\hat{\ddot\omega}_{H,jt} = \widetilde{\omega}_{jt} - \frac{\hat\eta}{1+\hat\eta}\hat{\zeta}_{jt}$ and the innovation, up to the deterministic Markov terms, via $\hat{\xi}_{H,jt} = \hat{\ddot\omega}_{H,jt} - \hat\rho_{H}\,\hat{\ddot\omega}_{H,jt-1}$.

\subsection{Robustness to the Calibration of \texorpdfstring{$\tau$}{tau}}\label{a:tau_sensitivity}

In the baseline specification of Table \ref{tab:structural}, we calibrate the normalization constant \( \tau \) to the ratio of geometric means \( \bar{E}_K / \bar{E}_M \) in the data. This calibration is consistent with capital being paid its marginal product at the geometric-mean input bundle. As a robustness check, we re-estimate the production function treating \( \tau \) as a free parameter in the Step-3 GMM objective. In this specification, \( \tau \) is identified from the Step-3 revenue moment conditions alongside \( \eta \) and the Hicks-neutral Markov-process parameters, augmenting the baseline instrument set with the current capital stock, which is predetermined with respect to the period-\(t\) productivity innovation and identifies the capital share from the revenue side. Table \ref{tab:structural_transformed} reports the corresponding structural estimates, and Table \ref{tab:lbe_transformed} reports the productivity Markov-process coefficients.

The free-\( \tau \) GMM recovers \( \hat{\tau} = 0.044 \), below the calibrated value of \( \bar{E}_K / \bar{E}_M = 0.092 \). The elasticities of substitution are unchanged by construction (\( \hat{\sigma}_{\rho} = 0.399 \), \( \hat{\sigma}_{\theta} = 0.344 \), numerically identical across the two specifications because \( \tau \) enters only the Step-3 moment condition), and the markup barely moves (\( \hat{\mu} = 1.093 \) versus \( 1.091 \) under the calibrated baseline). The capital share is the parameter most sensitive to the treatment of \( \tau \): \( \hat{\alpha}_K \) falls from \( 0.065 \) to \( 0.032 \), with offsetting movements in \( \hat{\alpha}_L \) (\( 0.233 \) to \( 0.241 \)) and \( \hat{\alpha}_M \) (\( 0.702 \) to \( 0.727 \)). This difference is not statistically significant. The Kalman-filter innovation standard deviations remain stable (\( \hat{\sigma}_H = 0.254 \) versus \( 0.263 \); \( \hat{\sigma}_\zeta = 0.114 \) versus \( 0.109 \)). The persistence and exporter/investment loadings in Table \ref{tab:lbe_transformed} are identical to those in Table \ref{tab:lbe} for the labor-augmenting and unskilled-relative processes; only the Hicks-neutral process shifts.

The Step-3 Hansen-J statistic under the free-\( \tau \) specification is 0.451 (seven moments, six parameters, one degree of freedom; plant-clustered \(p\)-value 0.502), so the overidentifying restrictions are not rejected.

\begin{table}[!h]
\caption{Estimates of Structural Parameters under Free-\texorpdfstring{$\tau$}{tau} Specification}
\resizebox{\textwidth}{!}{%
\begin{tabular}{ccccccccccccc}
\toprule\toprule
 $\mu$ & & $\sigma_{\rho}$ & $\sigma_{\theta}$ & & $\alpha_L$ & $\alpha_M$ & $\alpha_K$ & $\alpha_S$ & $\alpha_U$ & & $\sigma_H$ & $\sigma_\zeta$ \\
\midrule
 1.093 & & 0.399 & 0.344 & & 0.241 & 0.727 & 0.032 & 0.317 & 0.683 & & 0.254 & 0.114 \\
 (0.030) & & (0.105) & (0.022) & & (0.022) & (0.067) & (0.089) & (0.002) & (0.002) & & (0.030) & (0.036) \\
\bottomrule
\end{tabular}

}
\note{As Table \ref{tab:structural}, but \( \tau \) is estimated jointly with the other Step-3 parameters rather than calibrated to \( \bar{E}_K / \bar{E}_M \).}
\label{tab:structural_transformed}
\end{table}

\begin{table}[!h]
\caption{Estimates of the Productivity Markov Process Coefficients under Free-\texorpdfstring{$\tau$}{tau} Specification}
\begin{tabular}{lcccc}
\toprule\toprule
 & $\ddot{\omega}_{-1}$ & $e_{-1}$ & $i_{-1}$ & $e_{-1}i_{-1}$ \\
\midrule
$\ddot{\omega}_H$ & 0.792 & 0.028 & 0.006 & 0.007 \\
 & (0.106) & (0.027) & (0.025) & (0.022) \\
$\ddot{\omega}_L$ & 0.864 & 0.039 & 0.031 & 0.023 \\
 & (0.004) & (0.040) & (0.011) & (0.041) \\
$\ddot{\omega}_R$ & 0.855 & 0.060 & 0.062 & 0.001 \\
 & (0.004) & (0.028) & (0.009) & (0.028) \\
\bottomrule
\end{tabular}

\note{As Table \ref{tab:lbe}, but under the free-\( \tau \) specification. All regressions include industry and year fixed effects.}
\label{tab:lbe_transformed}
\end{table}

\subsection{Propensity Score Matching: Supplementary Figures}\label{a:appendix_matching}

\begin{figure}[!ht]
\caption{Descriptive Patterns of Matching}
\label{fig:matching_descriptive}
\centering
\begin{subfigure}[t]{0.47\textwidth}
\centering
\subcaption{Common Support}\label{fig:combined}
\vspace{4pt}
\includegraphics[width=\linewidth]{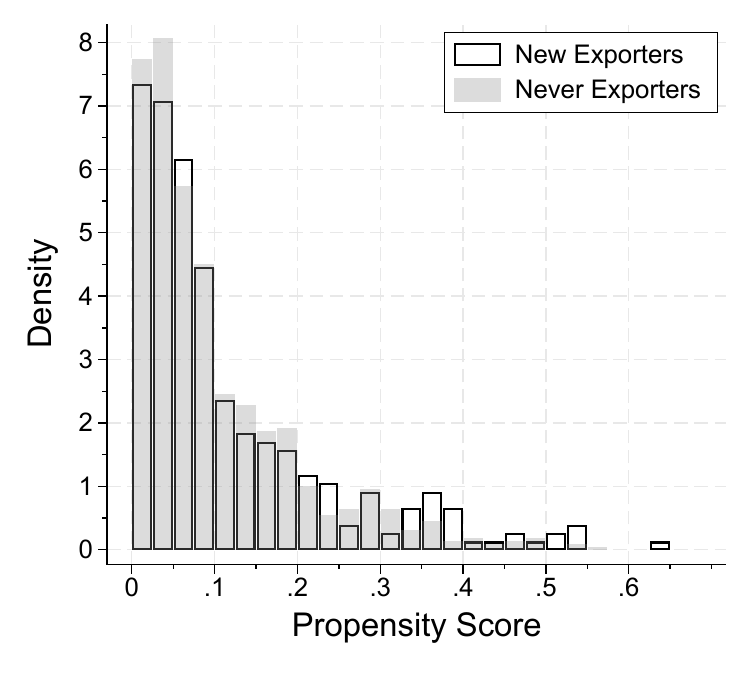}
\end{subfigure}\hfill
\begin{subfigure}[t]{0.51\textwidth}
\centering
\subcaption{Matching Balancedness}\label{tab:matching}
\vspace{4pt}
\renewcommand{\arraystretch}{1.15}
\begin{tabular*}{\linewidth}{@{\extracolsep{\fill}} lccccc}
\toprule\toprule
 & Treat. & Contr. & Diff. & $t$ & $p$ \\
\midrule
$\log\ddot{K}_{jt-1}$ &   0.442 &   0.442 &   0.000 &    0.00 &    1.00 \\
$\log\ddot{W}_{S,jt-1}$ &  -0.007 &  -0.022 &   0.015 &    0.34 &    0.73 \\
$\log\ddot{W}_{U,jt-1}$ &  -0.082 &  -0.079 &  -0.003 &   -0.08 &    0.94 \\
$\ddot{\omega}_{H,jt-1}$ &   0.169 &   0.162 &   0.007 &    0.14 &    0.89 \\
$\ddot{\omega}_{S,jt-1}$ &   0.248 &   0.304 &  -0.057 &   -0.45 &    0.66 \\
$\ddot{\omega}_{R,jt-1}$ &  -0.017 &  -0.006 &  -0.011 &   -0.10 &    0.92 \\
$\log\ddot{Q}_{It-1}$ &  -0.218 &  -0.218 &   0.000 &    0.00 &    1.00 \\
$\log\ddot{P}_{It-1}$ &  -0.159 &  -0.159 &   0.000 &    0.00 &    1.00 \\
$\log\ddot{P}_{M,jt-1}$ &  -0.110 &  -0.110 &   0.000 &    0.00 &    1.00 \\
\midrule
Plants &     305 &     655 &        &        &        \\
\bottomrule
\end{tabular*}

\end{subfigure}

\note{Panel A: estimated propensity-score densities, with scores from the entry-cohort-by-industry matching strata, for new exporters one year before entry and for the pooled set of their three matched nearest-neighbor controls. Clear bars: new-exporter observations; shaded bars: matched never-exporter observations. Panel B: matched-sample averages of normalized state variables by treatment status; last three columns show the difference in means, $t$-statistic, and $p$-value.}
\end{figure}

\begin{figure}[h!]
\caption{PSM Common Support across $k$-th Nearest Neighbors}
\label{fig:combined_neighbors}
\centering
\begin{subfigure}[t]{0.48\textwidth}
\centering
\subcaption{1st Nearest Neighbor}\label{fig:combined_first}
\includegraphics[width=\linewidth]{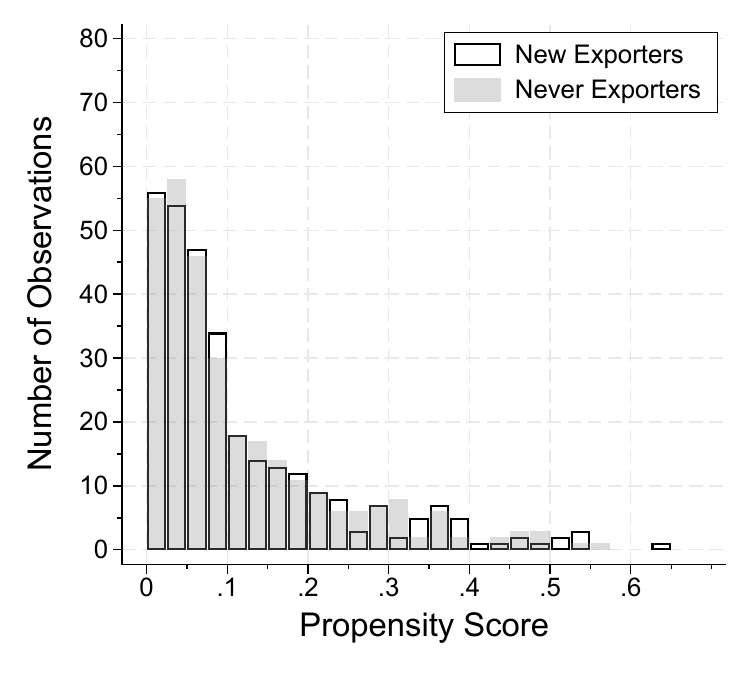}
\end{subfigure}\hfill
\begin{subfigure}[t]{0.48\textwidth}
\centering
\subcaption{2nd Nearest Neighbor}\label{fig:combined_second}
\includegraphics[width=\linewidth]{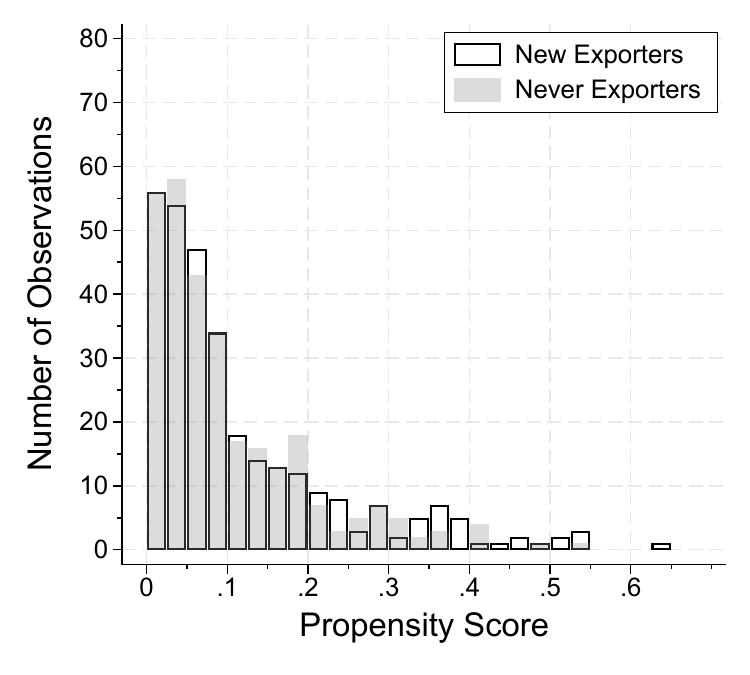}
\end{subfigure}

\vspace{1em}

\begin{subfigure}[t]{0.48\textwidth}
\centering
\subcaption{5th Nearest Neighbor}\label{fig:combined_fifth}
\includegraphics[width=\linewidth]{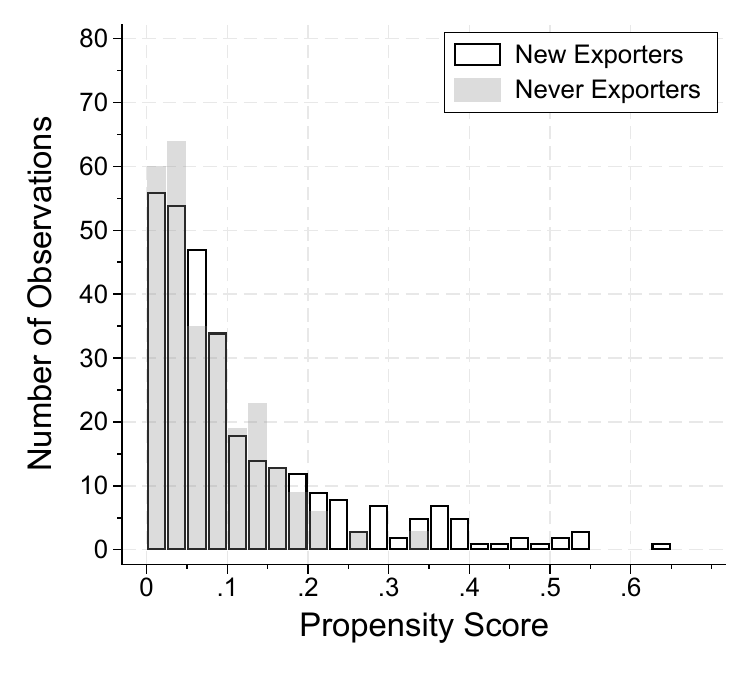}
\end{subfigure}\hfill
\begin{subfigure}[t]{0.48\textwidth}
\centering
\subcaption{10th Nearest Neighbor}\label{fig:combined_tenth}
\includegraphics[width=\linewidth]{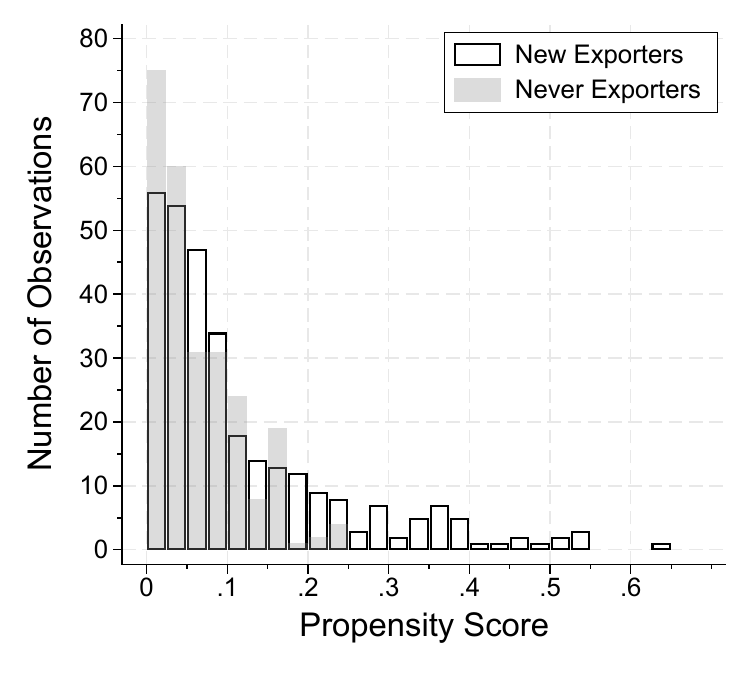}
\end{subfigure}

\note{Histograms of estimated propensity scores for new exporters one year before entry and for their $k$-th nearest-neighbor control plants, where $k$ ranges over $\{1, 2, 5, 10\}$. The histograms come from a ten-nearest-neighbor variant of the baseline 1:3 match, which retains the match ranks needed to display all four panels. Clear bars: new-exporter observations; shaded bars: matched never-exporter observations.}
\end{figure}

Figure \ref{fig:matching_descriptive} reports the two pre-treatment diagnostics for the baseline matched sample. Panel \subref{fig:combined} compares the estimated propensity-score distribution of new exporters one year before entry to that of their nearest-neighbor controls. The two densities overlap closely, consistent with common support, which also holds for higher-order nearest neighbors (Figure \ref{fig:combined_neighbors}). Panel \subref{tab:matching} reports treatment-control means on the matching covariates. We cannot reject equality at conventional levels for any covariate, consistent with balance on observed state variables. Because matching is carried out within each entry cohort and SIC3 stratum, a treated plant and its matched controls share the same industry and pre-entry year, so any matching covariate defined at the industry-year level, such as the price indices $\ddot{P}_I$ and $\ddot{P}_M$, coincides within the stratum and balances exactly.

\subsection{First-Order Counterfactual Loading Matrix for the Aggregate Skill-Intensity Analysis}\label{a:cf_inversion}

This appendix derives the first-order responses of the counterfactual factor quantities $S_{jt}^{cf}$, $U_{jt}^{cf}$, $M_{jt}^{cf}$ used in our analysis of aggregate skill intensity to the export-induced productivity gains. The counterfactual is a partial-equilibrium experiment: it holds capital $K_{jt}$, export status $e_{jt}$, factor prices $(W_{S,jt}, W_{U,jt}, P_{M,jt})$, and the industry price and quantity indices $(P_{It}, Q_{It})$ at their observed values, and lets each plant's own output $Q_{jt}$ adjust along its demand curve. We linearize the plant's static input first-order conditions around the observed allocation; the resulting loading matrix maps the productivity shocks $(\mathrm{d}\omega_H, \mathrm{d}\omega_L, \mathrm{d}\omega_R)$ into the log-changes of $(S, U, M)$.

\subsubsection{Static Allocation and First-Order Conditions}\label{a:cf_foc}

Conditional on the state, plant $j$ chooses $(S_{jt}, U_{jt}, M_{jt})$ to maximize profit $P_{jt}Q_{jt} - W_{S,jt}S_{jt} - W_{U,jt}U_{jt} - P_{M,jt}M_{jt}$, where $Q_{jt}$ is the nested-CES output of equations~\eqref{eq:prod}--\eqref{eq:labor} and the plant faces the inverse demand $P_{jt}=P_{It}(Q_{jt}/Q_{It})^{1/\eta}$. Because this demand curve has constant elasticity $\eta$, marginal revenue is $P_{jt}/\mu$, where $\mu=\eta/(\eta+1)$ is the markup of Table~\ref{tab:structural}. The plant therefore equates each variable input's marginal revenue product to its price:
\begin{equation}
\frac{P_{jt}}{\mu}\,\frac{\partial Q_{jt}}{\partial X_{jt}} = p_{X,jt}, \qquad X \in \{S, U, M\}, \quad (p_S, p_U, p_M) = (W_S, W_U, P_M).
\label{eq:a_foc}
\end{equation}
The relative first-order condition for $(S, U)$ gives the inner-nest skill ratio,
\begin{equation}
\frac{S_{jt}}{U_{jt}} = \left(\frac{W_{U,jt}\,\tilde\alpha_S}{W_{S,jt}\,\tilde\alpha_U}\right)^{\sigma_\theta}\exp\!\big(-(\sigma_\theta-1)\,\omega_{R,jt}\big),
\label{eq:a_skillratio}
\end{equation}
so removing the relative-unskilled gain scales the skill ratio by $a_{jt} = \exp((\sigma_\theta-1)\beta_R(h_{jt}))$. With output free to adjust, the conditions in~\eqref{eq:a_foc} additionally pin down the input \emph{levels}, which the loading matrix below delivers.

\subsubsection{Log-Linearization}\label{a:cf_loglin}

Let a circumflex denote a log-deviation, $\hat{x} \equiv \mathrm{d}\log x$, with productivity shocks $(\mathrm{d}\omega_H,\allowbreak \mathrm{d}\omega_L,\allowbreak \mathrm{d}\omega_R)$ and every other state variable held fixed. Write the plant-year outer-nest output elasticities of capital, materials, and effective labor,
\begin{equation}
\varepsilon_{K,jt} = \frac{\tilde\alpha_K K_{jt}^{\rho}}{G_{jt}^{\rho}}, \quad \varepsilon_{M,jt} = \frac{\tilde\alpha_M M_{jt}^{\rho}}{G_{jt}^{\rho}}, \quad \varepsilon_{L,jt} = \frac{\tilde\alpha_L (\mathcal{L}_{jt})^{\rho}}{G_{jt}^{\rho}}, \qquad \varepsilon_{K,jt} + \varepsilon_{M,jt} + \varepsilon_{L,jt} = 1,
\label{eq:a_epsilonL}
\end{equation}
where $G_{jt}^{\rho} = \tilde\alpha_K K_{jt}^{\rho} + \tilde\alpha_M M_{jt}^{\rho} + \tilde\alpha_L (\mathcal{L}_{jt})^{\rho}$ is the outer CES aggregator (so output is $Q_{jt} = \exp(\omega_{H,jt})\,G_{jt}$), $\mathcal{L}_{jt} \equiv \exp(\omega_{L,jt})L_{jt}$ is the effective labor bundle, and $\rho = (\sigma_\rho-1)/\sigma_\rho$; the three elasticities sum to one because the aggregator is homogeneous of degree one. The inner nest is likewise homogeneous of degree one, so $\varepsilon_{L,jt} = \varepsilon_{S,jt} + \varepsilon_{U,jt}$, and we write $\ell_{U,jt} = \varepsilon_{U,jt}/\varepsilon_{L,jt}$, $\ell_{S,jt} = \varepsilon_{S,jt}/\varepsilon_{L,jt}$ for the unskilled and skilled shares of the wage bill, all recovered plant by plant from the structural estimation.

Log-differentiating the building blocks of the static problem yields the demand curve $\hat P_{jt} = \hat Q_{jt}/\eta$, the inner labor aggregator $\hat L = \ell_S \hat S + \ell_U \hat U + \ell_U\,\mathrm{d}\omega_R$, the outer aggregator $\hat Q = \varepsilon_M \hat M + \varepsilon_L(\hat L + \mathrm{d}\omega_L) + \mathrm{d}\omega_H$, and the three first-order conditions in~\eqref{eq:a_foc}. Combining these reduces the system to the skill-mix differential of~\eqref{eq:a_skillratio},
\begin{equation}
\hat S_{jt} - \hat U_{jt} = (1-\sigma_\theta)\,\mathrm{d}\omega_{R,jt},
\label{eq:a_mix}
\end{equation}
together with a scale block that solves for output,
\begin{equation}
\hat Q_{jt} = \frac{\big(1+(\sigma_\rho-1) (1-\varepsilon_{K,jt})\big)\,\mathrm{d}\omega_H + \sigma_\rho \varepsilon_{L,jt}\,\mathrm{d}\omega_L + \sigma_\rho \varepsilon_{U,jt}\,\mathrm{d}\omega_R}{\Delta_{jt}},
\label{eq:a_yhat}
\end{equation}
where $b \equiv 1 + \sigma_\rho/\eta$ and $\Delta_{jt} \equiv 1 - b\,(1-\varepsilon_{K,jt}) = \sigma_\rho/|\eta| + b\,\varepsilon_{K,jt} > 0$. Scale is pinned down jointly by the demand markup and the fixed-capital share, and the small denominator $\Delta_{jt}$ governs how strongly the productivity gains expand the variable inputs.

\subsubsection{The Loading Matrix}\label{a:cf_loading}

Back-substituting~\eqref{eq:a_yhat} into the input first-order conditions and splitting the labor composite with~\eqref{eq:a_mix} expresses the counterfactual log-changes as a linear map from the productivity shocks. With $q_{H,jt} = (1+(\sigma_\rho-1)(1-\varepsilon_{K}))/\Delta$, $q_{L,jt} = \sigma_\rho \varepsilon_L/\Delta$, and $q_{R,jt} = \sigma_\rho \varepsilon_U/\Delta$ (plant-year subscripts on the elasticities suppressed),
\begin{equation}
\begin{pmatrix}\hat S_{jt}\\[2pt] \hat U_{jt}\\[2pt] \hat M_{jt}\end{pmatrix}
=
\begin{pmatrix}
b q_H+(\sigma_\rho-1) & b q_L+(\sigma_\rho-1) & b q_R+(\sigma_\rho-\sigma_\theta)\ell_U\\[3pt]
b q_H+(\sigma_\rho-1) & b q_L+(\sigma_\rho-1) & b q_R+\sigma_\rho\ell_U-1+\sigma_\theta\ell_S\\[3pt]
b q_H+(\sigma_\rho-1) & b q_L & b q_R
\end{pmatrix}
\begin{pmatrix}\mathrm{d}\omega_{H,jt}\\[2pt] \mathrm{d}\omega_{L,jt}\\[2pt] \mathrm{d}\omega_{R,jt}\end{pmatrix}.
\label{eq:a_loading}
\end{equation}
The Hicks-neutral column is common to all three rows, so $\omega_H$ shifts scale without tilting any input ratio. The $\omega_L$ column differs between labor and materials by $(\sigma_\rho-1) < 0$ but is common to $S$ and $U$: labor-augmenting efficiency saves labor relative to materials yet is neutral within the labor nest. Only the $\omega_R$ column separates $S$ from $U$, by $(1-\sigma_\theta)$ as in~\eqref{eq:a_mix}, so the relative-unskilled channel is the sole driver of the skill mix, whereas the two scale channels move $S$ and $U$ in common proportion.

\subsubsection{Counterfactual Levels}\label{a:cf_levels}

The counterfactual subtracts the export-induced gains, $\mathrm{d}\omega_v = -\beta_v(h_{jt})$ for $v \in \{H, L, R\}$, at new-exporter plant-years in their post-entry window and zero otherwise, with $\beta_S = \beta_L$ and $\beta_U = \beta_L + \beta_R$ following the recasting $\omega_{S} \equiv \omega_{L}$, $\omega_{U} \equiv \omega_{L} + \omega_{R}$. The counterfactual quantities follow from the loading matrix~\eqref{eq:a_loading} as
\begin{equation}
X_{jt}^{cf} = X_{jt}\,\exp\!\big(\hat X_{jt}\big), \qquad X \in \{S, U, M\},
\label{eq:a_cf_levels}
\end{equation}
and aggregating the level responses across plants delivers the head-count effects reported in the aggregate skill-intensity analysis. The counterfactual skill share weights each plant's exact inner-nest intensity shift, from equation~\eqref{eq:a_mix}, by the employment implied by the level responses in~\eqref{eq:a_cf_levels}.

Two properties carry through to the body. First, the skill ratio responds only to $\beta_R$: equation~\eqref{eq:a_mix} gives $\log(S_{jt}^{cf}/U_{jt}^{cf}) - \log(S_{jt}/U_{jt}) = -(1-\sigma_\theta)\,\beta_R(h_{jt})$, so under gross complementarity ($\sigma_\theta < 1$) a positive $\beta_R$ raises the observed skill ratio above the counterfactual, and the within-plant skill-intensity counterfactual is invariant to the Hicks-neutral and labor-augmenting channels. Second, because the level responses are governed by the small denominator $\Delta_{jt}$, the head-count magnitudes are sensitive to the demand elasticity $\eta$. The within-plant skill intensity is not, since it depends only on $\sigma_\theta$ and $\beta_R$, but the aggregate skill share weights plants by their employment, so its cross-plant reallocation component, and with it the aggregate gap, also depends on $\eta$.

\subsection{Sub-Decomposition of the FHK Export Effect on the Skill Share}\label{a:fhk_subdecomp}

This appendix derives the explicit forms of the three channels in equation~\eqref{eq:fhk_subdecomp}. For each year-pair $(t-1, t)$ in which a plant is present, \citet{foster2001aggregate} (FHK) decompose the year-over-year change in the aggregate skill share $\bar{s}_t = \sum_j \theta_{jt} s_{jt}$ (with $\theta_{jt}$ the plant employment share and $s_{jt}$ the plant skill share) into Within ($W_t$), Between ($B_t$), Cross ($C_t$), Entry ($E_t$), and Exit ($X_t$) terms:
\begin{equation}
\bar{s}_t - \bar{s}_{t-1} \;=\; W_t + B_t + C_t + E_t - X_t,
\label{eq:a_fhk}
\end{equation}
with the survivor-side terms
\begin{equation*}
W_t = \sum_{j\in\mathcal{S}_t} \theta_{j,t-1}\,\Delta s_{jt}, \quad
B_t = \sum_{j\in\mathcal{S}_t} \Delta\theta_{jt}\,(s_{j,t-1} - \bar{s}_{t-1}), \quad
C_t = \sum_{j\in\mathcal{S}_t} \Delta\theta_{jt}\,\Delta s_{jt},
\end{equation*}
where $\mathcal{S}_t$ is the set of plants present at both $t-1$ and $t$. Entrants and exiters contribute to $E_t$ and $X_t$. Applying equation~\eqref{eq:a_fhk} separately to the observed and counterfactual aggregates and summing over years gives the cumulative export effect at $t$:
\begin{equation*}
\bar{s}_t - \bar{s}^{cf}_t \;=\; \sum_{\tau=2}^{t}\Big[W^{export}_\tau + B^{export}_\tau + C^{export}_\tau + E^{export}_\tau - X^{export}_\tau\Big],
\end{equation*}
where each $D^{export}_\tau \equiv D_\tau - D^{cf}_\tau$ for $D \in \{W, B, C, E, X\}$.

We regroup the three survivor terms so that the observed-minus-counterfactual gaps in the within-plant change $\Delta s_{jt}$ and the employment-weight change $\Delta\theta_{jt}$ form the two channels $\Phi^{\Delta s}_\tau$ and $\Phi^{\Delta\theta}_\tau$, leaving the lagged-level gaps to the residual. Adding and subtracting $\theta_{j,\tau-1}\Delta s^{cf}_{j\tau}$ in $W^{export}_\tau$, $\Delta\theta_{j\tau}(s^{cf}_{j,\tau-1} - \bar{s}^{cf}_{\tau-1})$ in $B^{export}_\tau$ (the lagged share demeaned to match equation~\eqref{eq:a_fhk}), and $\Delta\theta_{j\tau}\Delta s^{cf}_{j\tau}$ in $C^{export}_\tau$:
\begin{align}
W^{export}_\tau &= \underbrace{\sum_j \theta_{j,\tau-1}\big(\Delta s_{j\tau} - \Delta s^{cf}_{j\tau}\big)}_{W^{(a)}_\tau:\;\text{pure }\Delta s} \;+\; \underbrace{\sum_j \big(\theta_{j,\tau-1} - \theta^{cf}_{j,\tau-1}\big)\Delta s^{cf}_{j\tau}}_{W^{(b)}_\tau:\;\text{interaction}}, \label{eq:a_W_subdecomp}\\[4pt]
B^{export}_\tau &= \underbrace{\sum_j \Delta\theta_{j\tau}\,\big(s^{dm}_{j,\tau-1} - s^{cf,dm}_{j,\tau-1}\big)}_{B^{(a)}_\tau:\;\text{interaction}} \;+\; \underbrace{\sum_j \big(\Delta\theta_{j\tau} - \Delta\theta^{cf}_{j\tau}\big)\,s^{cf,dm}_{j,\tau-1}}_{B^{(b)}_\tau:\;\text{pure }\Delta\theta},\label{eq:a_B_subdecomp}\\[4pt]
C^{export}_\tau &= \underbrace{\sum_j \Delta\theta_{j\tau}\,\big(\Delta s_{j\tau} - \Delta s^{cf}_{j\tau}\big)}_{C^{(a)}_\tau:\;\text{pure }\Delta s\text{ side}} \;+\; \underbrace{\sum_j \big(\Delta\theta_{j\tau} - \Delta\theta^{cf}_{j\tau}\big)\Delta s^{cf}_{j\tau}}_{C^{(b)}_\tau:\;\text{pure }\Delta\theta\text{ side}},\label{eq:a_C_subdecomp}
\end{align}
where the $dm$ superscript denotes demeaning by the lagged aggregate skill share, $s^{dm}_{j,\tau-1} = s_{j,\tau-1} - \bar{s}_{\tau-1}$ in the observed world and $s^{cf,dm}_{j,\tau-1} = s^{cf}_{j,\tau-1} - \bar{s}^{cf}_{\tau-1}$ in the counterfactual, matching the convention in equation~\eqref{eq:a_fhk}.

The two structural channels collect the sub-pieces carrying the gap in their own change ($\Delta s$ or $\Delta\theta$), and the remaining interaction sub-pieces join the plant-turnover channel ($\Phi^{NE}_\tau \equiv E^{export}_\tau - X^{export}_\tau$) in a single residual:
\begin{equation}
\Phi^{\Delta s}_\tau = W^{(a)}_\tau + C^{(a)}_\tau, \qquad \Phi^{\Delta\theta}_\tau = B^{(b)}_\tau + C^{(b)}_\tau, \qquad \Phi^{\mathrm{res}}_\tau = W^{(b)}_\tau + B^{(a)}_\tau + \Phi^{NE}_\tau.
\label{eq:a_channel_grouping}
\end{equation}
By construction, $\Phi^{\Delta s}_\tau + \Phi^{\Delta\theta}_\tau + \Phi^{\mathrm{res}}_\tau$ equals the year-$\tau$ export effect on $\Delta\bar{s}_\tau$, so summing over $\tau$ yields equation~\eqref{eq:fhk_subdecomp} exactly, using that the counterfactual coincides with the observed aggregate in the initial year. Like the underlying FHK accounting, the construction restricts to year-pair survivors: a plant contributes to $\Phi^{\Delta s}_\tau$ or $\Phi^{\Delta\theta}_\tau$ whenever it is present at both $\tau-1$ and $\tau$, whether or not it appears in every panel year.

\subsection{Aggregate TFP Counterfactual}\label{a:tfp_cf}

The compositional analysis of aggregate skill intensity characterizes how export-induced technical change reshapes the skill mix and scales input demand. The complementary question is how the same productivity gains move aggregate total factor productivity (the efficiency with which plants convert inputs into output), which we measure here with a growth-accounting counterfactual that mirrors the skill-share decomposition.

Plant-level log-TFP \emph{growth} is approximated by
\begin{equation}
   \Delta\log\text{TFP}_{jt} = \Delta\omega_{H,jt} + \varepsilon_{S,j,t-2}\,\Delta\omega_{S,jt} + \varepsilon_{U,j,t-2}\,\Delta\omega_{U,jt},
   \label{eq:tfp_decomp}
\end{equation}
where $\varepsilon_S$ and $\varepsilon_U$ are the plant-year output elasticities of skilled and unskilled labor obtained from the structural estimation.\footnote{The labor-augmenting outer-nest productivity $\omega_L$ does not appear because Solow accounting uses raw $S$ and $U$ rather than the effective labor bundle $\mathcal{L}_{jt}$, so the labor-augmenting contribution is folded into $\omega_S$ and $\omega_U$.} We work in growth rather than levels (consistent with the matched DiD identifying changes relative to the pre-entry year) because differencing a level index would pick up movements in the elasticities, not just in productivity.

The counterfactual zeroes the matched-DiD effects $\beta_H(h_{jt})$, $\beta_S(h_{jt})$, $\beta_U(h_{jt})$ for new exporters in their post-entry window and aggregates the cumulated plant-level log TFP to the sales-weighted level $\overline{\log\text{TFP}}_t = \sum_j \theta_{jt}\,\log\text{TFP}_{jt}$. The weights $\theta_{jt}$ are sales shares, $R_{jt}/R_t$ in the observed world and $R^{cf}_{jt}/R^{cf}_t$ in the counterfactual. The counterfactual sales $R^{cf}_{jt}$ come from the same first-order output response that underlies the skill-share counterfactual, so the export-induced movement in sales enters the reallocation term exactly as it does there. The series runs from 1983 onward.

\begin{figure}[!ht]
\caption{Aggregate Log TFP: Counterfactual and Decomposition}
\label{fig:tfp_aggregate}
\centering
\begin{subfigure}[t]{0.49\textwidth}
\centering
\subcaption{Observed vs.\ counterfactual aggregate log TFP}\label{fig:tfp_aggregate_a}
\includegraphics[width=\linewidth]{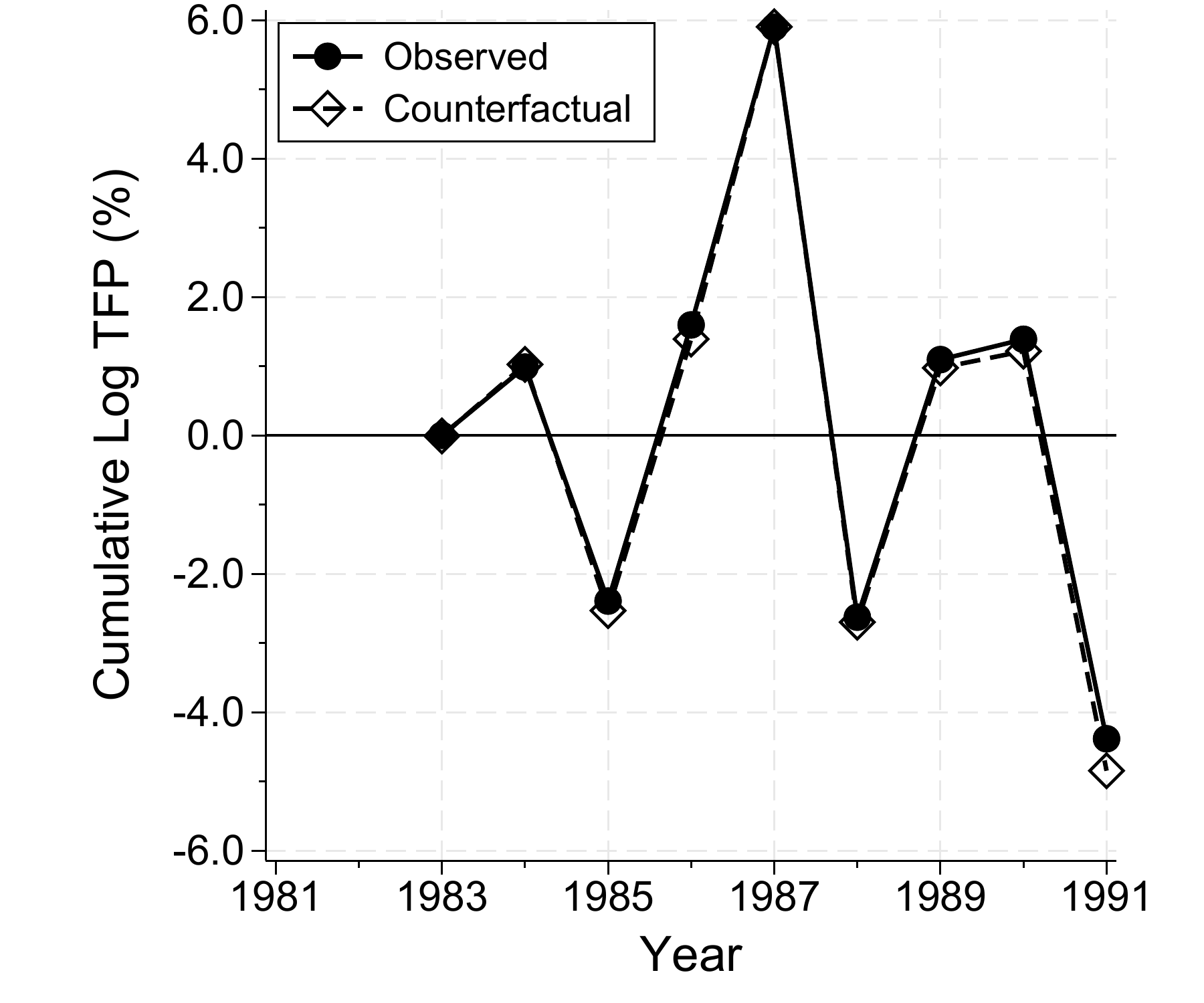}
\end{subfigure}\hfill
\begin{subfigure}[t]{0.49\textwidth}
\centering
\subcaption{FHK decomposition of the cumulative export effect}\label{fig:tfp_aggregate_b}
\includegraphics[width=\linewidth]{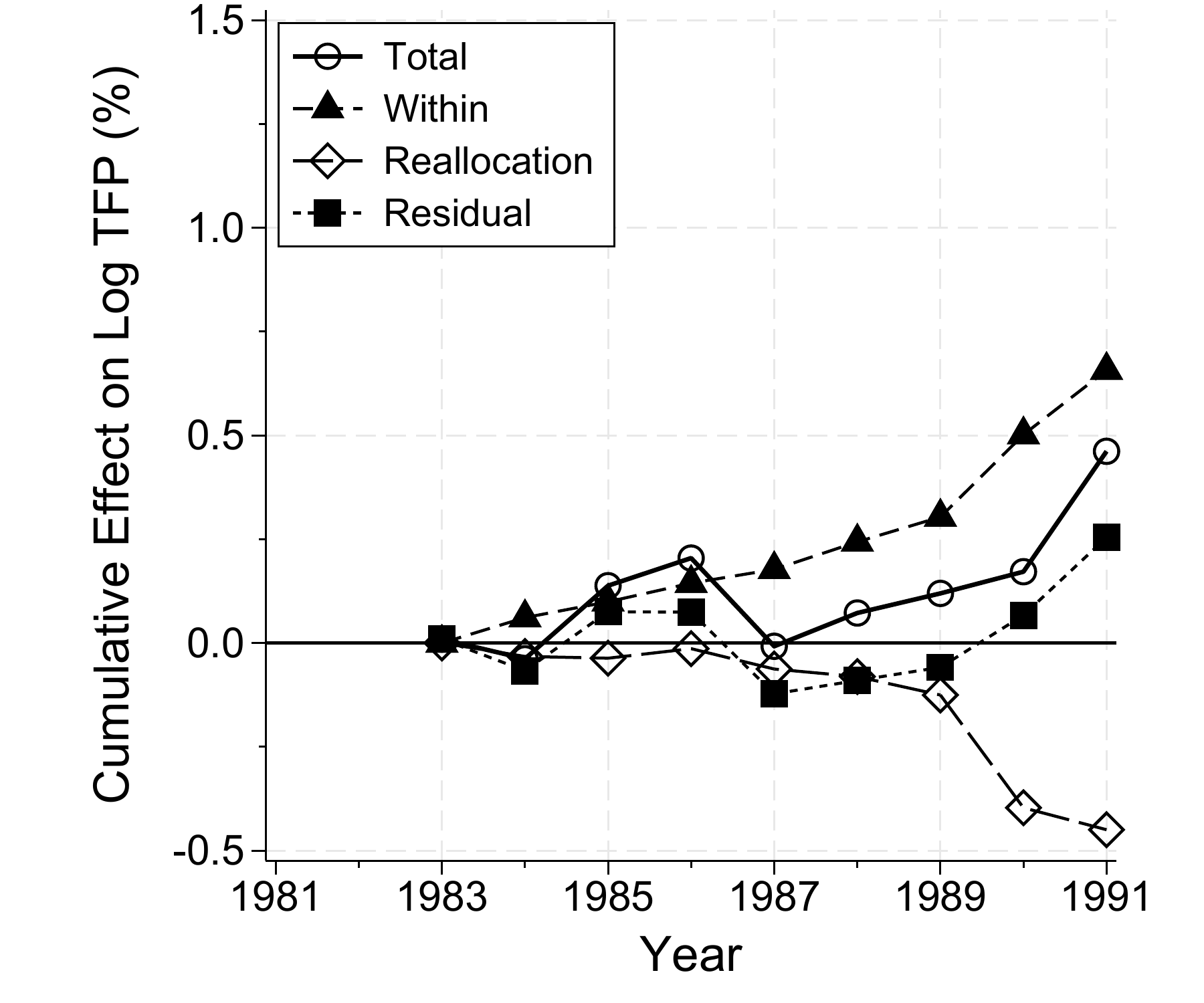}
\end{subfigure}
\note{Panel A: observed and counterfactual aggregate log TFP, both expressed as cumulative log change from the 1983 base. Firm-level log-TFP growth follows the Solow identity in equation~\eqref{eq:tfp_decomp}; the counterfactual subtracts the matched stacked-DiD post-entry effects on $(\omega_H, \omega_S, \omega_U)$ from each new-exporter plant-year and re-aggregates with sales weights: observed shares for the observed world and counterfactual shares $R^{cf}$ for the counterfactual. Panel B: cumulative export effect on aggregate log TFP (Total = observed minus counterfactual), in percent, decomposed via the \citet{foster2001aggregate} accounting into Within, Reallocation, and a Residual (the FHK interaction sub-pieces plus net plant entry-exit). The $y$-axis range matches that of Figure~\ref{fig:aggregate_cf}'s Panel B, so the two FHK decompositions are visually directly comparable.}
\end{figure}

Figure~\ref{fig:tfp_aggregate} plots the result. Panel A shows a small positive cumulative export effect on aggregate log TFP, reaching +0.46\% by 1991, about two-thirds of the +0.70\% effect on the aggregate skill share documented in Figure~\ref{fig:aggregate_cf}. Panel B decomposes it with the same Foster-Haltiwanger-Krizan sub-decomposition used for the skill share: the within-plant channel contributes +0.66\%, a cross-plant reallocation of sales toward initially less productive new exporters subtracts 0.45 percentage points, and a residual (the FHK interaction plus net plant entry-exit) adds +0.25\%. The within-plant productivity gain thus dominates, partly offset by the reallocation margin.

\subsection{Unfiltered Hicks-Neutral Productivity}\label{a:unfiltered}

\begin{figure}[!ht]
\caption{Descriptive Patterns of Matching: Unfiltered $\ddot{\omega}_H$}
\label{fig:matching_descriptive_noisy}
\centering
\begin{subfigure}[t]{0.47\textwidth}
\centering
\subcaption{Common Support}\label{fig:combined_noisy}
\vspace{4pt}
\includegraphics[width=\linewidth]{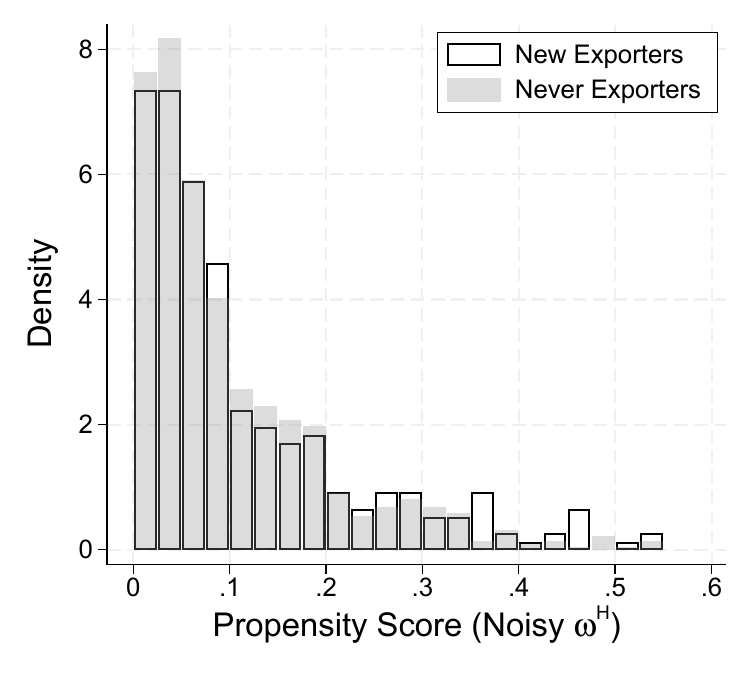}
\end{subfigure}\hfill
\begin{subfigure}[t]{0.51\textwidth}
\centering
\subcaption{Matching Balancedness}\label{tab:matching_unfiltered}
\vspace{4pt}
\renewcommand{\arraystretch}{1.15}
\begin{tabular*}{\linewidth}{@{\extracolsep{\fill}} lccccc}
\toprule\toprule
 & Treat. & Contr. & Diff. & $t$ & $p$ \\
\midrule
$\log\ddot{K}_{jt-1}$ &   0.442 &   0.425 &   0.016 &    0.13 &    0.90 \\
$\log\ddot{W}_{S,jt-1}$ &  -0.003 &  -0.025 &   0.022 &    0.52 &    0.60 \\
$\log\ddot{W}_{U,jt-1}$ &  -0.085 &  -0.089 &   0.004 &    0.12 &    0.91 \\
$\ddot{\omega}_{H,jt-1}$ &   0.166 &   0.168 &  -0.002 &   -0.05 &    0.96 \\
$\ddot{\omega}_{S,jt-1}$ &   0.246 &   0.233 &   0.013 &    0.10 &    0.92 \\
$\ddot{\omega}_{R,jt-1}$ &  -0.031 &  -0.046 &   0.016 &    0.14 &    0.89 \\
$\log\ddot{Q}_{It-1}$ &  -0.234 &  -0.234 &   0.000 &    0.00 &    1.00 \\
$\log\ddot{P}_{It-1}$ &  -0.155 &  -0.155 &   0.000 &    0.00 &    1.00 \\
$\log\ddot{P}_{M,jt-1}$ &  -0.108 &  -0.108 &   0.000 &    0.00 &    1.00 \\
\midrule
Plants &     305 &     675 &        &        &        \\
\bottomrule
\end{tabular*}

\end{subfigure}

\note{Panel A: estimated propensity-score densities for new exporters one year before entry and for the pooled set of their three matched nearest-neighbor controls, where the propensity score is estimated using the unfiltered measure of Hicks-neutral productivity. Clear bars: new-exporter observations; shaded bars: matched never-exporter observations. Panel B: matched-sample averages of normalized state variables by treatment status; last three columns show the difference in means, $t$-statistic, and $p$-value.}
\end{figure}

\subsection{Cross-Validation Through an Alternative ACF Estimator}\label{a:acf}

In this appendix we cross-validate the baseline estimate of Hicks-neutral productivity, $\ddot{\omega}_{H,jt}$, by replacing the GMM of Step~3 and the Kalman filter of Step~4 with a proxy-variable inversion in the spirit of \citet{olley1996dynamics}, \citet{levinsohn2003estimating}, and \citet{ackerberg2015identification}. Steps~1 and 2 of the baseline remain unchanged: $\sigma_{\theta}$ and $\sigma_{\rho}$ are pinned by their own moment conditions and identical to the baseline by construction. Steps~3 and 4 are jointly replaced by a two-stage proxy-variable procedure: a sparse approximation of the predicted output implied by the proxy-variable inversion (first stage), followed by GMM estimation of $\eta$ and the Markov process for $\ddot{\omega}_{H}$ (second stage).

The proxy-variable inversion requires plant-level investment $\ddot{I}_{jt}$ to be strictly monotonic in $\ddot{\omega}_{H,jt}$ conditional on other state variables. We follow \citet{olley1996dynamics} and restrict the sample to plant-years with positive investment, on which the monotonicity assumption is defensible. Within this subsample we recompute all demeaned variables, geometric means, and CES factor share parameters for internal consistency. The Step~1--2 estimates of $\sigma_{\theta}$ and $\sigma_{\rho}$ are held fixed; the rescaled efficiency residuals $\widetilde{\ddot{\omega}}_{R,jt}$ and $\widetilde{\ddot{\omega}}_{L,jt}$ are recomputed by re-evaluating equations~\eqref{eq:omega_US_char} and~\eqref{eq:omega_S_char} at the subsample-demeaned variables.\footnote{We deliberately keep Steps~1 and 2 on the full panel: the $\ddot{I}_{jt}>0$ subsample has too little within-plant variation in the skill premium to identify $\sigma_{\theta}$.}

\par\medskip
\noindent\textit{First stage.}\hspace{0.5em} Combining the production function with the closed-form expressions for $\widetilde{\ddot{\omega}}_{R,jt}$ and $\widetilde{\ddot{\omega}}_{L,jt}$ derived in Steps~1 and 2 yields the log-output equation
\begin{equation}
\log\ddot{Q}_{jt} = \log f\!\left(\tau, \widehat{\sigma}_{\rho}, \ddot{X}_{jt}\right) + \ddot{\omega}_{H,jt} + \zeta_{jt},
\label{eq:acf_lnQ}
\end{equation}
where $\ddot{X}_{jt} = (\ddot{K}_{jt}, \ddot{M}_{jt}, \ddot{L}_{jt}, \ddot{\omega}_{L,jt})$ is recovered from the previous steps. We assume that the normalized investment policy function~\eqref{eq:investment} is strictly monotonic in $\exp(\ddot{\omega}_{H,jt})$ conditional on the other normalized state variables. Inverting it gives $\ddot{\omega}_{H,jt} = h\!\left(\log\ddot{I}_{jt}, \ddot{\Omega}_{jt}\right)$ for some function $h$ and a vector of conditioning state variables $\ddot{\Omega}_{jt}$. Substituting back into~\eqref{eq:acf_lnQ} expresses log output as a function $\phi_{jt}$ of observable inputs and prices plus the measurement error:
\begin{equation}
\log\ddot{Q}_{jt} = \phi_{jt} + \zeta_{jt}, \qquad \phi_{jt} = \log f\!\left(\tau, \widehat{\sigma}_{\rho}, \ddot{X}_{jt}\right) + h\!\left(\log\ddot{I}_{jt}, \ddot{\Omega}_{jt}\right).
\label{eq:acf_phi}
\end{equation}

We approximate $\phi_{jt}$ nonparametrically by a third-degree polynomial in
\begin{align*}
\bigl(&\log\ddot{I}_{jt},\ \log\ddot{K}_{jt},\ \log\ddot{M}_{jt},\ \log\ddot{S}_{jt},\ \log\ddot{U}_{jt},\ \log\ddot{W}_{S,jt},\ \log\ddot{W}_{U,jt},\\
&\log\ddot{P}_{M,jt},\ \log\ddot{P}_{It},\ \log\ddot{Q}_{It},\ \ddot{\omega}_{R,jt},\ \ddot{\omega}_{L,jt}\bigr),
\end{align*}
together with all pairwise and three-way interactions, polynomial-by-export-status cross terms, and a full set of year and industry dummies. To address the resulting high dimensionality, we sparsify the design matrix via LASSO. The fit, $\widehat{\phi}_{jt}$, isolates the predicted component of $\log\ddot{Q}_{jt}$; its residual $\widehat{\zeta}_{jt}$ yields an estimate of the measurement-error variance, $\widehat{\sigma}_{\zeta}^{\,2} = \mathrm{Var}(\widehat{\zeta}_{jt})$. The first stage does not identify $\eta$: identification is deferred to the second-stage moment condition below, which uses inputs predetermined with respect to the AR(1) innovation $\xi_{H,jt+1}$ and thereby addresses the functional-dependence concern in \citet{ackerberg2015identification}.

\par\medskip
\noindent\textit{Second stage.}\hspace{0.5em} Following \citet{klette1996inconsistency}, the price-deflated revenue equation~\eqref{eq:revenue_generating_eq} can be rearranged to express Hicks-neutral productivity as a function of the demand-elasticity parameter $\eta$ and the first-stage fit:
\begin{equation}
\ddot{\omega}_{H,jt}(\eta) = \frac{\eta}{\eta + 1}\,\widehat{\phi}_{jt} - \log f\!\left(\tau, \widehat{\sigma}_{\rho}, \ddot{X}_{jt}\right) + \frac{1}{\eta+1}\log\frac{\ddot{R}_{It}}{\ddot{P}_{It}},
\label{eq:acf_omegaH}
\end{equation}
where $\widehat{\phi}_{jt}$ is the first-stage fit, $\log f(\cdot)$ is the CES production component, and $\tau = \bar{E}_K/\bar{E}_M$ is calibrated as in equation~\eqref{eq:tau_calibration} on the $\ddot{I}_{jt}>0$ subsample.

To estimate $\eta$ alongside the Markov process for $\ddot{\omega}_{H,jt}$, we form the AR(1) innovation
\begin{align*}
\xi_{H,jt+1}(\eta)
&= \ddot{\omega}_{H,jt+1}(\eta)
- \rho_{H}\,\ddot{\omega}_{H,jt}(\eta)
- \beta_{H}^{e}\, e_{jt}
- \beta_{H}^{i}\, i_{jt}
- \beta_{H}^{ei}\, e_{jt}\, i_{jt}
- \iota_{H,t} - \iota_{H,s},
\end{align*}
where $\iota_{H,t}$ and $\iota_{H,s}$ denote year and industry fixed effects. We concentrate out the Markov coefficients and the fixed effects, so $\eta$ is the single free parameter, and estimate it from the moment condition
\begin{align}
\mathbb{E}\!\left[\xi_{H,jt+1}(\eta) \times Z_{H,jt+1}\right] = 0,
\label{eq:moment_condition_acf}
\end{align}
where the instrument vector is
\[
Z_{H,jt+1} = \left(\frac{\ddot{K}_{jt+1}}{\ddot{M}_{jt}},\ E_{L,jt} + E_{M,jt},\ \log\ddot{Q}_{It}\right)'.
\]
The baseline estimates $\eta$ and the Markov process jointly, with the lagged $\ddot{K}/\ddot{M}$ ratio helping identify $\rho_H$; the ACF concentrates out the Markov coefficients, so $\eta$ alone is free and $\rho_H$ needs no instrument.

\par\medskip
The $\ddot{I}_{jt}>0$ subsample is not random: plant-years with zero investment differ systematically from the full panel, so the cross-validation compares the two estimators of $\ddot{\omega}_H$ on this subsample only. Even so, the markup, the variance components $\sigma_H$ and $\sigma_\zeta$, and the AR(1) persistence of $\ddot{\omega}_H$ remain close to their baseline counterparts in Tables~\ref{tab:structural} and~\ref{tab:lbe}. The ACF's second-stage estimate of $\eta$ implies a markup of $\widehat{\mu} = 1.075$. The ACF-implied standard deviations of the Hicks-neutral innovation and the measurement error are $\widehat{\sigma}_H = 0.235$ and $\widehat{\sigma}_\zeta = 0.156$. These estimates, together with the factor share parameters, appear in Table~\ref{tab:structural_acf}; Table~\ref{tab:structural} reports the baseline counterparts. 

The proxy-variable inversion changes only the parameter estimates for $\ddot{\omega}_H$ in the productivity Markov process; those for $\ddot{\omega}_{L}$ and $\ddot{\omega}_{R}$ are unchanged from Steps~1 and 2 of the baseline and appear in Table~\ref{tab:lbe}. Table~\ref{tab:lbe_acf} reports the alternative ACF estimates for $\ddot{\omega}_H$, with AR(1) persistence $\widehat{\rho}_H = 0.831$; Table~\ref{tab:lbe} reports the baseline counterpart.

\begin{table}[!ht]
\caption{Estimates of the Productivity Markov Process Coefficients: Alternative ACF Specification}
\centering
\begin{tabular}{lcccc}
\toprule\toprule
 & $\ddot{\omega}_{-1}$ & $e_{-1}$ & $i_{-1}$ & $e_{-1}i_{-1}$ \\
\midrule
$\ddot{\omega}_H$ & 0.831 & 0.022 & 0.002 & -0.007 \\
 & (0.027) & (0.035) & (0.011) & (0.035) \\
\bottomrule
\end{tabular}

\note{This table presents coefficients of the productivity Markov process for $\ddot{\omega}_H$ under the alternative ACF specification. The regression includes industry and year fixed effects. Standard errors are computed using a nonparametric bootstrap clustered at the establishment level. We draw 1,000 bootstrap samples with replacement and re-estimate the structural model for each sample; 998 replications converged successfully.}
\label{tab:lbe_acf}
\end{table}

\begin{figure}[!ht]
\caption{Cross-Validation: Distribution of $\ddot{\omega}_H$ under Alternative ACF vs.\ Baseline}
\label{fig:comparison_density}
\centering
\includegraphics[width=0.75\linewidth]{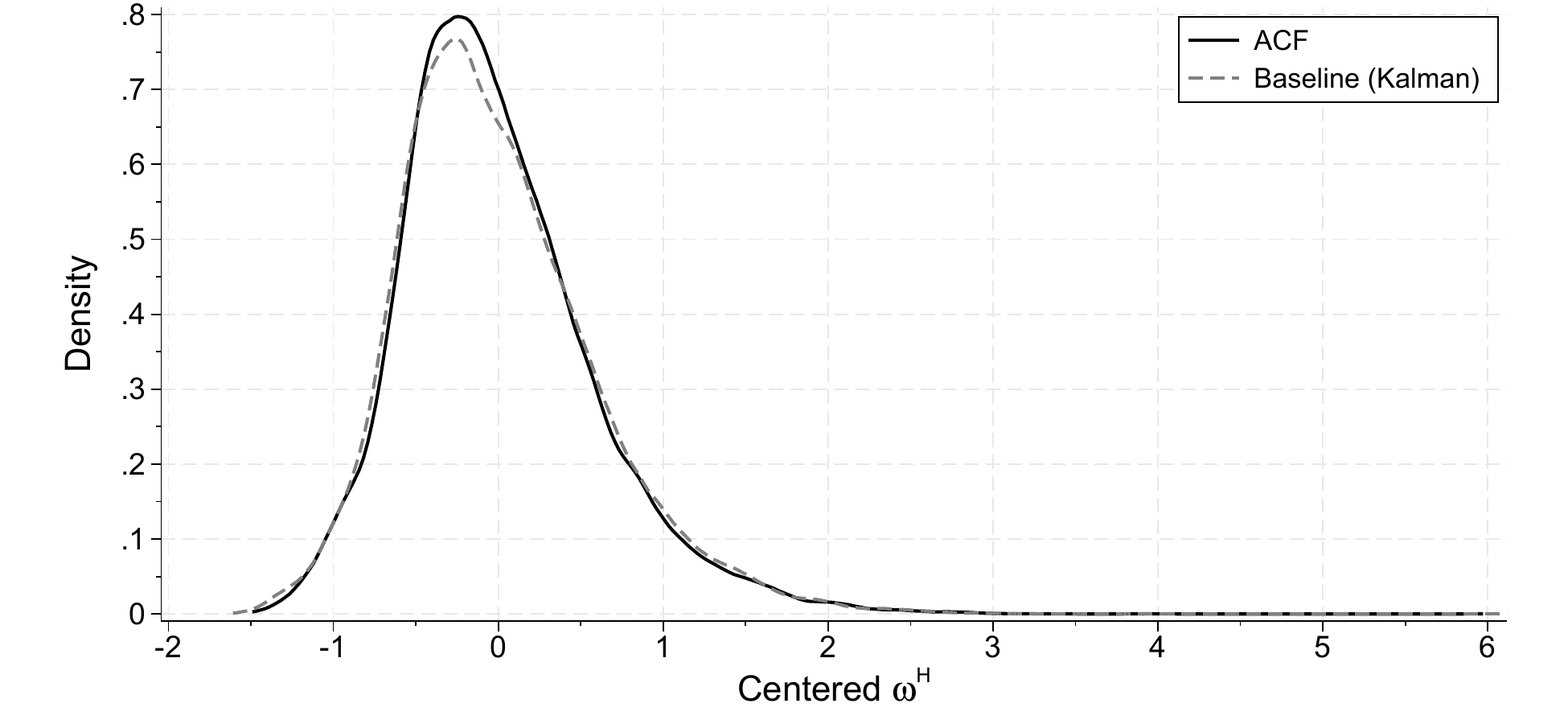}
\note{Kernel densities of the alternative ACF $\ddot{\omega}_H$ and the baseline Kalman-filtered $\ddot{\omega}_H$, on the subsample of plant-years with positive investment and gap-free panel coverage, the intersection on which both estimators are defined ($N = 30{,}552$). Because the baseline and ACF $\ddot{\omega}_H$ are normalized at different geometric means (the full panel and the $\ddot{I}_{jt}>0$ subsample respectively), the two series differ by a constant; we center each at its own mean so the comparison reflects shape only.}
\end{figure}

\begin{figure}[!ht]
\caption{Cross-Validation: Alternative ACF $\ddot{\omega}_H$ vs.\ Baseline (Kalman) $\ddot{\omega}_H$}
\label{fig:comparison}
\centering
\includegraphics[width=0.75\linewidth]{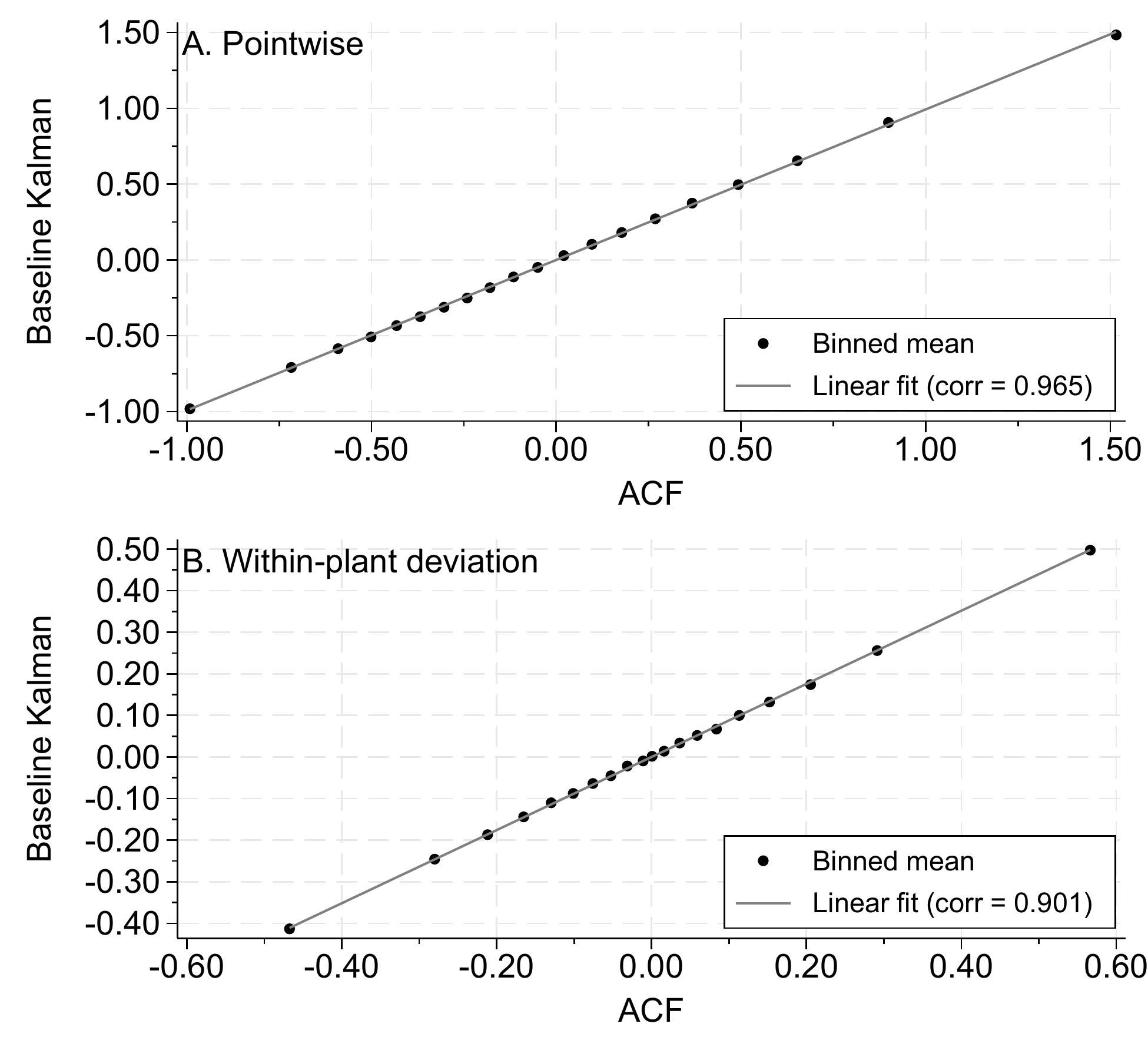}
\note{Binned-mean scatter of the alternative ACF $\ddot{\omega}_H$ (horizontal axis) against the baseline Kalman-filtered $\ddot{\omega}_H$ (vertical axis), on the subsample of plant-years with positive investment and gap-free panel coverage, the intersection on which both estimators are defined ($N = 30{,}552$). Panel A: pointwise alignment in levels. Panel B: within-plant alignment, residualizing both series by plant fixed effects. The grey line is the linear fit; the legend reports the corresponding correlation.}
\end{figure}

Figure~\ref{fig:comparison_density} overlays the kernel densities of the alternative ACF $\ddot{\omega}_H$ and the baseline Kalman-filtered $\ddot{\omega}_H$ restricted to the $\ddot{I}_{jt}>0$ subsample; the two have similar dispersion and shape. Figure~\ref{fig:comparison} reports the corresponding binned-mean scatter plots, with pointwise correlation $0.965$ and within-plant correlation $0.901$. Taken together, the figures show that the ACF and baseline $\ddot{\omega}_H$ series track each other closely both in distribution and at the plant-year level.

\subsection{Robustness: Supplementary Figures}\label{a:robustness_figures}

This appendix collects the figures for the unfiltered-productivity, matching-design, placebo, and import-liberalization checks discussed in Section~\ref{s:robustness}, together with the corresponding point estimates.

\begin{figure}[!ht]
\caption{Unfiltered $\ddot{\omega}_{H,jt}$}
\label{fig:unfiltered}
\centering
\begin{subfigure}[t]{0.49\textwidth}
\centering
\subcaption{Hicks-Neutral Productivity}
\includegraphics[width=\linewidth]{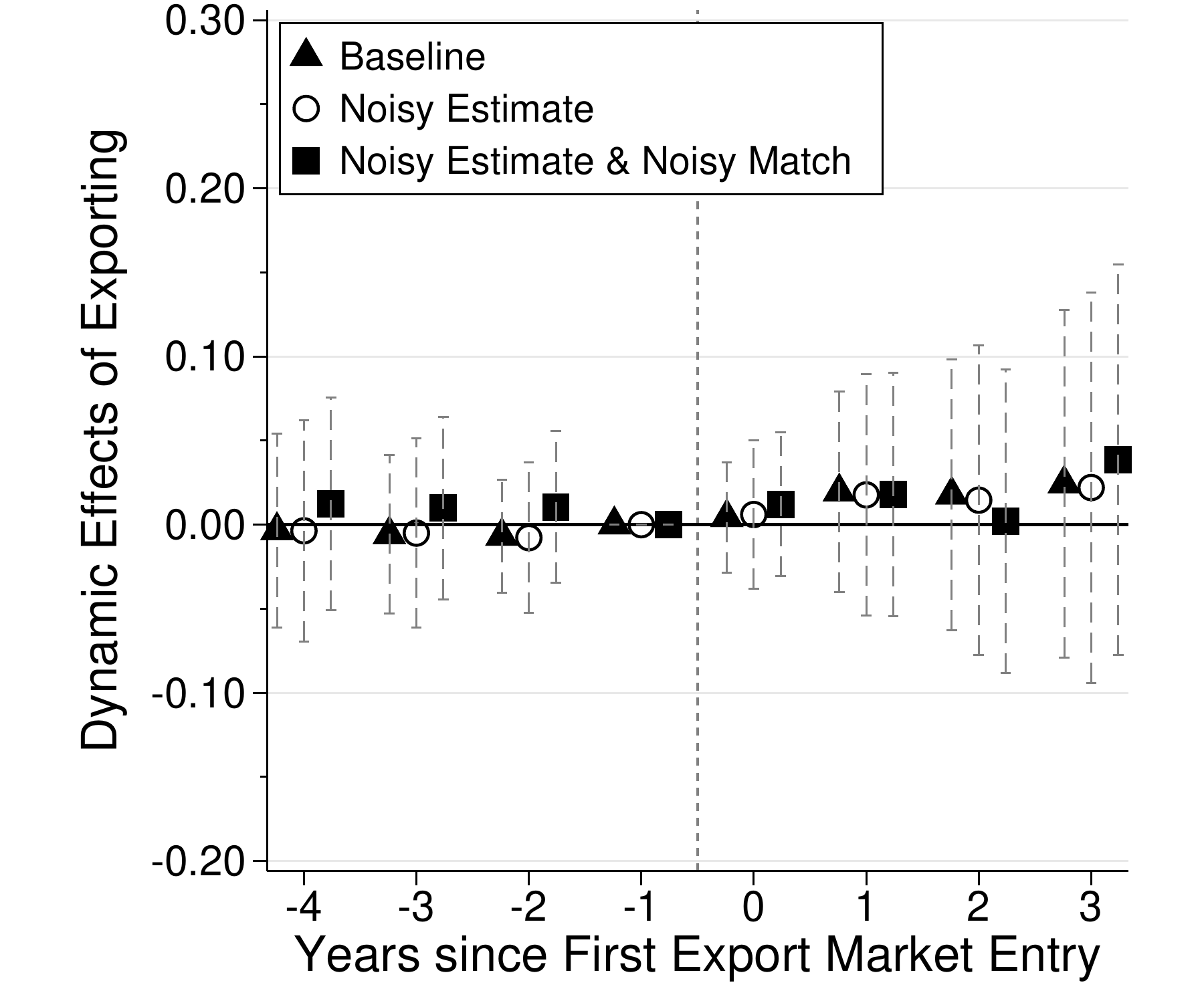}
\end{subfigure}\hfill
\begin{subfigure}[t]{0.49\textwidth}
\centering
\subcaption{TFP}
\includegraphics[width=\linewidth]{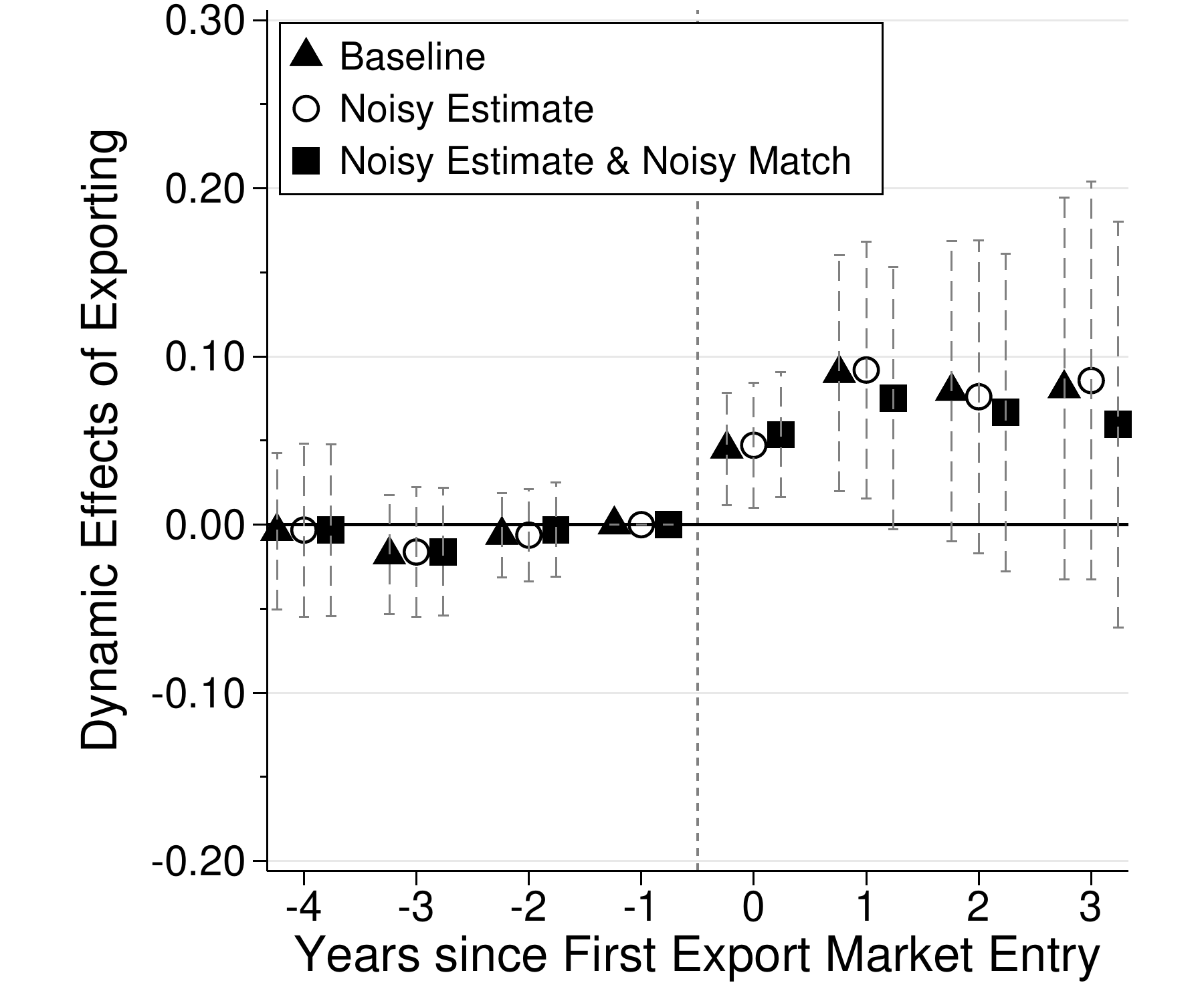}
\end{subfigure}
\note{This figure reports the local effects of exporting on Hicks-neutral productivity and TFP. It compares the baseline results (``Baseline'') with two alternative scenarios: one using unfiltered values of Hicks-neutral productivity with the same matched sample (``Noisy Estimate'') and another additionally matching plants based on unfiltered productivity values (``Noisy Estimate \& Noisy Match''). 90\% confidence intervals are constructed from 998 nonparametric bootstrap replications.}
\end{figure}

Figure~\ref{fig:unfiltered} reports the unfiltered-productivity comparison. Using the unfiltered measure as the outcome widens the bootstrap confidence intervals by roughly 20\% on average across horizons for the Hicks-neutral effects and by roughly 7\% for the TFP effects.

\begin{figure}[!ht]
\caption{Dynamic Effects of Exporting on Productivity across Matching Designs}
\label{fig:robustness_matching}
\centering
\includegraphics[width=0.95\linewidth]{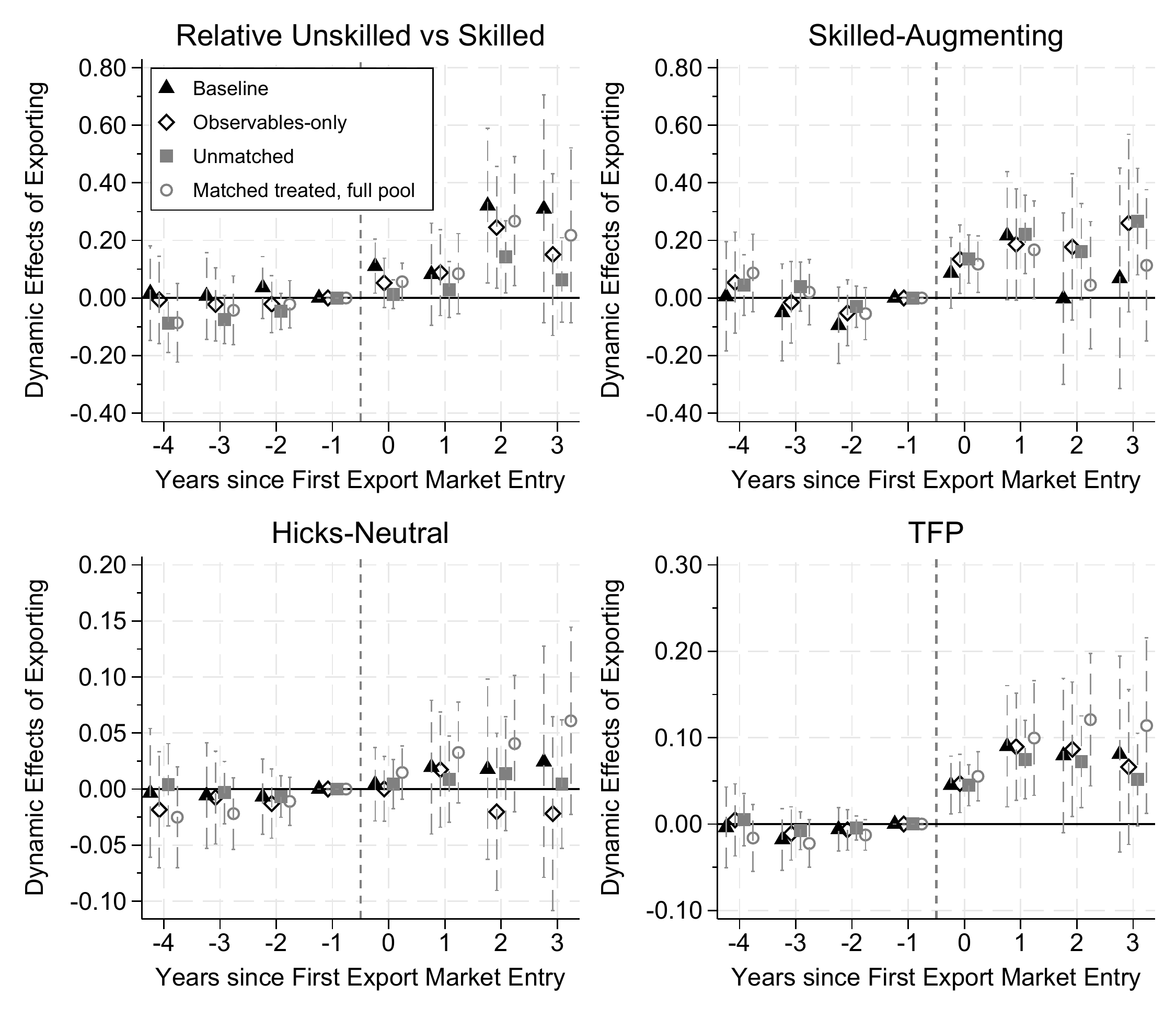}
\note{This figure reports cohort-stacked event-study estimates of the dynamic effects of exporting on four productivity components: relative unskilled-versus-skilled, skilled-augmenting, Hicks-neutral, and total factor productivity. Each panel plots four comparisons: the baseline propensity-score match on the model-implied covariates (baseline); an observables-only match that drops the three estimated productivity states (observables-only); the unmatched comparison against the full never-exporter pool (unmatched); and the baseline-matched new exporters set against the full never-exporter pool (matched treated, full pool), which separates the treated-composition and control-group channels. The omitted reference period is $h=-1$. Bars are 90\% confidence intervals from 998 nonparametric bootstrap replications.}
\end{figure}

Figure~\ref{fig:robustness_matching} reports the matching-design comparison. At $h=2$, the effect of export entry on relative-unskilled productivity is about $0.32$ under the baseline match, $0.25$ under the observables-only variant, and $0.14$ without matching. Setting the baseline-matched new exporters against the full never-exporter pool splits the gap between the baseline and unmatched estimates into about two-thirds from the composition of the treated group and one-third from the control group.

\begin{figure}[!ht]
\caption{Placebo Distributions from 500 Pseudo-Entry Draws}
\label{fig:placebo_hist}
\centering
\includegraphics[width=0.95\linewidth]{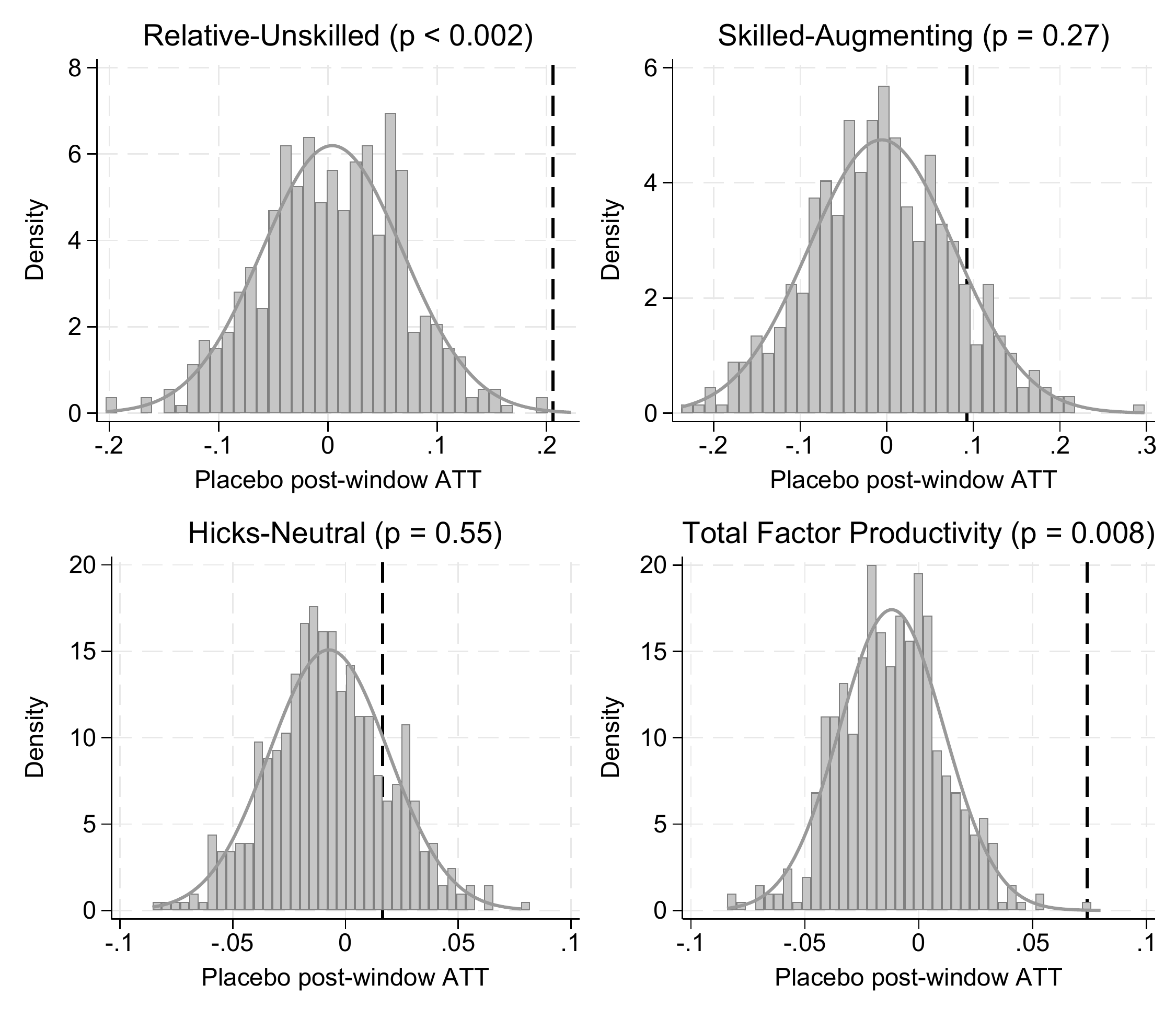}
\note{Each panel plots the distribution of the placebo effect across 500 pseudo-entry draws for one outcome. Each draw removes all new exporters from the sample, assigns pseudo entry years to never-exporters that replicate the true cohort-by-industry composition of entry, and re-runs the full procedure, from the propensity-score matching to the stacked regressions; the placebo effect is the average of the event-time coefficients over the post-entry window. The dashed vertical line marks the effect estimated in the data; the solid curve is a normal density with the mean and standard deviation of the placebo draws. The $p$-value in each panel title is the share of draws at least as large in magnitude as the estimate; $p < 0.002$ denotes zero exceedances among the 500 draws.}
\end{figure}

Figure~\ref{fig:placebo_hist} reports the placebo distributions. Across the 500 pseudo-entry draws, none produces an effect on relative-unskilled productivity, unskilled-augmenting productivity, or equipment as large in magnitude as the estimate in the data, so the randomization $p$-values fall below $0.002$. The estimated effect of export entry on relative-unskilled productivity lies 3.2 placebo standard deviations from zero. Four of the 500 draws match the magnitude of the estimated effect on TFP, giving a $p$-value of $0.008$. The $p$-values for the Hicks-neutral and skilled-augmenting effects are $0.55$ and $0.27$.

\begin{figure}[!ht]
\caption{Dynamic Effects of Exporting under the Import-Liberalization Checks}
\label{fig:apertura_compare}
\centering
\includegraphics[width=0.95\linewidth]{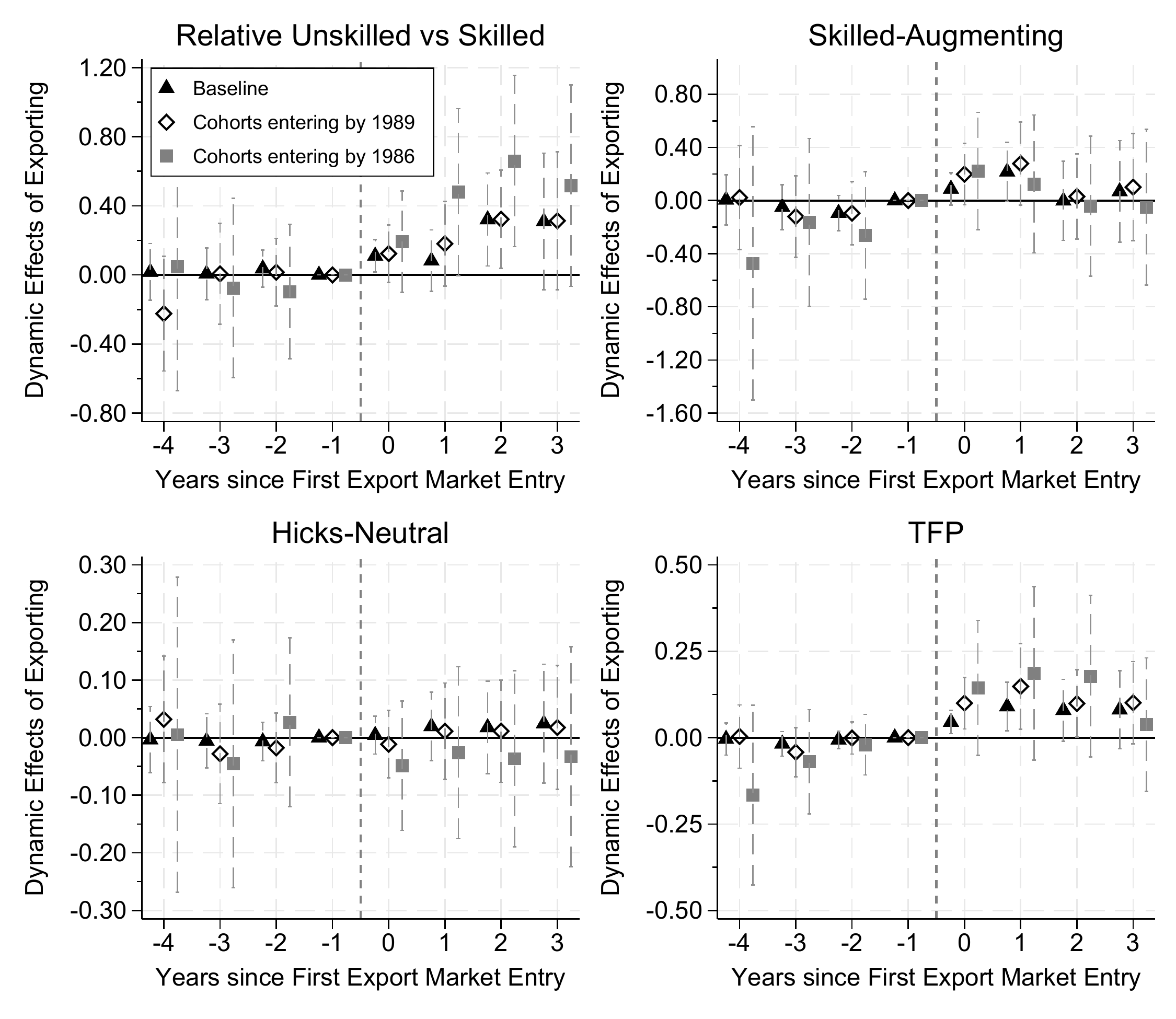}
\note{This figure reports cohort-stacked event-study estimates of the dynamic effects of exporting on relative-unskilled productivity, skilled-augmenting productivity, Hicks-neutral productivity, and total factor productivity under the baseline design and the two import-liberalization cohort restrictions: entry cohorts 1989 or earlier and entry cohorts 1986 or earlier. The omitted reference period is $h = -1$. Bars are 90\% confidence intervals from 998 nonparametric bootstrap replications.}
\end{figure}

Figure~\ref{fig:apertura_compare} reports the import-liberalization checks. The effect of exporting on relative-unskilled productivity averages $0.24$ over the post-entry window when the 1990 and 1991 entry cohorts are dropped and $0.46$ when only plants entering by 1986 are kept, against $0.21$ in the baseline.

\end{document}